\documentclass[10pt,psamsfonts]{amsart}
\usepackage{amsmath}
\usepackage{amsthm}
\usepackage{amssymb}
\usepackage{amscd}
\usepackage{amsfonts}
\usepackage{amsbsy}
\usepackage{booktabs}
\usepackage{color}
\usepackage{calc}
\usepackage{graphicx}
\usepackage{lineno}
\usepackage{subfigure}
\usepackage{setspace}
\usepackage{tikz}
\usetikzlibrary{arrows, patterns}

\advance\textwidth by +1.6in \advance\textheight by +1.0in
\advance\topmargin by -0.6in \advance\oddsidemargin by -0.8in
\advance\evensidemargin by -0.8in
\begin{document}
%\linenumbers
\title[]
{Global dynamics of the Ho\v{r}ava-Lifshitz cosmological model in a non-flat universe with non-zero cosmological constant}

\author[F. Gao and J. Llibre]{Fabao~Gao$^{1,2}$ and Jaume~Llibre$^{2}$}

\address{${}^1$School of Mathematical Science, Yangzhou University, Yangzhou 225002, China}
\address{\textnormal{E-mail: gaofabao@sina.com (Fabao Gao, ORCID 0000-0003-2933-1017)}}

\address{${}^2$ Departament de Matem$\grave{\text{a}}$tiques, Universitat Aut$\grave{\text{o}}$noma de Barcelona, Bellaterra 08193, Barcelona, Catalonia, Spain}
\address{\textnormal{E-mail: jllibre@mat.uab.cat (Jaume Llibre, ORCID 0000-0002-9511-5999)}}


\begin{abstract}
When the cosmological constant is non-zero the dynamics of the cosmological model based on Ho\v{r}ava-Lifshitz gravity in a non-flat universe is characterized by using the qualitative theory of differential equations.
\end{abstract}

\maketitle
\hrulefill
\tableofcontents
\hrulefill

%section 1
\section{Introduction}

After the Newtonian era, Einstein put forward the general theory of relativity, and this made our understanding of gravity once again a huge leap forward. In 2009 Ho\v{r}ava proposed a non-relativistic theory of renormalizable gravity \cite{Horava}, which can be reduced to Einstein's general theory of relativity on a large scale. It is named Ho\v{r}ava-Lifshitz gravity together with the scalar field theory of Lifshitz. This theory has inspired many studies and applications in length renormalization \cite{Antoniadis2021}, entropy argument \cite{Bajardi2021}, cosmology \cite{Abreu}-\cite{Casalino2019}, dark energy \cite{Saridakis2010}-\cite{Sheykhi2019}, black holes \cite{Cai2009}-\cite{Poshteh2021}, gravitational waves \cite{Paez2021}, and electromagnetics \cite{Lin2011}-\cite{Restucciaa2020}. More information can also be found from the review articles \cite{Wang2017}-\cite{Sotiriou} and the references therein.

In the past ten years Leon et al. \cite{Leon2009}-\cite{Paliathanasis} have conducted several excellent studies on the Ho\v{r}ava-Lifshitz cosmological model whether the curvature $k$ of the universe is zero and whether the cosmological constant $\Lambda$ is considered. They divided the cosmological model into four types in the Friedmann-Lema\^{i}tre-Robertson-Walker (FLRW) background spacetime: (1) $\Lambda = 0, k = 0$; (2) $\Lambda\neq 0, k = 0$; (3) $\Lambda = 0, k\neq 0$; (4) $\Lambda\neq 0, k\neq 0$. By using the phase-space analysis, they either discussed the two-dimensional dynamics of the Ho\v{r}ava-Lifshitz cosmological model under the usual exponential potential, and partially studied its three-dimensional dynamics.

For the important cosmological constant $\Lambda$ that many researchers have been paying attention to, this constant put the Ho\v{r}ava-Lifshitz gravitational theory with detailed balance, which led to a conflict between its cosmology and observations. Appignani et al. \cite{Appignani2010} showed that the huge difference between the standard predictions from quantum field theory and the observed value of $\Lambda$ may have a solution in the Ho\v{r}ava-Lifshitz gravity framework. Akarsu et al. \cite{Akarsu2020} investigated the $\Lambda$ in the standard cold-dark matter model by introducing graduated dark energy. Their results provided a high probability that the sign of $\Lambda$ could be spontaneously converted, and inferred that the universe had transformed from anti-de Sitter vacua to de Sitter vacua and triggered late acceleration. Carlip \cite{Carlip} proposed that the vacuum fluctuations produce a huge $\Lambda$ and produce a high curvature $k$ on the Planck scale under the standard effective field theory. Although the debate about the shape of the universe has not yet reached an agreement or the boundaries of the universe are blurred, we have a 50$:$1 odds to conclude that the universe is closed if the Planck CMB data is correct \cite{Handley2019}. Besides, Valentino et al. \cite{Valentino} also believed that the universe can be a closed three-dimensional sphere compared with the prediction in the standard $\Lambda$ cold dark matter model. The curvature can be positive according to the enhanced lensing amplitude in the cosmic microwave background power spectrum confirmed by the Planck Legacy 2018.

For the case $\Lambda=0$ the global dynamics of the Ho\v{r}ava-Lifshitz scalar field cosmological model under the background of FLRW was described in \cite{Gao20191}-\cite{Gao20201}, and the case of $\Lambda\neq0$ with zero curvature has also been addressed in \cite{Gao20192}. In the present paper we will discuss the global dynamics of a non-flat universe with $\Lambda\neq0$. We will provide the detailed information on obtaining cosmological equations in section 2.

\section{The cosmological equations}

In this section we first briefly recall the classic Ho\v{r}ava-Lifshitz gravitational theory, where the field content can be represented by the space vector $N_i$ and scalar $N$, and they are common `lapse' and `shift' variables in general relativity \cite{Horava}, \cite{Leon2009}, \cite{Kiritsis}. Then the full metric can be defined as
\begin{equation}
\begin{array}{rl}
ds^2 = -N^2dt^2 + g_{ij}(dx^i + N^idt)(dx^j + N^jdt),\ \ \ N_i=g_{ij}N^j,
\end{array}
\end{equation}
where $g_{ij}\ (i,j=1,2,3)$ is a spatial metric. The rescaling conversion meets the conditions $t\to l^3t,\ x^i\to lx^i$,
under which $g_{ij}$ and $N$ remain unchanged, but $N^i$ is scaled to $N^i\to l^{-2}N_i$.

Under the detailed-balance condition, the full gravitational action of Ho\v{r}ava-Lifshitz is written as
\begin{equation}
\begin{array}{rl} \vspace{2mm}
 S_g=&\displaystyle\int dtd^3x\sqrt{g}N\left\{\dfrac{2}{\kappa^2}\left(K_{ij}K^{ij}-\lambda K^2\right)-\dfrac{\kappa^2}{2w^4}C_{ij}C^{ij}\right.\\\vspace{2mm}
& \ \ \ +\dfrac{\mu\kappa^2}{2w^2}\dfrac{\epsilon^{ijm}}{\sqrt{g}}R_{il}\nabla_jR^l_k-\dfrac{\mu^2\kappa^2}{8}R_{ij}R^{ij}\\
& \ \ \ -\left.\dfrac{\mu^2\kappa^2}{8(3\lambda-1)}\left(\dfrac{1-4\lambda}{4}R^2+\Lambda R-3\Lambda^2\right)\right\},
\end{array}
\end{equation}
where $K_{ij}=(\dot{g}_{ij}-\nabla_iN_j-\nabla_jN_i)/(2N)$ denotes the extrinsic curvature, $C^{ij}=\epsilon^{ijm}\nabla_k\left(4R^j_i-R\delta^j_i\right)/(4\sqrt{g})$ represents the Cotton tensor, and $\epsilon^{ijm}/\sqrt{g}$ is the standard general covariant antisymmetric tensor, the indices are to raise and lower with the metric $g_{ij}$. The parameters $\lambda$, $\mu$ and $w$ are constants (see \cite{Horava} for more details).

For the potential $V(\phi)$ we take into account the gravitational action term as follows
\begin{equation}
\begin{array}{rl} \vspace{2mm}
 S=\displaystyle\int dtd^3x\sqrt{g}N\left(\dfrac{3\lambda-1}{4}\dfrac{\dot{\phi}^2}{N^2}-V(\phi)\right),
\end{array}
\end{equation}
and the metric $N^i=0$, $g_{ij}=a^2(t)\gamma_{ij}$, $\gamma_{ij}dx^idx^j=r^2d\Omega^2_2+dr^2/(1-kr^2)$. Here the function $a(t)$ is the dimensionless rescaling factor of the expanding universe, and $\gamma_{ij}$ is a constant curvature metric of maximally symmetric. Without loss of generality we normalize $\kappa^2$ and $N$ to the number one, and then we can describe the cosmological model as
\begin{equation}
\begin{array}{rl} \vspace{2mm}
&H^2=\dfrac{\dot{\phi}^2}{24}+\dfrac{V(\phi)}{6(3\lambda-1)}-\dfrac{1}{16(3\lambda-1)^2}\left[\dfrac{\mu^2k^2}{a^4}+\mu^2\Lambda^2-\dfrac{2\mu^2\Lambda k}{a^2}\right],\\ \vspace{2mm}
&\dot{H}+\dfrac{3}{2}H^2=-\dfrac{\dot{\phi}^2}{16}+\dfrac{V(\phi)}{4(3\lambda-1)}+\dfrac{1}{32(3\lambda-1)^2}\left[\dfrac{\mu^2k^2}{a^4}-3\mu^2\Lambda^2-\dfrac{2\mu^2\Lambda k}{a^2}\right],\\ \vspace{2mm}
&\ddot{\phi}+3H\dot{\phi}+\dfrac{2V'(\phi)}{3\lambda-1}=0,
\end{array}
\label{5}
\end{equation}
where $H=\dot{a}(t)/a(t)$ is the Hubble parameter which gives the expansion rate of the universe.

According to \cite{Leon2009}-\cite{Paliathanasis} we perform the dimensionless transformation
\begin{equation}
\begin{array}{rl}
x=\dfrac{\dot{\phi}}{2\sqrt{6}H},\ y=\dfrac{\sqrt{V(\phi)}}{\sqrt{6}H\sqrt{3\lambda-1}},\ z=\dfrac{\mu}{4(3\lambda-1)a^2H},\ u=\dfrac{\Lambda\mu}{4(3\lambda-1)H}.
\end{array}
\label{6}
\end{equation}
Then it can be followed from equations (\ref{5}) and (\ref{6}) that
\begin{equation}
\begin{array}{rl}
x^2+y^2-(u-kz)^2=1,\ \dfrac{H'}{H}=2z(z-u)-3x^2.
\end{array}
\label{7}
\end{equation}
Therefore the field equations become the form of the following autonomous dynamical system
\begin{equation}
\begin{array}{rl} \vspace{2mm}
\dfrac{dx}{dt}&=\sqrt{6} s \left[-x^2 + (u - z)^2 + 1\right] + x \left[3 x^2 + 2 (u - z) z - 3\right],\\ \vspace{2mm}
\dfrac{dz}{dt}&=z \left[3 x^2 + 2 (u - z) z - 2\right] ,\\ \vspace{2mm}
\dfrac{du}{dt}&=u \left[3 x^2 + 2 (u - z) z\right],
\end{array}
\label{1}
\end{equation}
where $s=-\dfrac{1}{\kappa V(\phi)}\dfrac{dV(\phi)}{d\phi}$, $\kappa$ is a constant, $V(\phi)$ is the usual scalar field potential, which admits various mathematical representations (see \cite{Leon2019,Fadragas2014,Escobar2014,Alho}), it can even be presented in a constant form in scalar cosmology (\cite{Kim2013,Chervon2019}), and $s$ is assumed to be a constant under the usual exponential potentials. For more details on system (\ref{1}) see the equation (5.72) of \cite{Leon2012}, or equations (287)-(290) of \cite{Leon2019}, or equations (59)-(61) of \cite{Paliathanasis}.

In this paper we study the global dynamics of system (\ref{1}) in the physical region of interest 
$$G=\left\{(x,z,u): x^2-(u-kz)^2\leq1,u,z\in\mathbb R\right\},$$ 
where $k=1,\ 0,\ -1$ corresponding to closed, flat, and open universe, respectively. For the case $k=1$ we note that $f_+(x,z,u)=x^2-(u-z)^2-1=0$ is an invariant surface because there is a polynomial $\mathcal{P}=-2\sqrt{6} s x + 6 x^2 + 4 (u - z) z$ such that
$$\frac{\partial f_+}{\partial x}\cdot\dfrac{dx}{dt}+\frac{\partial f_+}{\partial z}\cdot\dfrac{dz}{dt}+\frac{\partial f_+}{\partial u}\cdot\dfrac{du}{dt}=\mathcal{P}f_+.$$ The invariant surface here is essential for understanding the complex dynamic behavior of the model (\ref{1}), because if a point on an orbit of system (\ref{1}) is located on the invariant surface, then the whole orbit is contained in the surface.
But $f_-(x,z,u)=x^2-(u+z)^2-1=0$ is not an invariant surface for the case $k=-1$. In addition it is also noted that system (\ref{1}) is invariant under the three symmetries $(x,z,u) \mapsto (x,-z,-u)$ and $(x,z,u)\mapsto(-x,-z,-u)$, $(x,z,u)\mapsto(-x,z,u)$ if $s = 0$, i.e. it is symmetric with respect to the $x$-axis and additionally with respect to the origin and the plane $x=0$ when $s = 0$. Therefore we divide the study of system (\ref{1}) into four cases taking into account the existence or not of the symmetric plane and of the invariant surface.\\
Case I: $s\neq0$ and $k=1$, so system (\ref{1}) is symmetric with respect to the $x$-axis and it has the invariant surface $f_+(x,z,u)=0$.\\
Case II: $s\neq0$ and $k=-1$, so system (\ref{1}) is symmetric with respect to the $x$-axis and it has not the invariant surface.\\
Case III: $s=0$ and $k=1$, so system (\ref{1}) is symmetric with respect to the origin and with respect to the $x$-axis, and it has the invariant surface  $f_+(x,z,u)=0$.\\
Case IV: $s=0$ and $k=-1$, so system (\ref{1}) is symmetric with respect to the origin and with respect to the $x$-axis, and it has not the invariant surface.

In section 3.1 we will investigate the phase portraits of case I of system (\ref{1}) on the invariant planes and surface, as well as the local phase portraits at the finite and infinite equilibrium points. In section 3.2 we will discuss the phase portraits of case I of system (\ref{1}) inside the Poincar\'{e} ball restricted to the region $G$. Based on these sections, considering the symmetry of system (\ref{1}), we will study the global dynamics of system (\ref{1}) adding its behavior at infinity in section 3.3. In section 4 we will study the case II in the same way as in case I. In sections 5 and 6 we will describe the global dynamics of system (\ref{1}) in the closed and open universe respectively when the field potential $V(\phi)$ of the system takes the form of constant, i.e. $s=0$ in cases III and IV. Moreover we will give the final discussion and summary in the last section 7.

\section{Case I: $s \neq 0, k=1$}

\subsection{Phase portraits on the invariant planes and surface}

In order to analyze in detail the local phase portraits at the finite and infinite equilibrium points and the global phase portrait of system (\ref{1}) in the region $G$  (refer to \cite{Leon2012, Leon2019, Leon2009} or \cite{Paliathanasis} again), we first study its phase portraits on the invariant planes $z=0$ and $u=0$, as well as on the invariant surface $x^2-(u-z)^2=1$, respectively.
%subsection 3.1.1
\subsubsection{The invariant plane $z=0$}
\par On this plane system (\ref{1}) becomes
\begin{equation}
\begin{array}{rl}\vspace{2mm}
\dfrac{dx}{dt}&=\sqrt{6} s \left(-x^2 + u^2 + 1\right) + 3x \left( x^2 - 1\right),\\
\dfrac{du}{dt}&=3 u x^2.
\end{array}
\end{equation}
There are three equilibrium points $e_{1}=(1,0)$, $e_{2}=(-1,0)$ and $e_{3}=(2s/\sqrt{6},0)$ of system (8), where $e_{1}$ has eigenvalues 3 and $6-2\sqrt{6}s$, $e_{2}$ has eigenvalues 3 and $6+2\sqrt{6}s$, and $e_{3}$ has eigenvalues $2s^2$ and $2s^2-3$.
Therefore the equilibrium point $e_{1}$ is a hyperbolic unstable node when $s<\sqrt{6}/2$, and it is a hyperbolic unstable saddle when $s>\sqrt{6}/2$. The equilibrium point $e_{2}$ is a hyperbolic unstable node when $s>-\sqrt{6}/2$, and it is a hyperbolic unstable saddle when $s<-\sqrt{6}/2$. The equilibrium point $e_{3}$ is a hyperbolic unstable node when $|s|>\sqrt{6}/2$, and it is a hyperbolic unstable saddle when $|s|<\sqrt{6}/2$. Moreover for $s=-\sqrt{6}/2$, $e_{2}=e_{3}$ is a semi-hyperbolic saddle-node, by using the semi-hyperbolic singular point theorem (see Theorem 2.19 in \cite{Dumortier} for more details). Similarly for $s=\sqrt{6}/2$, $e_{1}=e_{3}$ is also a semi-hyperbolic saddle-node.
\par Since the types and stability of these three finite equilibrium points vary with the different values of $s$, we summarize them in Table 1.
%Table 1
\begin{table}[!htb]
\newcommand{\tabincell}[2]{\begin{tabular}{@{}#1@{}}#2\end{tabular}}
\centering
\caption{\label{opt}Equilibrium points for the different values of $s$, where $e_{1}=(1,0)$, $e_{2}=(-1,0)$ and $e_{3}=(2s/\sqrt{6},0)$.}
\footnotesize
\rm
\centering
\begin{tabular}{@{}*{12}{l}}
\specialrule{0em}{2pt}{2pt}
 \toprule
\hspace{2mm}\textbf{Values of $s$}&\hspace{25mm}\textbf{Equilibrium points}\\
\specialrule{0em}{2pt}{2pt}
\toprule
\tabincell{l}{\hspace{2mm}$-\infty<s<-\dfrac{\sqrt{6}}{2}$}&\tabincell{l}{$e_{1}$ and $e_{3}$ are unstable nodes, $e_{2}$ is an unstable saddle}\hspace{2mm}\\
\specialrule{0em}{2pt}{2pt}
\hline
\specialrule{0em}{2pt}{2pt}
\tabincell{l}{\hspace{2mm}$s=-\dfrac{\sqrt{6}}{2}$}&\tabincell{l}{$e_{1}$ is an unstable node, $e_{2}=e_{3}$ is an unstable saddle-node}\hspace{2mm}\\
\specialrule{0em}{2pt}{2pt}
\hline
\specialrule{0em}{2pt}{2pt}
\tabincell{l}{\hspace{2mm}$-\dfrac{\sqrt{6}}{2}<s<\dfrac{\sqrt{6}}{2}$}&\tabincell{l}{$e_{1}$ and $e_{2}$ are unstable nodes, $e_{3}$ is an unstable saddle}\hspace{2mm}\\
\specialrule{0em}{2pt}{2pt}
\hline
\specialrule{0em}{2pt}{2pt}
\tabincell{l}{\hspace{2mm}$s=\dfrac{\sqrt{6}}{2}$}&\tabincell{l}{$e_{1}=e_{3}$ is an unstable saddle-node, $e_{2}$ is an unstable node}\hspace{2mm}\\
\specialrule{0em}{2pt}{2pt}
\hline
\specialrule{0em}{2pt}{2pt}
\tabincell{l}{\hspace{2mm}$\dfrac{\sqrt{6}}{2}<s<+\infty$}&\tabincell{l}{$e_{1}$ is an unstable saddle, $e_{2}$ and $e_{3}$ are unstable nodes}\hspace{2mm}\\
\specialrule{0em}{2pt}{2pt}
\toprule
\end{tabular}
\end{table}

\par By using the Poincar\'e compactification (see Chapter 5 of \cite{Dumortier} for more details), it will help us realize how to draw the vector field of system (8) in the local charts $U_1$ and $U_2$, and then we can determine how the orbits come from or go to infinity. On the local chart $U_1$ we let $x=1/V$ and $u=U/V$, then system (8) becomes
\begin{equation}
\begin{array}{rl}\vspace{2mm}
\dfrac{dU}{dt}&=-U V \left[-3 V + \sqrt{6} s (U^2 + V^2-1)\right],\\
\dfrac{dV}{dt}&=-V \left[3 - 3 V^2 + \sqrt{6} s V (U^2 + V^2-1)\right].
\end{array}
\end{equation}
Then all the points of system (9) at infinity $V=0$ are equilibrium points, we rescale the time $d\tau_1=Vdt$, so this system writes
\begin{equation}
\begin{array}{rl}\vspace{2mm}
\dfrac{dU}{d\tau_1}&=3 UV - \sqrt{6} s U(U^2 + V^2-1),\\
\dfrac{dV}{d\tau_1}&=-3 + 3 V^2 - \sqrt{6} s V (U^2 + V^2-1).
\end{array}
\end{equation}
However this system has no equilibrium points at infinity $V=0$.
\par Similarly on the local chart $U_2$ we have $x=U/V$ and $u=1/V$, then system (8) reduces to
\begin{equation}
\begin{array}{rl}\vspace{2mm}
\dfrac{dU}{dt}&=V \left[-3 U V + \sqrt{6} s (- U^2 + V^2+1)\right],\\
\dfrac{dV}{dt}&=-3 U^2 V.
\end{array}
\end{equation}

On the local chart $U_2$ we only need to study the origin $e_o = (0,0)$ of system (11). Obviously $e_o$ is an equilibrium point. Since its linear part is identically zero, we cannot use the usual eigenvalue method to determine the type of the equilibrium point and its local phase portrait, but we note that the straight line $V=0$ of system (11) is full of equilibrium points, and if the common factor $V$ in system (11) is eliminated, there are no other equilibrium points in $V=0$. When $s>0$, on the positive semi-axis of $V$, $dV/dt=-3U^2V<0$ indicates that $V$ decreases monotonically, and on the negative semi-axis of $V$, $dV/dt=-3U^2V>0$ means that $V$ increases monotonically. Near the straight line $U= 0$, $dU/dt=\sqrt {6}sV(1 + V^2) $, $U$ increases monotonically when $V>0$, and $U$ decreases monotonically when $V<0$. Therefore the local phase portrait of the equilibrium point $e_4$ of system (11) is shown in Figure 1(a) when $s>0$. Similarly the local phase portrait of $e_4$ is illustrated in Figure 1(b) when $s <0$.

%Figure 1
\begin{figure}[]
  \begin{minipage}{130mm}
\centering\subfigure[]{\label{fig:subfig:a}
    \includegraphics[width=6cm]{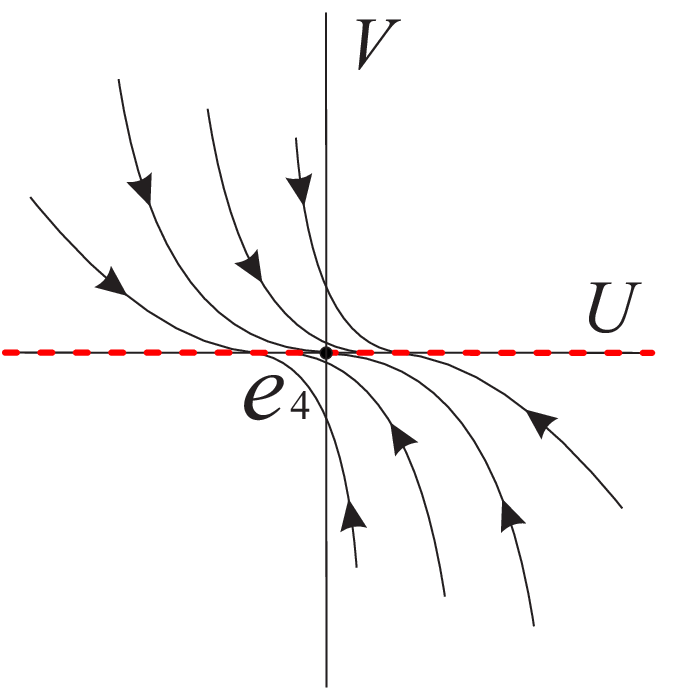}}
  \subfigure[]{\label{fig:subfig:b}
    \includegraphics[width=6cm]{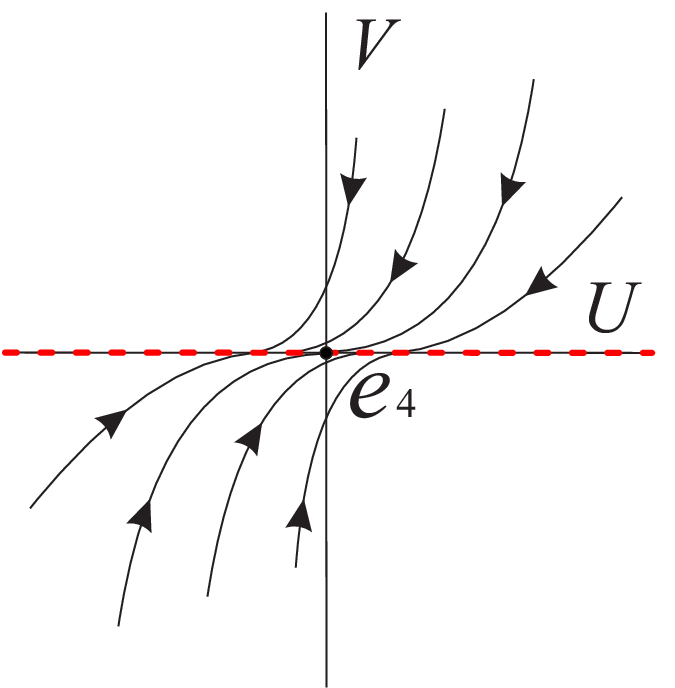}}
    \caption{The local phase portrait at the equilibrium point $e_o=(0,0)$ of system (11) in (a) when $s>0$, and in (b) when $s<0$.}
  \label{fig:subfig}
  \end{minipage}
 \end{figure}

Therefore the corresponding global phase portrait of system (8) restricted to the region $x^2-u^2\leq1$ can be summarized in Figure 2.

%Figure 2
\begin{figure}[htbp]
  \begin{minipage}{130mm}
\centering\subfigure[]{
    \includegraphics[width=4.1cm]{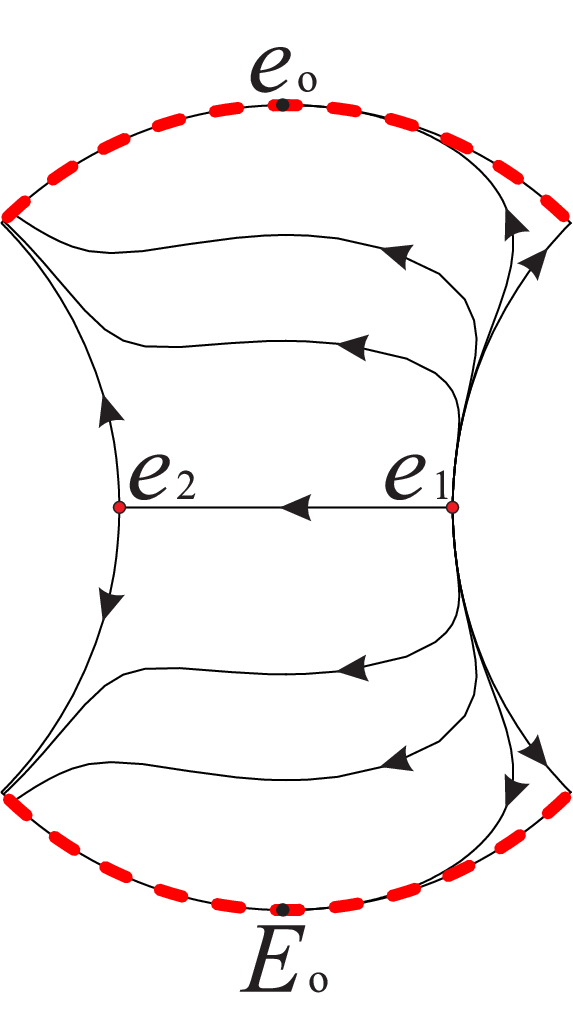}}
  \subfigure[]{
    \includegraphics[width=4.1cm]{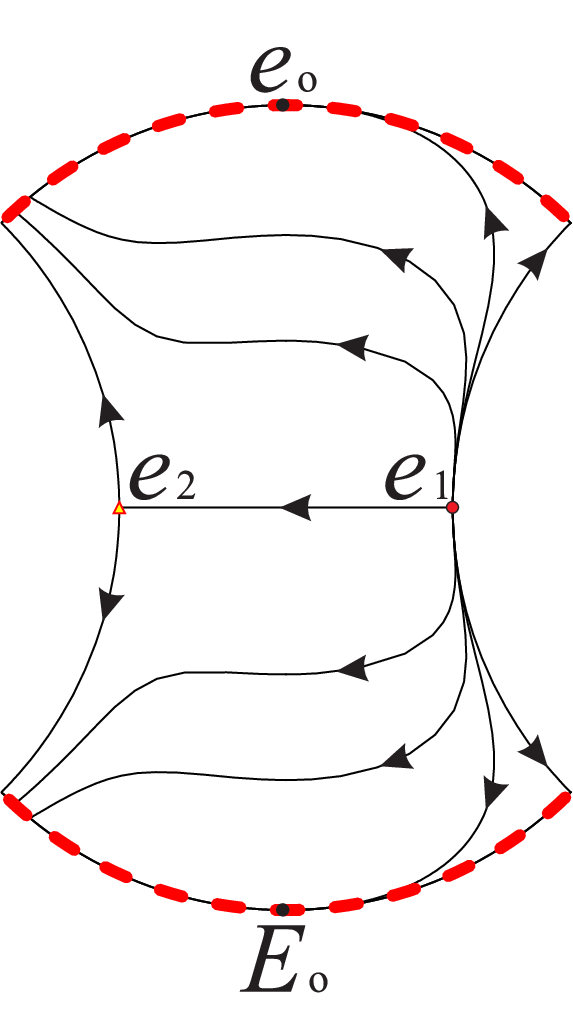}}
  \subfigure[]{
    \includegraphics[width=4.1cm]{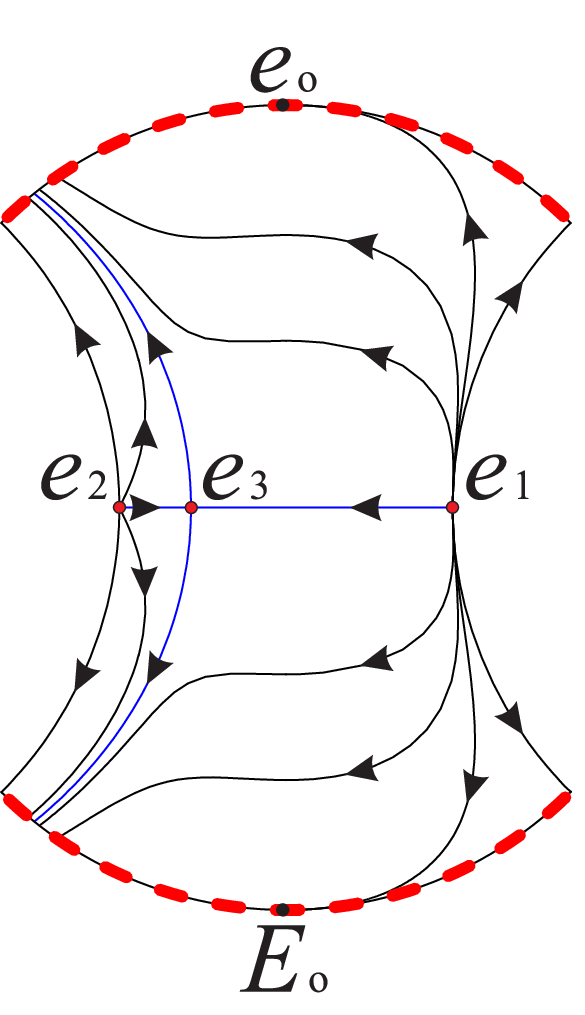}}\\
      \subfigure[]{
    \includegraphics[width=4.1cm]{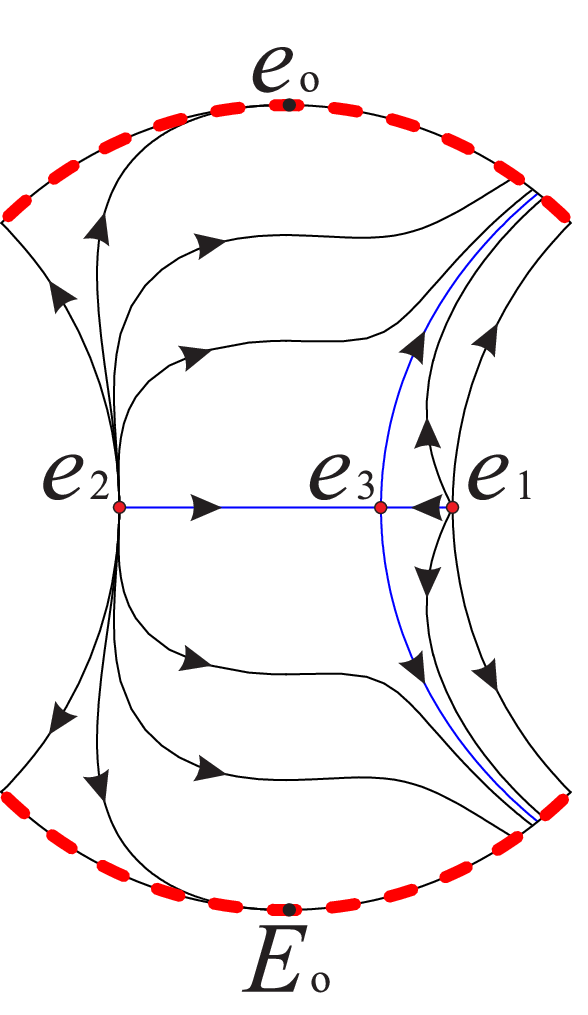}}
  \centering\subfigure[]{
    \includegraphics[width=4.1cm]{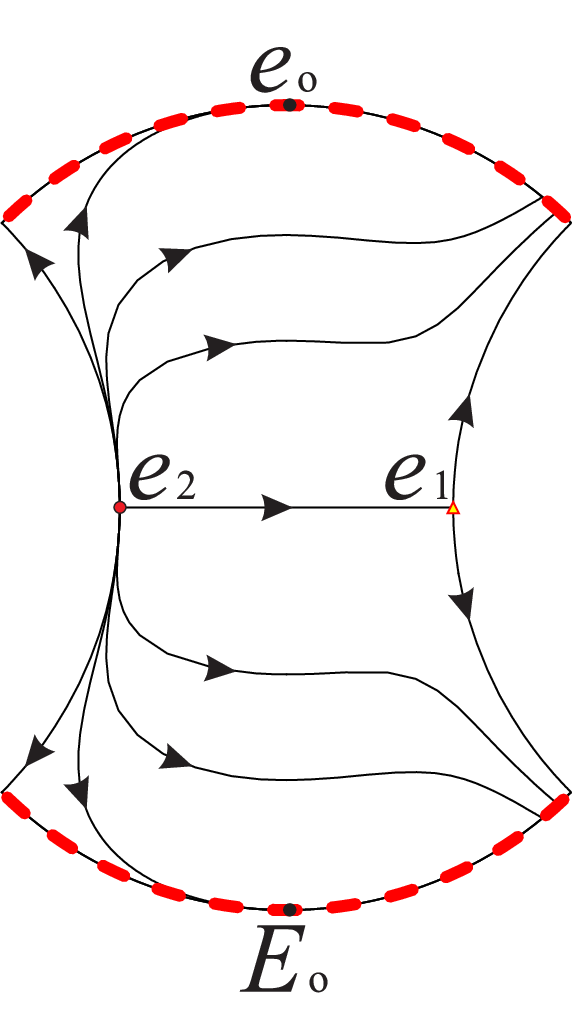}}
      \subfigure[]{
    \includegraphics[width=4.1cm]{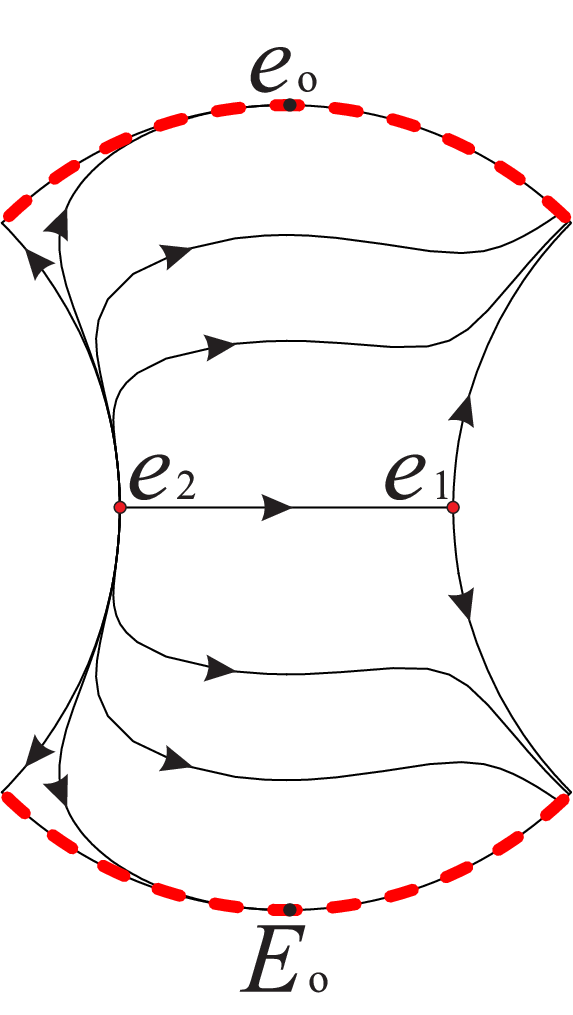}}
    \hspace{0cm}
     \subfigure{\includegraphics[width=2cm]{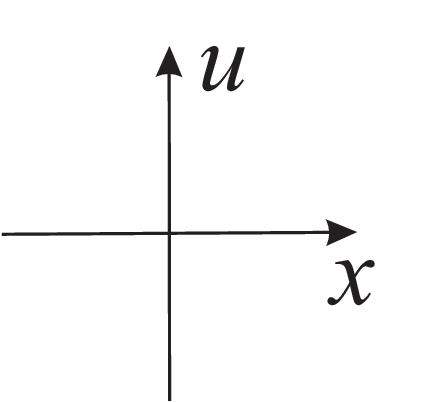}}
    \caption{The phase portrait on the invariant plane $z=0$ of system (8) restricted to the region $x^2-u^2\leq1$ inside the Poincar\'e disc for different values of $s$: (a) $s<-\sqrt{6}/2$, (b) $s=-\sqrt{6}/2$, (c) $-\sqrt{6}/2<s<0$, (d) $0<s<\sqrt{6}/2$, (e) $s=\sqrt{6}/2$, (f) $s>\sqrt{6}/2$. Here $E_o$ is the diametrically opposite equilibrium point of $e_o$ at infinity.}
  \label{fig:subfig}
  \end{minipage}
 \end{figure}

%subsection 3.1.2
\subsubsection{The invariant plane $u=0$}
On this plane system (\ref{1}) writes
\begin{equation}
\begin{array}{rl} \vspace{2mm}
\dfrac{dx}{dt}&=\sqrt{6} s \left(-x^2 + z^2 + 1\right) + x \left(3 x^2 - 2 z^2 - 3\right),\\ \vspace{2mm}
\dfrac{dz}{dt}&=z \left(3 x^2 - 2 z^2 - 2\right).
\end{array}
\label{u1}
\end{equation}

There are five equilibrium points $ e_1= (1,0) $, $ e_2 = (-1,0) $, $ e_3 = (2s / \sqrt{6}, 0) $, $ e_5 = (2 / (\sqrt{6} s),-\sqrt{1-s^2} / s) $ and $ e_6 = (2 / (\sqrt{6} s), \sqrt {1-s^2} / s) $ in system (12) when $ s ^ 2 \neq1 $, and the latter three equilibrium points $ e_3$, $ e_5 $ and $ e_6$ coincide when $ s^2 =1 $.
The equilibrium point $e_1$ has eigenvalues $1$ and $6-2 \sqrt{6}s$, it is a hyperbolic unstable node when $s <\sqrt {6} / 2$, and it is a hyperbolic unstable saddle when $s> \sqrt{6} / 2$.
The equilibrium point $e_2 $ has eigenvalues $1$ and $6+2\sqrt{6}s$, it is a hyperbolic unstable node when $ s>-\sqrt{6} / 2$, and it is a hyperbolic unstable saddle when $ s <-\sqrt{6} / 2$.
The equilibrium point $e_3$ has eigenvalues $2s^2-2$ and $2s^2-3$, it is a hyperbolic unstable node when $ | s | <1$ or $ | s |> \sqrt{6} / 2$, and it is a hyperbolic unstable saddle when $ 1 < | s | <\sqrt{6} / 2$.
The equilibrium points $ e_5 $ and $e_6$ have the same eigenvalues $-1/ 2- \sqrt {16-15s^2} / (2 | s |)$ and $ -1 / 2 + \sqrt{16-15s^2} / (2 | s |)$, they are hyperbolic unstable saddles when $ | s | <1$, and they are not real singularities when $ | s |> 1x$.

In addition, for the value $ s =-\sqrt{6} / 2 $, the equilibrium point $ e_2 = e_3$ is a semi-hyperbolic saddle-node by using the semi-hyperbolic singular point theorem. Similarly for the value $ s = \sqrt{6} / 2 $, the equilibrium point $ e_1= e_3$ is also a semi-hyperbolic saddle-node. For the value $ s = \pm1 $, the equilibrium point $ e_3 = e_5= e_6$ is a semi-hyperbolic unstable saddle.

Since the types and stability of these five finite equilibrium points change with different values of $ s $, we summarize them in Table 2.

%Table 2
\begin{table}[!htb]
\newcommand{\tabincell}[2]{\begin{tabular}{@{}#1@{}}#2\end{tabular}}
\centering
\caption{\label{opt}Equilibrium points for the different values of $s$, where $e_{1}=(1,0)$, $e_{2}=(-1,0)$, $e_{3}=(2s/\sqrt{6},0)$, $e_{5}=(2/(\sqrt{6}s),-\sqrt{1-s^2}/s)$ and $e_{6}=(2/(\sqrt{6}s),\sqrt{1-s^2}/s)$.}
\footnotesize
\rm
\centering
\begin{tabular}{@{}*{12}{l}}
\specialrule{0em}{2pt}{2pt}
 \toprule
\hspace{2mm}\textbf{Values of $s$}&\hspace{25mm}\textbf{Equilibrium points}\\
\specialrule{0em}{2pt}{2pt}
\toprule
\tabincell{l}{\hspace{2mm}$-\infty<s<-\dfrac{\sqrt{6}}{2}$}&\tabincell{l}{$e_{1}$ and $e_{3}$ are unstable nodes, $e_{2}$ is an unstable saddle}\hspace{2mm}\\
\specialrule{0em}{2pt}{2pt}
\hline
\specialrule{0em}{2pt}{2pt}
\tabincell{l}{\hspace{2mm}$s=-\dfrac{\sqrt{6}}{2}$}&\tabincell{l}{$e_{1}$ is an unstable node, $e_{2}=e_{3}$ is an unstable saddle-node}\hspace{2mm}\\
\specialrule{0em}{2pt}{2pt}
\hline
\specialrule{0em}{2pt}{2pt}
\tabincell{l}{\hspace{2mm}$-\dfrac{\sqrt{6}}{2}<s<-1$}&\tabincell{l}{$e_{1}$ and $e_{2}$ are unstable nodes, $e_{3}$ is an unstable saddle}\hspace{2mm}\\
\specialrule{0em}{2pt}{2pt}
\hline
\specialrule{0em}{2pt}{2pt}
\tabincell{l}{\hspace{2mm}$s=-1$}&\tabincell{l}{$e_{1}$ and $e_{2}$ are unstable nodes,\\ $e_{3}=e_{5}=e_{6}$ is a semi-hyperbolic unstable saddle}\hspace{2mm}\\
\specialrule{0em}{2pt}{2pt}
\hline
\specialrule{0em}{2pt}{2pt}
\tabincell{l}{\hspace{2mm}$-1<s<1$}&\tabincell{l}{$e_{1}$ and $e_{2}$ are unstable nodes, $e_{3}$ is stable node,\\ $e_{5}$ and $e_{6}$ are unstable saddles}\hspace{2mm}\\
\specialrule{0em}{2pt}{2pt}
\hline
\specialrule{0em}{2pt}{2pt}
\tabincell{l}{\hspace{2mm}$s=1$}&\tabincell{l}{$e_{1}$ and $e_{2}$ are unstable nodes,\\ $e_{3}=e_{5}=e_{6}$ is a semi-hyperbolic unstable saddle}\hspace{2mm}\\
\specialrule{0em}{2pt}{2pt}
\hline
\specialrule{0em}{2pt}{2pt}
\tabincell{l}{\hspace{2mm}$1<s<\dfrac{\sqrt{6}}{2}$}&\tabincell{l}{$e_{1}$ and $e_{2}$ are unstable nodes, $e_{3}$ is an unstable saddle}\hspace{2mm}\\
\specialrule{0em}{2pt}{2pt}
\hline
\specialrule{0em}{2pt}{2pt}
\tabincell{l}{\hspace{2mm}$s=\dfrac{\sqrt{6}}{2}$}&\tabincell{l}{$e_{1}=e_{3}$ is an unstable saddle-node, $e_{2}$ is an unstable node}\hspace{2mm}\\
\specialrule{0em}{2pt}{2pt}
\hline
\specialrule{0em}{2pt}{2pt}
\tabincell{l}{\hspace{2mm}$\dfrac{\sqrt{6}}{2}<s<+\infty$}&\tabincell{l}{$e_{1}$ is an unstable saddle, $e_{2}$ and $e_{3}$ are unstable nodes}\hspace{2mm}\\
\specialrule{0em}{2pt}{2pt}
\toprule
\end{tabular}
\end{table}

\par By using the Poincar\'e compactification again on the local chart $U_1$, then system (\ref{u1}) becomes
\begin{equation}
\begin{array}{rl}\vspace{2mm}
\dfrac{dU}{dt}&=-U V \left[-V + \sqrt{6} s ( U^2 + V^2-1)\right],\\
\dfrac{dV}{dt}&=-V \left[U^2 (\sqrt{6} s V-2) + ( \sqrt{6} s V-3) ( V^2-1)\right].
\end{array}
\label{u2}
\end{equation}
Hence all the infinite points of system (\ref{u2}) are filled with equilibrium points at $V=0$, doing the time rescaling $d\tau_2=Vdt$ yields
\begin{equation}
\begin{array}{rl}\vspace{2mm}
\dfrac{dU}{d\tau_2}&=UV - \sqrt{6} sU ( U^2 + V^2-1),\\
\dfrac{dV}{d\tau_2}&=-U^2 (\sqrt{6} s V-2) - ( \sqrt{6} s V-3) ( V^2-1).
\end{array}
\label{u3}
\end{equation}
However this system does not admit any equilibrium point on $V=0$.
\par Similarly on the local chart $U_2$ system (\ref{u1}) reduces to
\begin{equation}
\begin{array}{rl}\vspace{2mm}
\dfrac{dU}{dt}&=V \left[-U V + \sqrt{6} s (- U^2 + V^2+1)\right],\\
\dfrac{dV}{dt}&=V \left[ - 3 U^2 + 2 V^2+2\right].
\end{array}
\label{u4}
\end{equation}

The origin $e_4 = (0,0)$ of system (\ref{u4}) is a equilibrium point with eigenvalues 2 and 0, but it is not semi-hyperbolic because it is not isolated in the set of all equilibrium points. It is noted that the axis $V=0$ is full of equilibrium points of system (\ref{u4}). For the positive semi-axis of $V$ near $e_4$, $dV/dt>0$ means that $V$ increases monotonically, and on the negative semi-axis of $V$, $dV/dt<0$ indicates that $V$ decreases monotonically. Moreover $dU/dt=\sqrt{6} sV (V^2+1)$ around the straight line $U= 0$, thus $U$ increases monotonically when $sV>0$, and $U$ decreases monotonically when $sV<0$. Therefore the local phase portrait of the semi-hyperbolic equilibrium point $e_4$ in system (\ref{u4}) is illustrated in Figure 3(a) when $s>0$. Similarly the local phase portrait of $e_4$ is shown in Figure 3(b) when $s <0$.

%Figure 3
\begin{figure}[]
  \begin{minipage}{130mm}
\centering\subfigure[]{\label{fig:subfig:a}
    \includegraphics[width=6cm]{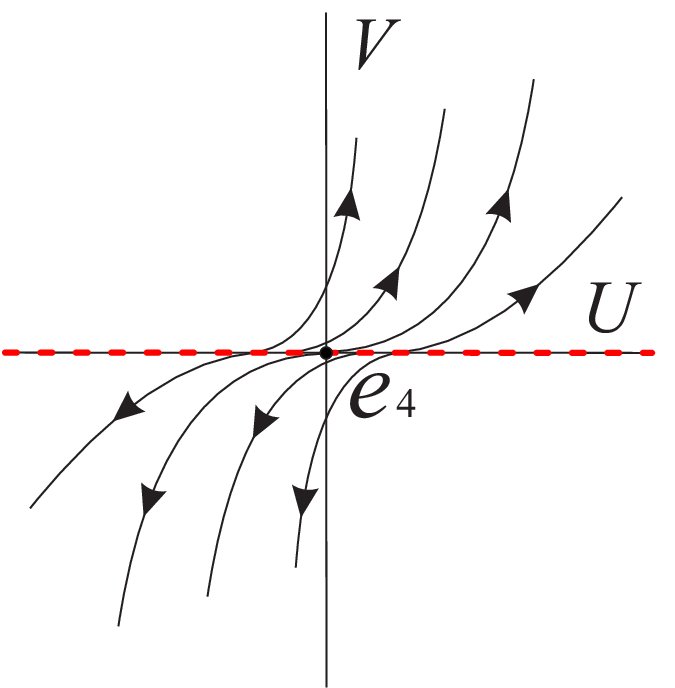}}
  \subfigure[]{\label{fig:subfig:b}
    \includegraphics[width=6cm]{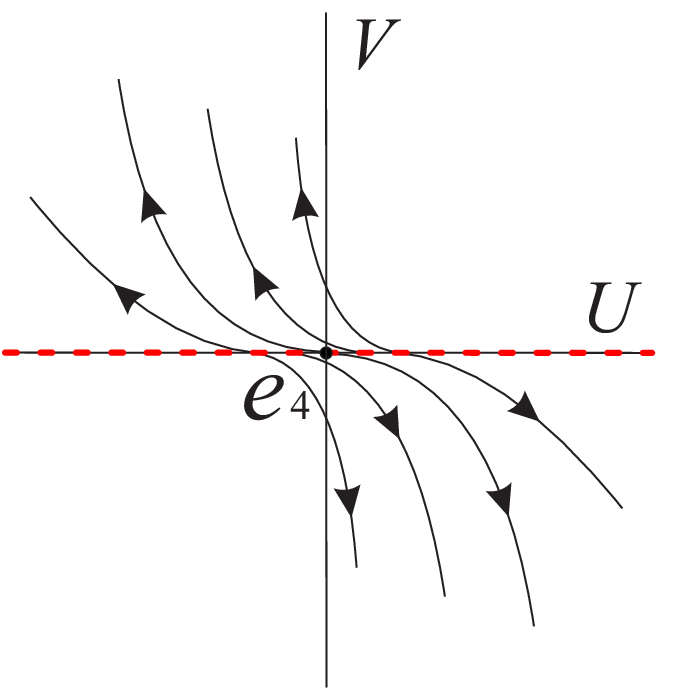}}
    \caption{The local phase portrait at the equilibrium point $e_4=(0,0)$ of system (\ref{u4}) in (a) when $s>0$, and in (b) when $s<0$.}
  \label{fig:subfig}
  \end{minipage}
 \end{figure}

Hence the corresponding global phase portrait of system (\ref{u1}) restricted to the region $x^2-z^2\leq1$ is illustrated in Figures 4 and 5. However we need to pay attention that the global phase portrait of system (\ref{u1}) is similar when $-\sqrt{6}/2<s<-1$ and $s=-1$ (See Figures 4(c) and 4(d)). The main difference is that the equilibrium point $e_3$ is a hyperbolic unstable saddle when $-\sqrt{6}/2<s<-1$, while $s=-1$ it is a semi-hyperbolic unstable saddle. The same situation occurs when $s=1$ and $1<s<\sqrt{6}/2$ (See Figures 5(b) and 5(c)).

%Figure 4
\begin{figure}[]
  \begin{minipage}{130mm}
\centering
\subfigure[]{\label{fig:subfig:a}
    \includegraphics[width=4.1cm]{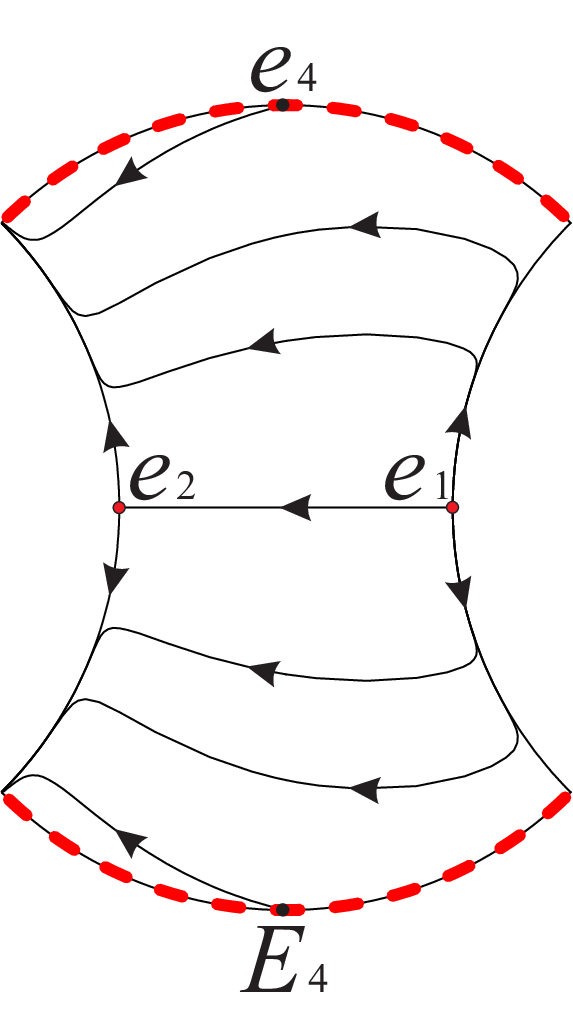}}
\subfigure[]{\label{fig:subfig:b}
    \includegraphics[width=4.1cm]{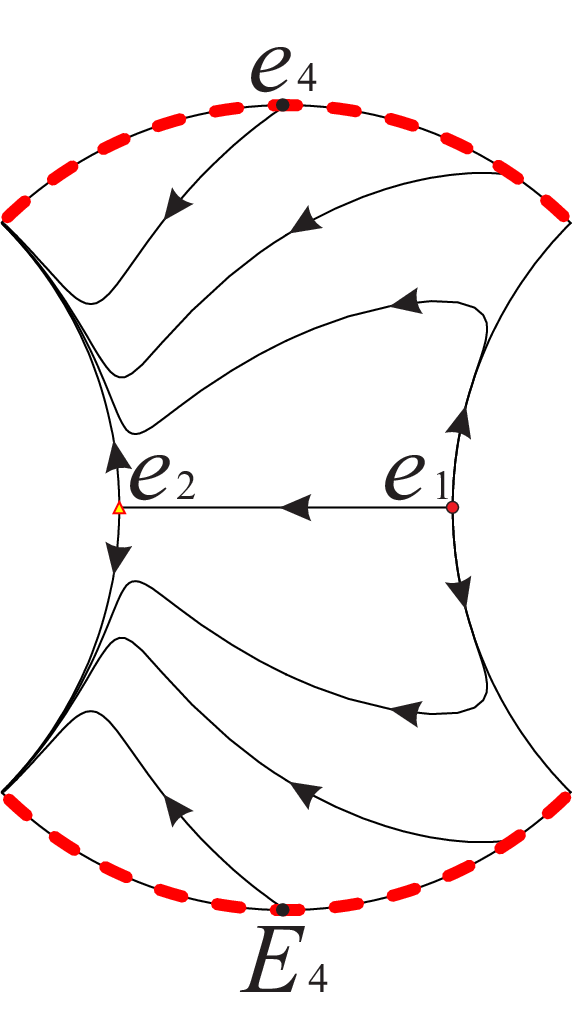}}
\subfigure[]{\label{fig:subfig:c}
    \includegraphics[width=4.1cm]{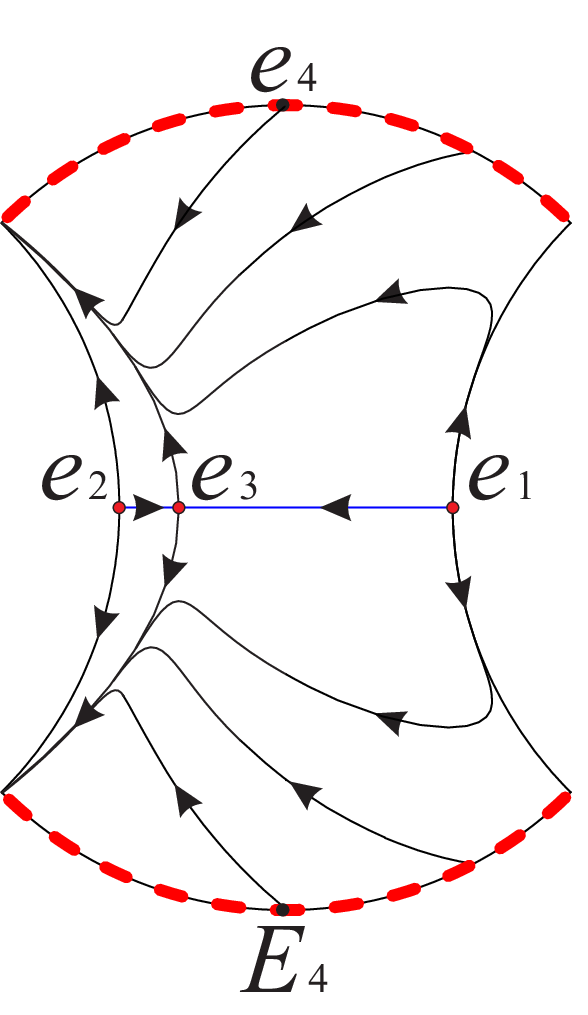}}\\
\subfigure[]{\label{fig:subfig:d}
    \includegraphics[width=4.1cm]{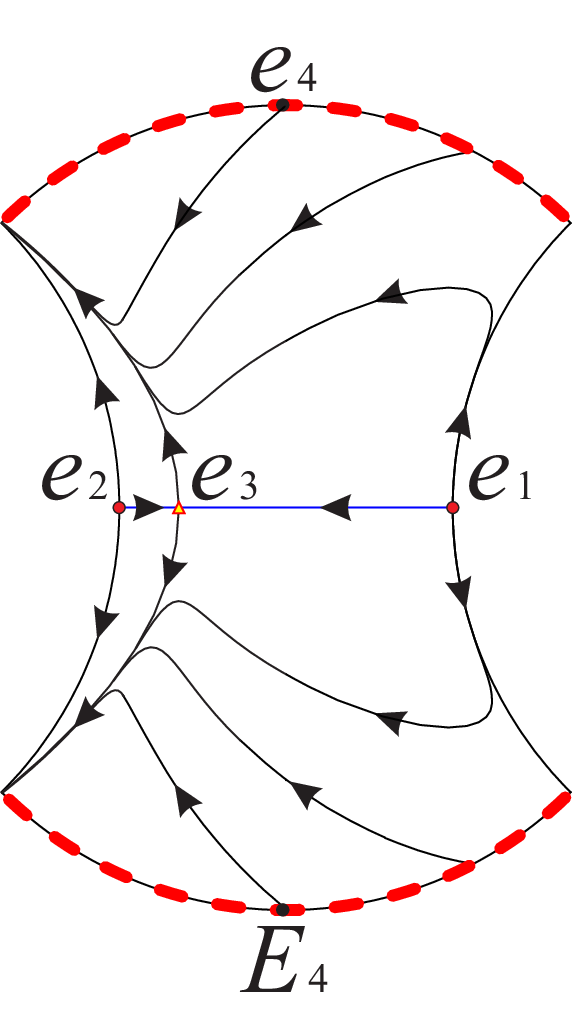}}
\subfigure[]{\label{fig:subfig:e}
    \includegraphics[width=4.05cm]{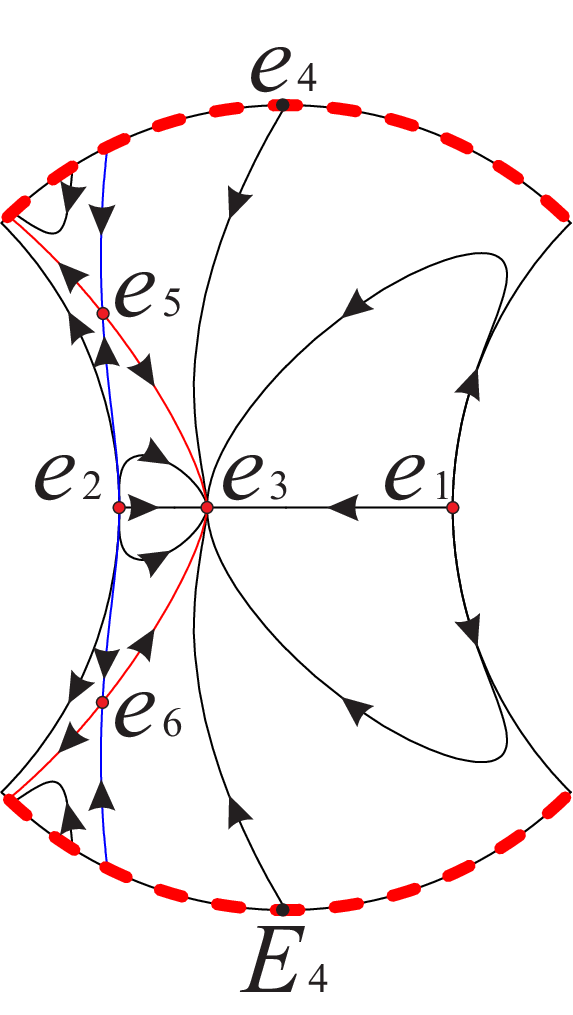}}
\subfigure{\includegraphics[width=2cm]{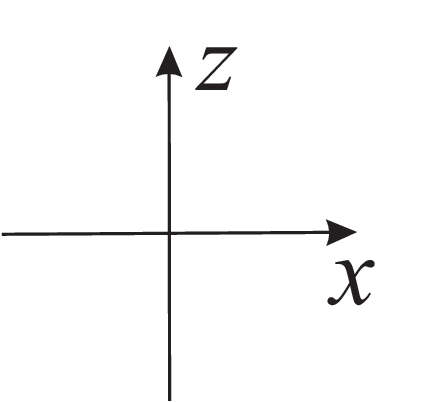}}
    \caption{The phase portrait on the invariant plane $u=0$ restricted to the region $x^2-z^2\leq1$ inside the Poincar\'e disc for different values of $s$: (a) $s<-\sqrt{6}/2$, (b) $s=-\sqrt{6}/2$, (c) $-\sqrt{6}/2<s<-1$, (d) $s=-1$, (e) $-1<s<0$. Here $E_4$ is the diametrically opposite equilibrium point of $e_4$ at infinity.}
  \label{fig:subfig}
  \end{minipage}
 \end{figure}

 %Figure 5
\begin{figure}[]
  \begin{minipage}{130mm}
\centering
\subfigure[]{\label{fig:subfig:a}
    \includegraphics[width=4.1cm]{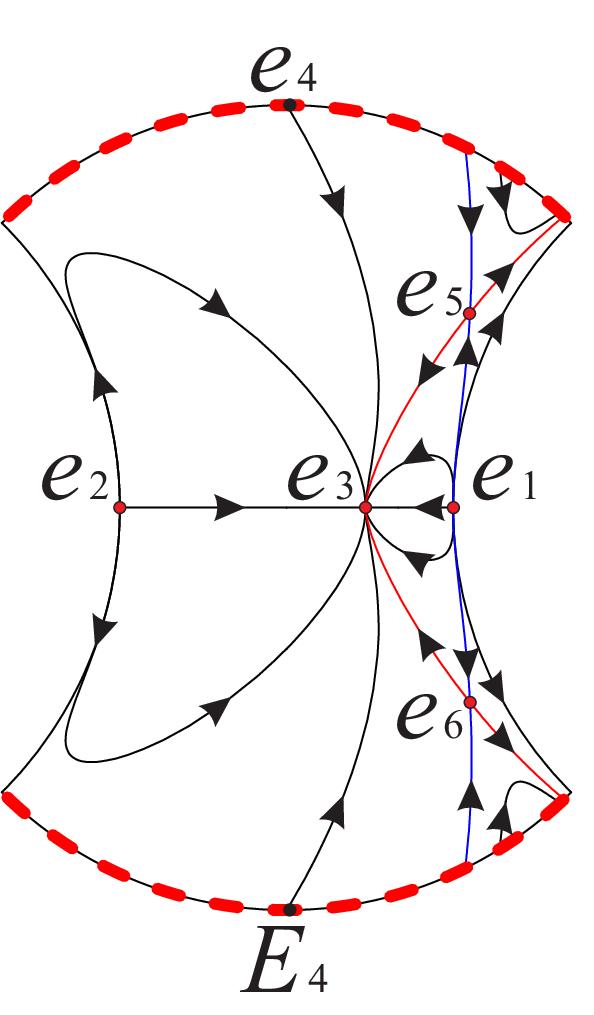}}
\subfigure[]{\label{fig:subfig:b}
    \includegraphics[width=4.1cm]{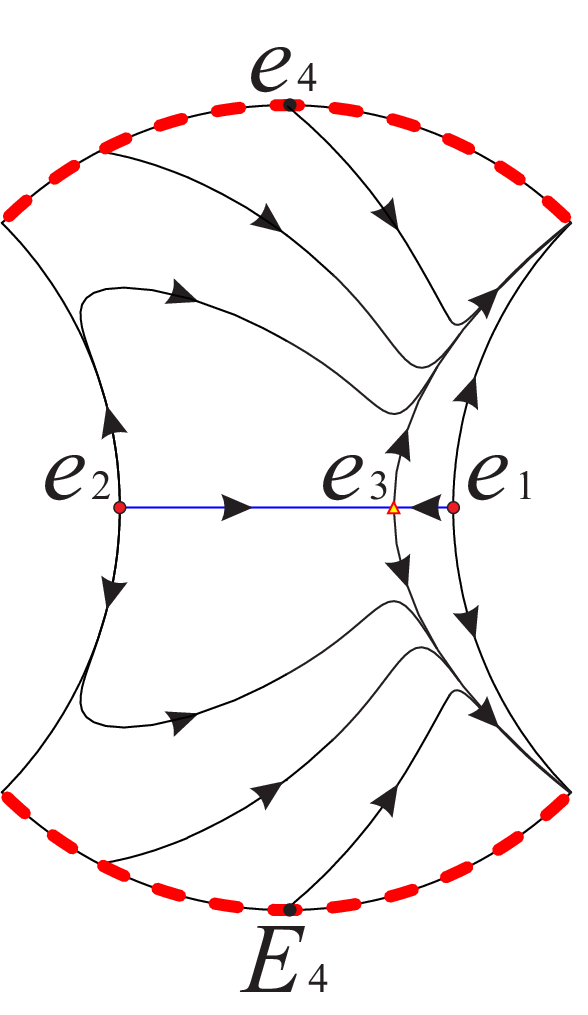}}
\subfigure[]{\label{fig:subfig:c}
    \includegraphics[width=4.1cm]{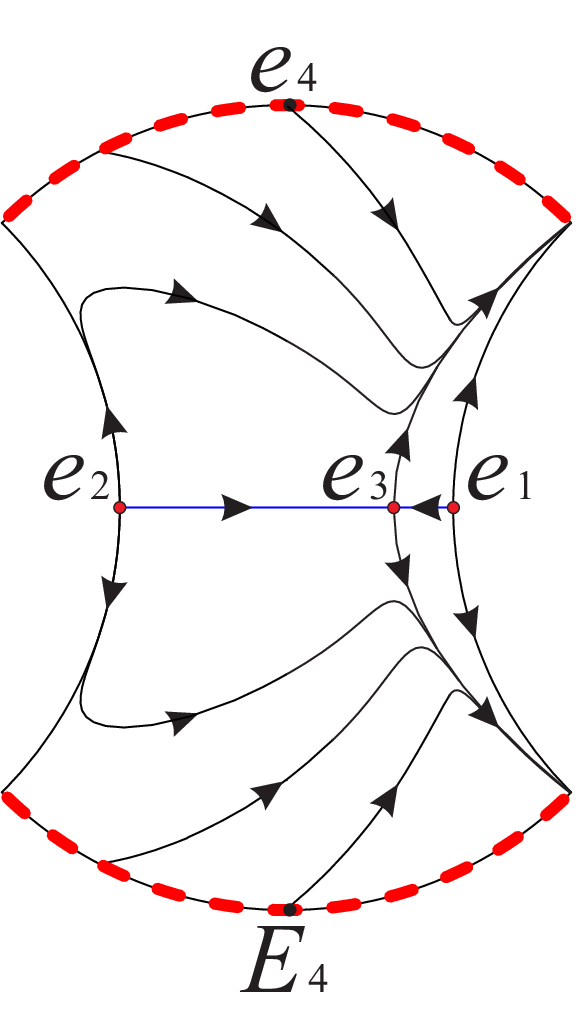}}\\
\subfigure[]{\label{fig:subfig:d}
    \includegraphics[width=4.1cm]{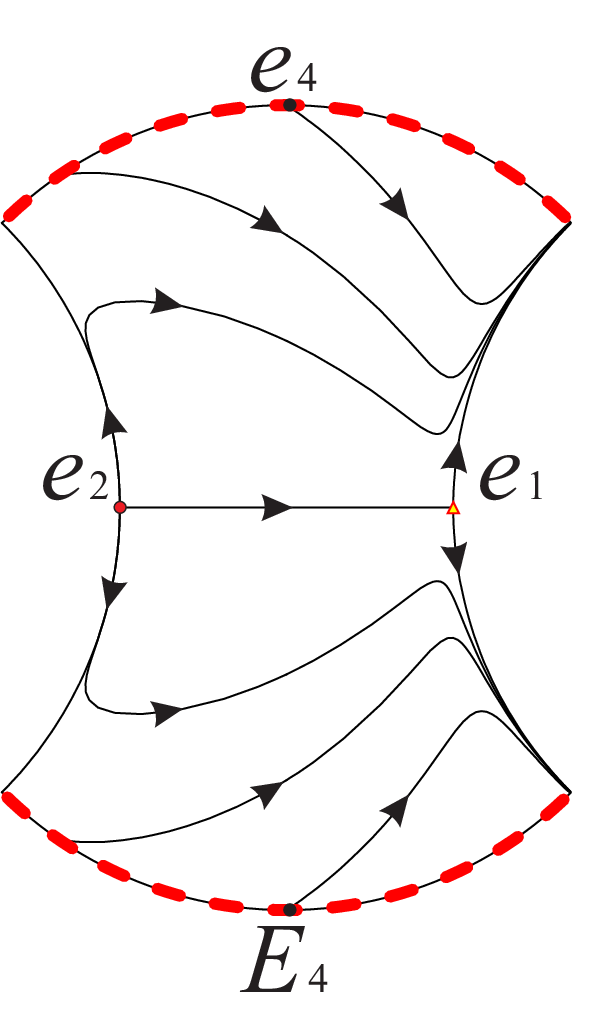}}
\subfigure[]{\label{fig:subfig:e}
    \includegraphics[width=3.9cm]{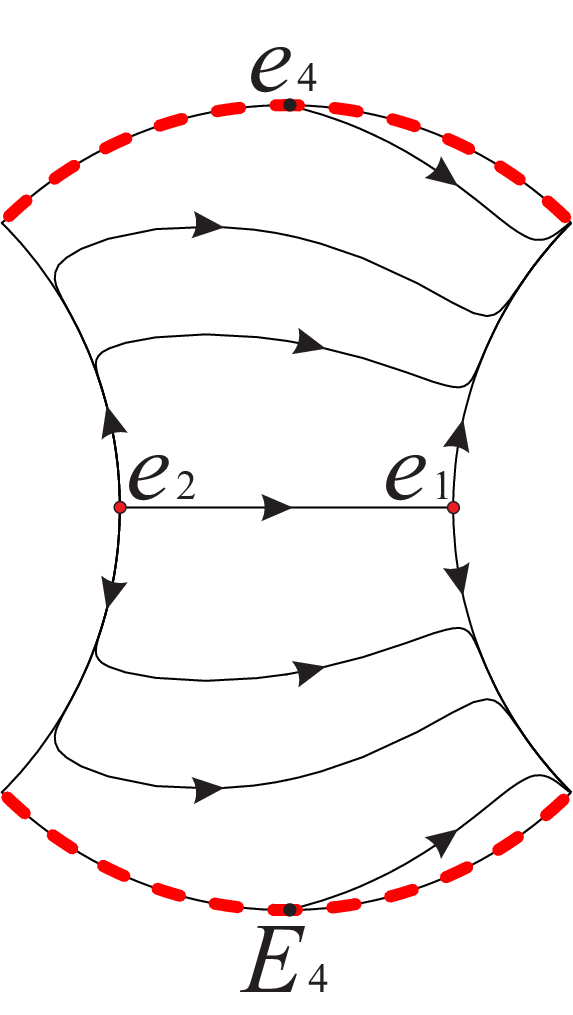}}
\subfigure{\includegraphics[width=2cm]{Fig4f5f29b_frame_x_z.eps}}
    \caption{The phase portrait on the invariant plane $u=0$ restricted to the region $x^2-z^2\leq1$ inside the Poincar\'e disc for different values of $s$: (a) $0<s<1$, (b) $s=1$, (c) $1<s<\sqrt{6}/2$, (d) $s=\sqrt{6}/2$, (e) $s>\sqrt{6}/2$.}
  \label{fig:subfig}
  \end{minipage}
 \end{figure}

%subsection 3.1.3
\subsubsection{The invariant surface $f_+(x,z,u)=0$}
\par On this surface system (\ref{1}) becomes
\begin{equation}
\begin{array}{rl}\vspace{2mm}
\dfrac{dx}{dt}&=x \left(x^2-1+2u\sqrt{x^2-1}\right),\\
\dfrac{du}{dt}&=u \left(x^2+2+2u\sqrt{x^2-1}\right),
\end{array}
\label{eq16}
\end{equation}
when $u\geq z$, and for the case $u\leq z$ system (\ref{1}) reads
\begin{equation}
\begin{array}{rl}\vspace{2mm}
\dfrac{dx}{dt}&=x \left(x^2-1-2u\sqrt{x^2-1}\right),\\
\dfrac{du}{dt}&=u \left(x^2+2-2u\sqrt{x^2-1}\right).
\end{array}
\label{eq17}
\end{equation}

Note that $|x|\geq1$, then for the case $u\geq z$, system (16) admits two equilibrium points $e_7=(1,0)$ and $e_8=(-1,0)$.
In addition we also note that system (16) is symmetric with respect to $u$-axis under the symmetry $(x, u)\mapsto (-x, u)$, so we only need to discuss the equilibrium point $e_7$. However the functions on the right side of system (16) have no derivative at $e_7$, which means that the local dynamics near $e_7$ cannot be studied by the methods in the previous sections 3.1.1 and 3.1.2.
When $x = 1$, the first equation in system (16) is always equal to zero, and the second equation is simplified to $du / dt = 3u$, i.e. on the invariant straight line $x=1$ the solution is $u(t) = ce^{3t}$ ($c$ is an arbitrary constant), which indicates that $u(t)$ tends to infinity in forward time and leads to $e_7$ in backward time.

We know the dynamics on $x=1$ near $e_7$, but we want to know the dynamics in a neighborhood of $e_7$. To this end we set $\sqrt{x^2-1}=y>0$ $(|x|>1)$ i.e. $x=\pm\sqrt{1+y^2}$ $(y>0)$, considering the aforementioned symmetry, we only discuss the case $x=\sqrt{1+y^2}$ $(y>0)$ here, then system (9) can be rewritten as follows
\begin{equation}
\begin{array}{rl}\vspace{2mm}
\dfrac{dy}{dt}&=(y+2u)\left(1+y^2\right),\\
\dfrac{du}{dt}&=u \left(3+2yu+y^2\right).
\end{array}
\end{equation}

\begin{figure}[]
\centering
\subfigure{\includegraphics[width=13cm]{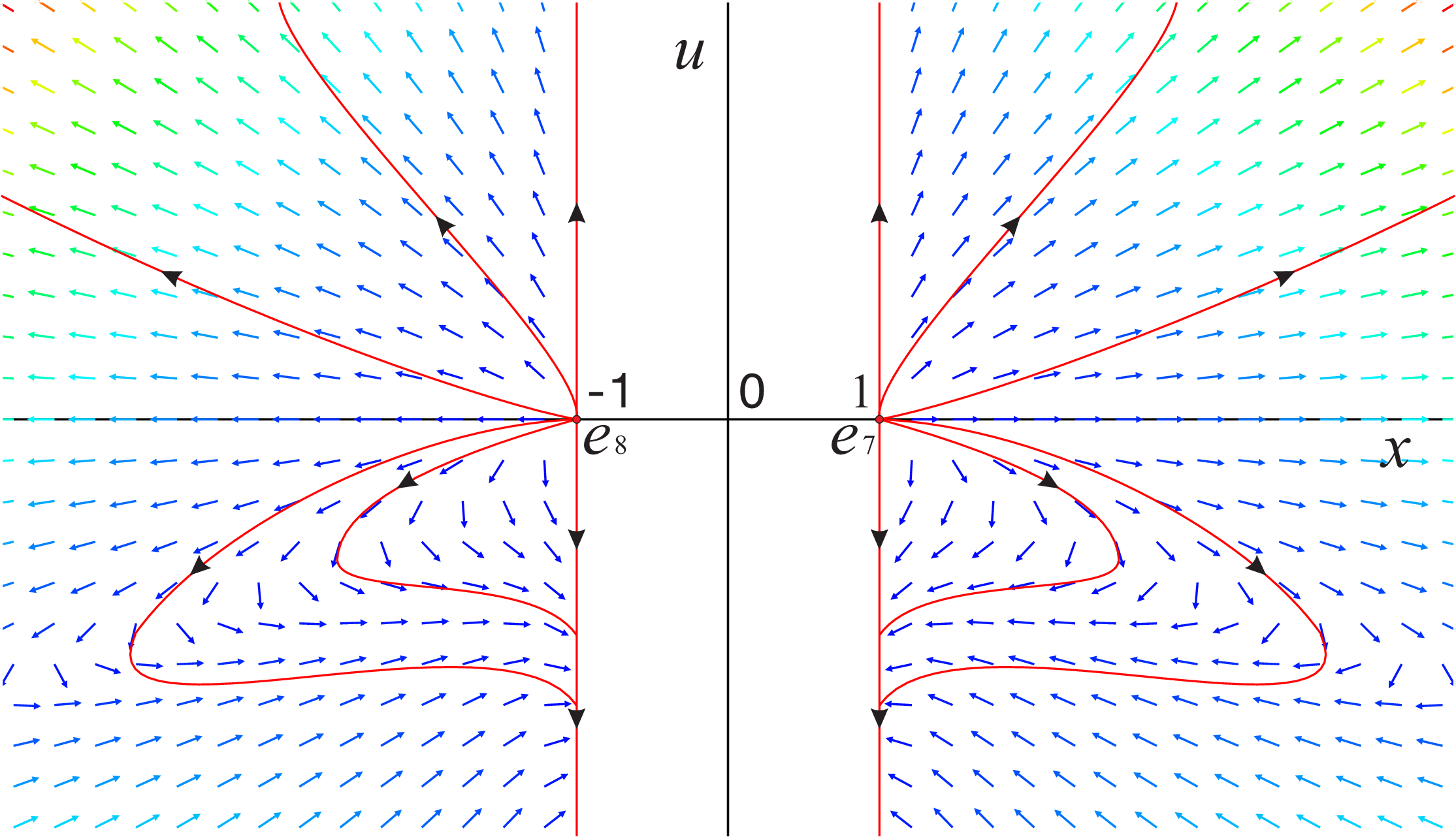}}
    \caption{The phase portrait on the invariant surface $f_+(x,z,u)=0$ restricted to the region $u\geq z$.}
  \label{fig6}
 \end{figure}

\begin{figure}[]
\centering
\subfigure{\includegraphics[width=13cm]{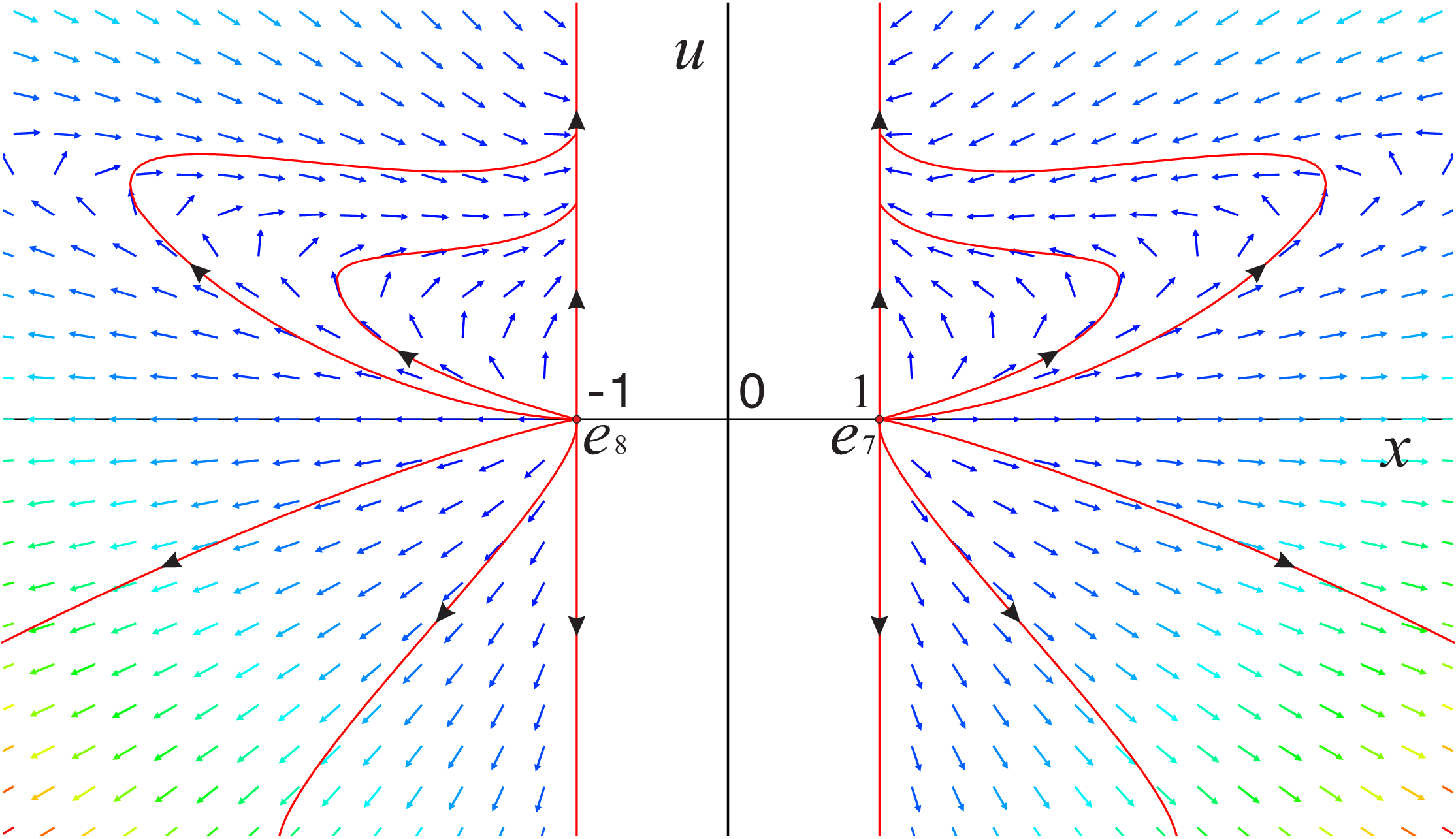}}
    \caption{The phase portrait on the invariant surface $f_+(x,z,u)=0$ restricted to the region $u\leq z$.}
  \label{fig7}
 \end{figure}

It is obvious that system (18) has a fictitious equilibrium point $(0, 0)$ because $y>0$, and its eigenvalues are 1 and 3 respectively, i.e. $e_9$ is a fictitious hyperbolic unstable node. The first equation of system (18) shows that when $y + 2u> 0$, so $y$ increases monotonically. In contrast if $y + 2u<0$, then $y$ decreases monotonically. Note that $x$ and $y$ have the same monotonicity when $y> 0$, so the local phase portrait of system (16) near $e_7$ and system (18) near $(0, 0)$ have the same local phase portrait. Then considering the symmetry $(y, u)\mapsto(-y, -u)$ of system (18), we can find that the local phase portrait of system (16) is shown in Figure \ref{fig6}.
 %Figure 6
Similarly the local phase portrait of system (17) is shown in Figure \ref{fig7}.
 %Figure 7

 Note that system (16) can be transformed into a polynomial differential system by letting $ y = \sqrt {x ^ 2-1} $ when $|x|\neq 1$. In order to study the dynamic behavior of the equilibrium points of system (16) at infinity, we first study the infinite equilibrium points of system (18). On the local chart $U_1$ system (18) writes
 \begin{equation}
\begin{array}{rl}\vspace{2mm}
\dfrac{dU}{dt}&=2U(1 - U)V^2,\\
\dfrac{dV}{dt}&=-(1 + 2 U) V \left(1 + V^2\right).
\end{array}
\label{13}
\end{equation}
It is easy to find that all the infinite points of system (19) at $V = 0$ are equilibrium points. Doing time scale transformation $d\tau_3=Vdt$ to eliminate the common factor $V$ in system (\ref{13}), then we have
  \begin{equation}
\begin{array}{rl}\vspace{2mm}
\dfrac{dU}{d\tau_3}&=2U(1 - U)V,\\
\dfrac{dV}{d\tau_3}&=-(1 + 2 U) \left(1 + V^2\right).
\end{array}
\label{14}
\end{equation}
This system has a unique equilibrium point in $V=0$, $e_9=(-1/2,0)$ with eigenvalues $\pm\sqrt{3}$, it is a hyperbolic saddle.

On the local chart $U_2$ system (18) reads
 \begin{equation}
\begin{array}{rl}\vspace{2mm}
\dfrac{dU}{dt}&=2 (1 - U) V^2,\\
\dfrac{dV}{dt}&=-U (2 + U) V - 3 V^3.
\end{array}
\label{15}
\end{equation}
Using time rescaling $d\tau_4=Vdt$ we obtain
 \begin{equation}
\begin{array}{rl}\vspace{2mm}
\dfrac{dU}{d\tau_4}&=2 (1 - U) V,\\
\dfrac{dV}{d\tau_4}&=-U (2 + U) - 3 V^2.
\end{array}
\label{16}
\end{equation}
The origin $e_{10}=(0,0)$ is an equilibrium point, which is a hyperbolic stable center with eigenvalue $\pm3i$ ($i$ is the imaginary unit). Hence $e_{10}$ is either a weak focus or a center, but since $\mathcal{H}_f=(2U-U^2+V^2-2/3)/(U-1)^3$ is a first integral of system (\ref{16}) defined at $(0,0)$, $e_{10}$ is a center. Then the global phase portrait of system (18) with $y>0$ is shown in Figure 8.
  %Figure 8
\begin{figure}[]
\centering
\subfigure{\includegraphics[width=3.5cm]{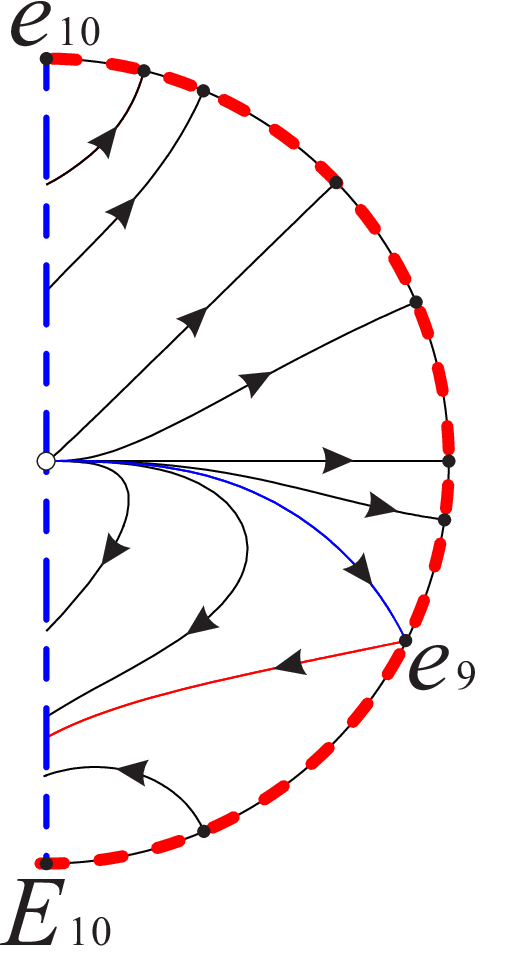}}
\subfigure{\includegraphics[width=2cm]{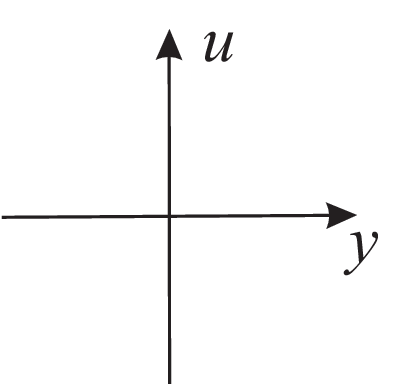}}
    \caption{The global phase portrait of system (18) with $y>0$. Here $E_{10}$ is the diametrically opposite equilibrium point of $e_{10}$ at infinity.}
  \label{fig8}
 \end{figure}

 Therefore, combining Figure 8 with Figures 6 and 7, we obtain the global phase portrait of systems (16) and (17) as shown in Figures 9 and 10, respectively.
%Figure 9
\begin{figure}[]
\centering
\subfigure{\includegraphics[width=6cm]{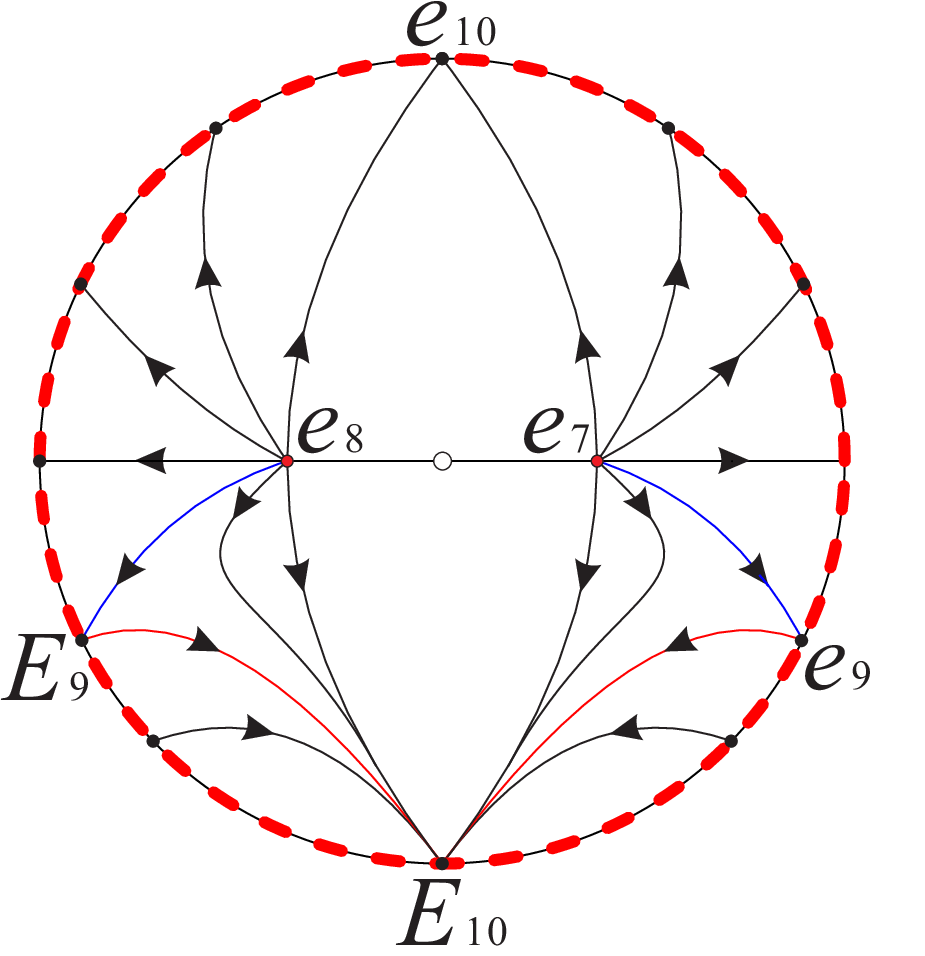}}
\subfigure{\includegraphics[width=2cm]{Fig2g9b10b28b_frame_x_u.eps}}
    \caption{The global phase portrait of system (16) on the invariant surface $f_+(x,z,u)=0$ restricted to the region $u\geq z$. Here $E_9$ is the symmetric point of $e_9$ with respect to the axis $x=0$, and $E_{10}$ is the diametrically opposite equilibrium point of $e_{10}$ at infinity.}
  \label{fig9}
 \end{figure}

%Figure 10
\begin{figure}[]
\centering
\subfigure{\includegraphics[width=6cm]{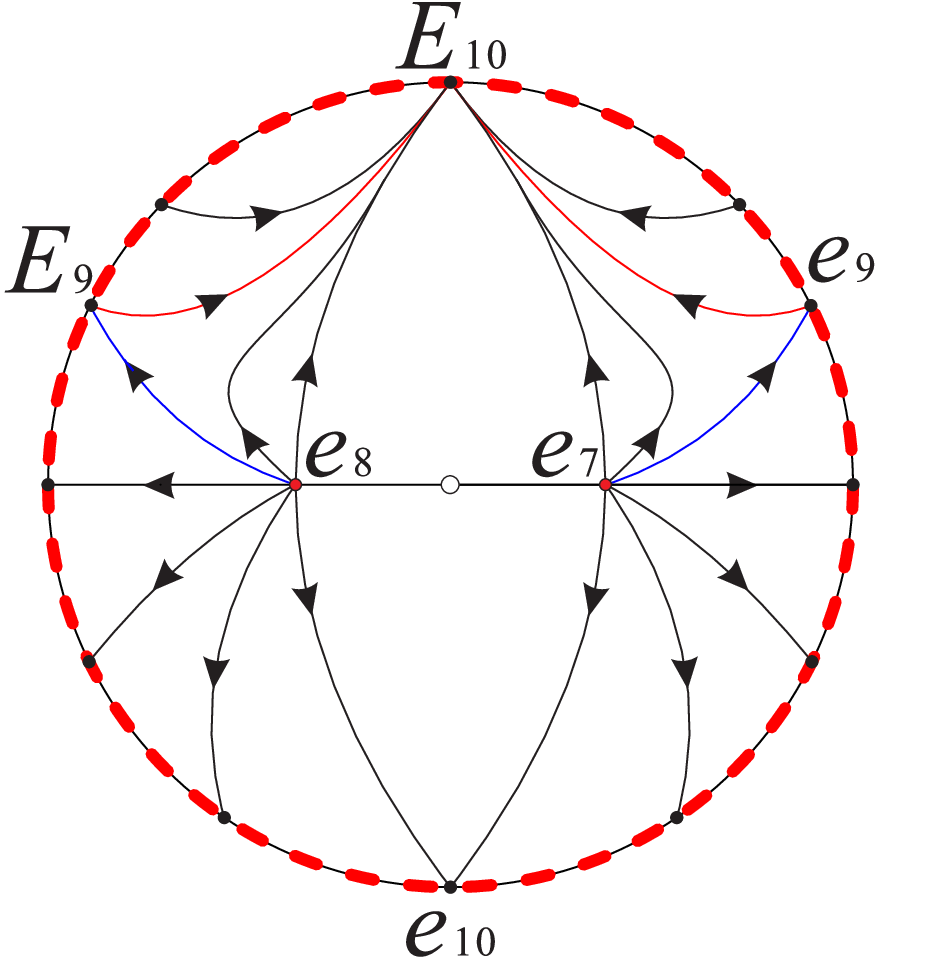}}
\subfigure{\includegraphics[width=2cm]{Fig2g9b10b28b_frame_x_u.eps}}
    \caption{The global phase portrait of system (17) on the invariant surface $f_+(x,z,u)=0$ restricted to the region $u\leq z$.}
  \label{fig10}
 \end{figure}

%subsection 3.1.4
\subsubsection{The finite equilibrium points}
It is easy to find that there are five finite equilibrium points of system (\ref{1}) because $s\neq0$ in case I. The equilibrium point $p_1=(1,0,0)$ with eigenvalues 3, 1 and $6-2\sqrt{6}s$, the equilibrium point $p_2=(-1,0,0)$ with eigenvalues 3, 1 and $6+2\sqrt{6}s$, the equilibrium points $p_3=(\sqrt{6}/(3s),-\sqrt{1-s^2}/s,0)$, $p_4=(\sqrt{6}/(3s),\sqrt{1-s^2}/s,0)$, $p_3$ and $p_4$ have the same eigenvalues 2, $-1/2- \sqrt{16 - 15 s^2})/(2 |s|)$ and $-1/2 +\sqrt{16 - 15 s^2})/(2 |s|)$, and the equilibrium point $p_5=(\sqrt{6}s/3,0,0)$ with eigenvalues $2s^2$, $2s^2-2,$ and $2s^2-3$.
Since different values of $s$ determine the types of these five equilibrium points, we list the relevant results more clearly in Table \ref{table3}.
%Table 3
\begin{table}[!htb]
\newcommand{\tabincell}[2]{\begin{tabular}{@{}#1@{}}#2\end{tabular}}
\centering
\caption{\label{opt}Finite equilibrium points for the different values of $s$, where $p_1=(1,0,0)$, $p_2=(-1,0,0)$, $p_3=(\sqrt{6}/(3s),-\sqrt{1-s^2}/s,0)$, $p_4=(\sqrt{6}/(3s),\sqrt{1-s^2}/s,0)$ and $p_5=(\sqrt{6}s/3,0,0)$.}
\footnotesize
\rm
\centering
\begin{tabular}{@{}*{12}{l}}
\specialrule{0em}{2pt}{2pt}
 \toprule
\hspace{2mm}\textbf{Values of $s$}&\hspace{25mm}\textbf{Equilibrium points}\\
\specialrule{0em}{2pt}{2pt}
\toprule
\tabincell{l}{\hspace{2mm}$-\infty<s<-\dfrac{\sqrt{6}}{2}$}&\tabincell{l}{$p_1$  and $p_5$ are unstable nodes,\\ $p_2$ is a saddle}\hspace{2mm}\\
\specialrule{0em}{2pt}{2pt}
\hline
\specialrule{0em}{2pt}{2pt}
\tabincell{l}{\hspace{2mm}$s=-\dfrac{\sqrt{6}}{2}$}&\tabincell{l}{$p_1$ is an unstable node,\\ $p_2$ and $p_5$ are non-hyperbolic equilibrium points}\hspace{2mm}\\
\specialrule{0em}{2pt}{2pt}
\hline
\specialrule{0em}{2pt}{2pt}
\tabincell{l}{\hspace{2mm}$-\dfrac{\sqrt{6}}{2}<s<-1$}&\tabincell{l}{$p_1$ and $p_2$ are unstable nodes,\\ $p_5$ is a saddle}\hspace{2mm}\\
\specialrule{0em}{2pt}{2pt}
\hline
\specialrule{0em}{2pt}{2pt}
\tabincell{l}{\hspace{2mm}$s=-1$}&\tabincell{l}{$p_1$ and $p_2$ are unstable nodes,\\ $p_3$, $p_4$ and $p_5$ are non-hyperbolic equilibrium points}\hspace{2mm}\\
\specialrule{0em}{2pt}{2pt}
\hline
\specialrule{0em}{2pt}{2pt}
\tabincell{l}{\hspace{2mm}$-1<s<1$}&\tabincell{l}{$p_1$ and $p_2$ are unstable nodes,\\ $p_3$, $p_4$ and $p_5$ are saddles}\hspace{2mm}\\
\specialrule{0em}{2pt}{2pt}
\hline
\specialrule{0em}{2pt}{2pt}
\tabincell{l}{\hspace{2mm}$s=1$}&\tabincell{l}{$p_1$ and $p_2$ are unstable nodes,\\ $p_3$, $p_4$ and $p_5$ are non-hyperbolic equilibrium points}\hspace{2mm}\\
\specialrule{0em}{2pt}{2pt}
\hline
\specialrule{0em}{2pt}{2pt}
\tabincell{l}{\hspace{2mm}$1<s<\dfrac{\sqrt{6}}{2}$}&\tabincell{l}{$p_1$ and $p_2$ are unstable nodes,\\ $p_5$ is a saddle}\hspace{2mm}\\
\specialrule{0em}{2pt}{2pt}
\hline
\specialrule{0em}{2pt}{2pt}
\tabincell{l}{\hspace{2mm}$s=\dfrac{\sqrt{6}}{2}$}&\tabincell{l}{$p_1$ and $p_5$ are non-hyperbolic equilibrium points,\\ $p_2$ is an unstable node}\hspace{2mm}\\
\specialrule{0em}{2pt}{2pt}
\hline
\specialrule{0em}{2pt}{2pt}
\tabincell{l}{\hspace{2mm}$\dfrac{\sqrt{6}}{2}<s<+\infty$}&\tabincell{l}{$p_1$ is a saddle,\\ $p_2$ and $p_5$ are unstable nodes}\hspace{2mm}\\
\specialrule{0em}{2pt}{2pt}
\toprule
\end{tabular}
  \label{table3}
\end{table}

%subsection 3.1.5
\subsubsection{Phase portrait on the Poincar\'{e} sphere at infinity}
According to the three-dimensional Poincar\'{e} compactification (see \cite{Cima} for more details), we set $x=1/z_3,\ z=z_1/z_3,\ u=z_2/z_3$, then on the local chart $U_1$ system (\ref{1}) is rewritten as
\begin{equation}
\begin{array}{rl}\vspace{2mm}
\dfrac{dz_1}{dt}&=-z_1 z_3 \left[\sqrt{6} s ( (z_1 - z_2)^2 + z_3^2-1)-z_3\right],\\\vspace{2mm}
\dfrac{dz_2}{dt}&=-z_2 z_3 \left[\sqrt{6} s ( (z_1 - z_2)^2 + z_3^2-1)-3 z_3\right],\\\vspace{2mm}
\dfrac{dz_3}{dt}&=z_3 \left[-3 + 2 z_1^2 - 2 z_1 z_2 - \sqrt{6} s ((z_1 - z_2)^2-1) z_3 \right.\\
&\ \ \ + \left.3 z_3^2 - \sqrt{6} s z_3^3\right].
\end{array}
\label{eq23}
\end{equation}

Since the infinity at the different local charts of Poincar\'{e} sphere corresponds to $z_3 = 0$, then for all $z_1,z_2\in\mathbb{R}$ system (\ref{eq23}) has the equilibrium point $(z_1,z_2,0)$ with eigenvalues ${\{0,0, 2 z_1(z_1-z_2)-3\}}$, which means that the local chart $U_1$ are full of equilibrium points at infinity. Note that the corresponding eigenvectors are
\begin{equation}
\begin{array}{rl}\vspace{2mm}
\{0, 1, 0\},& \{1, 0, 0\},\\
\left\{\dfrac{\sqrt{6} z_1 s[(z_1 - z_2)^2-1]}{3 - 2 z_1 (z_1 - z_2)}\right.,& \left.\dfrac{ \sqrt{6} z_2 s[(z_1 - z_2)^2-1]}{3 - 2 z_1 (z_1 - z_2)},\ 1\right\}.
\end{array}
\label{eq24}
\end{equation}
By using normally hyperbolic submanifold theorem (see Appendix A for details), the equilibrium point $(z_1,z_2,0)$ has a one-dimensional stable manifold when $2 z_1 (z_1 - z_2)< 3$ and unstable when $2 z_1 (z_1 - z_2)> 3$. More details are shown in region I as well as in regions II and III of Figure 11, respectively. However for the equilibrium points on the hyperbola $2 z_1 (z_1 - z_2)=3$ there are six local phase portraits in Figure 12. Note that there is one-dimensional stable manifold in region I and one-dimensional unstable manifold in regions II and III filled with infinite equilibrium points. Since the orbits arriving or ending at equilibrium points at infinity in the different regions cannot collide into finite equilibrium points when they tend to equilibrium points on the hyperbola $2 z_1 (z_1 - z_2)=3$ coming from both sides of this hyperbola, there is a hyperbolic sector on the equilibrium points of the branches $L_1$ and $L_2$ of the hyperbola, with the exception of two points $p_6$ and $p_7$, see the first two pictures in Figure 12 for details.

 %Figure 11
\begin{figure}[]
  \begin{minipage}{130mm}
\centering
\subfigure[]{\label{fig:subfig:a}
    \includegraphics[width=10cm]{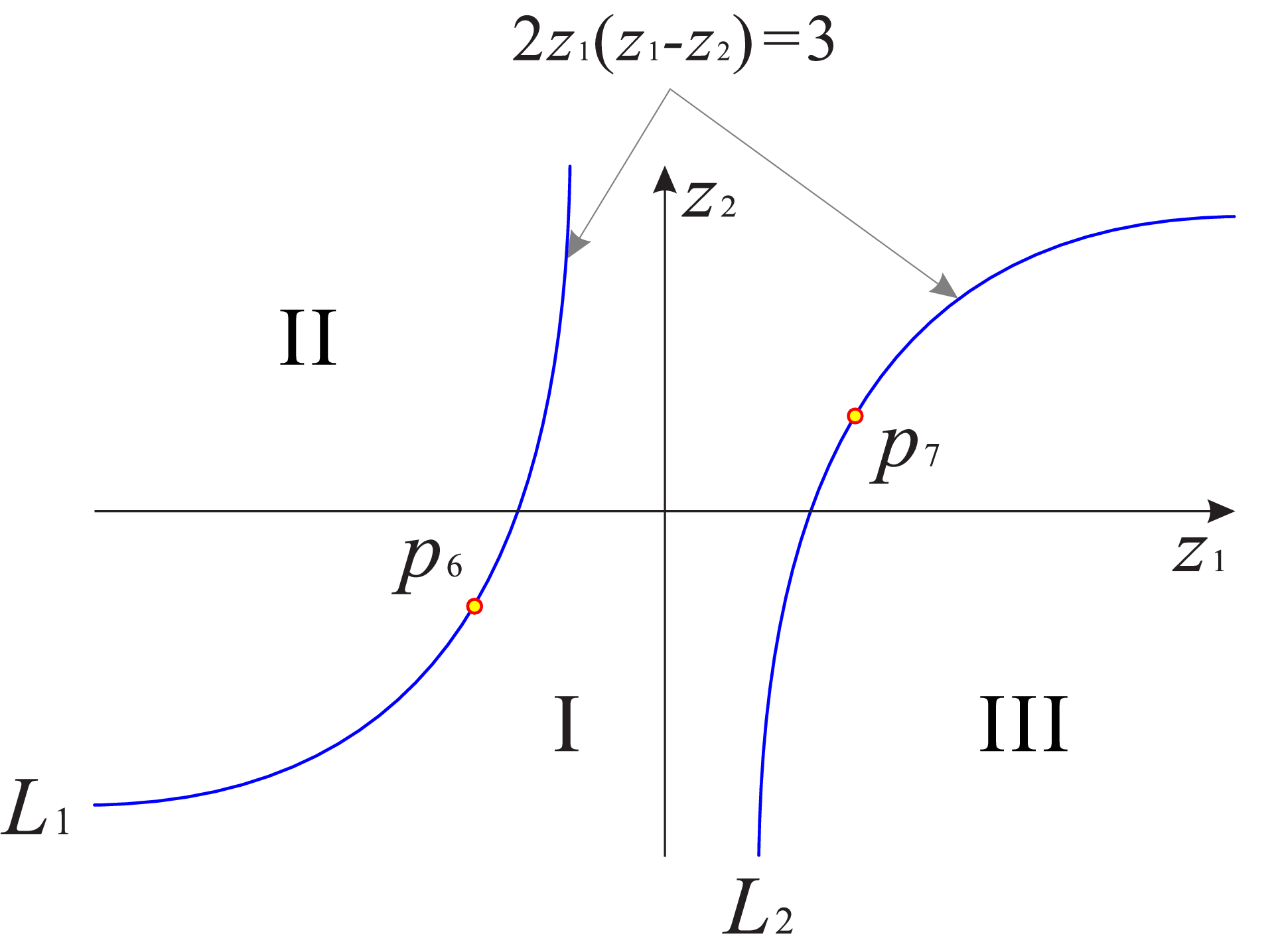}}\\
\subfigure[]{\label{fig:subfig:b}
    \includegraphics[width=3cm]{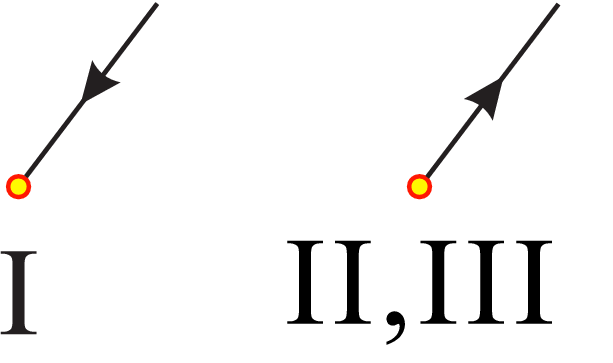}}
    \caption{There is an one-dimensional stable manifold in the region I and one-dimensional unstable manifold in regions II and III on the local chart $U_1$.}
  \label{fig:subfig}
  \end{minipage}
 \end{figure}

  %Figure 12
\begin{figure}[]
\begin{minipage}{130mm}
\centering
\subfigure{
    \includegraphics[width=12cm]{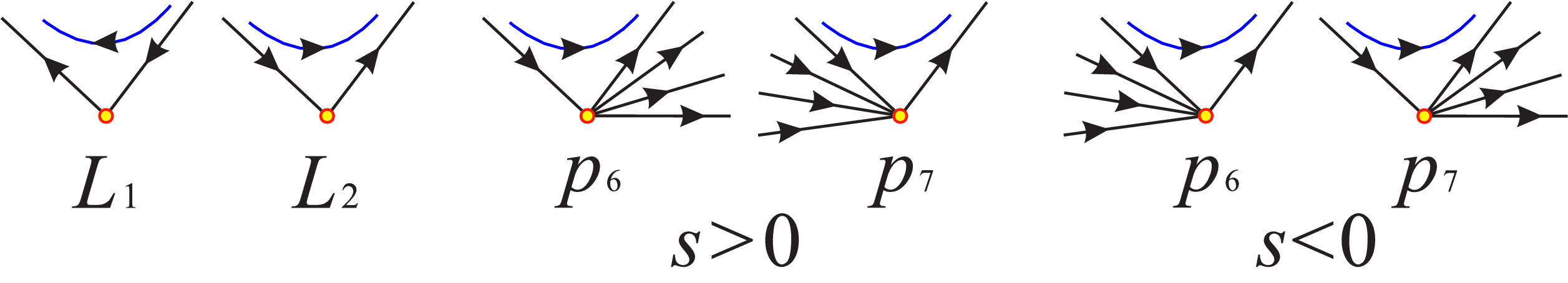}}
    \caption{There are six kinds of equilibrium points on the hyperbola $2 z_1 (z_1 - z_2)=3$.}
  \label{fig:subfig}
  \end{minipage}
 \end{figure}

Doing time rescaling $d\tau_5=z_3dt$ system (\ref{eq23}) becomes
  \begin{equation}
\begin{array}{rl}\vspace{2mm}
\dfrac{dz_1}{d\tau_5}&=-\sqrt{6} s z_1\left[(z_1 - z_2)^2 + z_3^2-1\right]+z_1z_3,\\\vspace{2mm}
\dfrac{dz_2}{d\tau_5}&=-\sqrt{6} s z_2\left[(z_1 - z_2)^2 + z_3^2-1\right]+3 z_2z_3,\\\vspace{2mm}
\dfrac{dz_3}{d\tau_5}&= -3 + 2 z_1^2 - 2 z_1 z_2 - \sqrt{6} s z_3 [(z_1 - z_2)^2-1] \\
&\ \ \ + 3 z_3^2 - \sqrt{6} s z_3^3.
\end{array}
\label{eq25}
\end{equation}
At $z_3=0$ system (\ref{eq25}) has two lines $|z_1-z_2|=1$ filled with infinite equilibrium points and these two lines intersect with hyperbola $2 z_1 (z_1 - z_2)=3$ at $p_6=(-3/2,-1/2,0)$ and $p_7=(3/2,1/2,0)$, respectively. Both of them are hyperbolic with the same eigenvalues ${\left\{-\sqrt{3}, \sqrt{3}, -2 \sqrt{6} s\right\}}$ and the corresponding eigenvectors have the following form
\begin{equation}
\begin{array}{rl}\vspace{2mm}
&\left(\dfrac{\sqrt{3}}{2}, \dfrac{\sqrt{3}}{2}, 1\right), \left(-\dfrac{\sqrt{3}}{2}, -\dfrac{\sqrt{3}}{2}, 1\right),
\left(\dfrac{\sqrt{6}(8s^2+1)}{16 s}, \dfrac{\sqrt{6}(8s^2+5)}{48s},\ 1\right).\\
&\left(-\dfrac{\sqrt{3}}{2}, -\dfrac{\sqrt{3}}{2}, 1\right), \left(\dfrac{\sqrt{3}}{2}, \dfrac{\sqrt{3}}{2}, 1\right),
\left(-\dfrac{\sqrt{6}(8s^2+1)}{16 s}, -\dfrac{\sqrt{6}(8s^2+5)}{48s},\ 1\right).
\end{array}
\label{eq26}
\end{equation}
Then $p_6$ and $p_7$ have an unstable manifold of dimension one (respectively two) and a stable manifold of dimension two (respectively one) if $s>0$ (respectively $s<0$), see the last four pictures in Figure 12 for details.

On the local chart $U_2$, we let $x=z_1/z_3,\ z=1/z_3,\ u=z_2/z_3$, then system (\ref{1}) writes
\begin{equation}
\begin{array}{rl}\vspace{2mm}
\dfrac{dz_1}{dt}&= z_3 \left[-z_1 z_3 + \sqrt{6} s (-z_1^2 + (z_2-1)^2 + z_3^2)\right],\\\vspace{2mm}
\dfrac{dz_2}{dt}&= 2 z_2 z_3^2,\\\vspace{2mm}
\dfrac{dz_3}{dt}&= z_3 \left(2 - 3 z_1^2 - 2 z_2 + 2 z_3^2\right).
\end{array}
\label{eq27}
\end{equation}
By eliminating the common factor $z_3$ in system (\ref{eq27}) through time scale transformation $d\tau_6=z_3dt$ we get
\begin{equation}
\begin{array}{rl}\vspace{2mm}
\dfrac{dz_1}{d\tau_6}&= -z_1 z_3 + \sqrt{6} s \left[-z_1^2 + (z_2-1)^2 + z_3^2\right],\\\vspace{2mm}
\dfrac{dz_2}{d\tau_6}&= 2 z_2 z_3,\\\vspace{2mm}
\dfrac{dz_3}{d\tau_6}&= 2 - 3 z_1^2 - 2 z_2 + 2 z_3^2.
\end{array}
\label{eq28}
\end{equation}
Note that system (\ref{eq28}) with $z_1=z_3=0$ has the unique infinite equilibrium point $p_8=(0,1,0)$ with eigenvalues \{2i, -2i, 0\} and eigenvectors \{0, -i, 1\}, \{0, i, 1\}, \{1, 0, 0\}, so there will be a fold-Hopf bifurcation at the infinite equilibrium point $p_8$, sometimes called a zero-pair bifurcation or a Gavrilov-Guckenheimer (see Chapter 5 of \cite{Kuznetsov2004} for more details). We will not continue to discuss other infinite equilibrium points of this system, because these are already included in the local chart $U_1$.

Similarly on the local chart $U_3$, we let $x=z_1/z_3,\ z=z_2/z_3,\ u=1/z_3$, then system (\ref{1}) becomes
\begin{equation}
\begin{array}{rl}\vspace{2mm}
\dfrac{dz_1}{dt}&= z_3 \left[-3 z_1 z_3 + \sqrt{6} s (-z_1^2 + (z_2-1)^2 + z_3^2)\right],\\\vspace{2mm}
\dfrac{dz_2}{dt}&= -2 z_2 z_3^2,\\\vspace{2mm}
\dfrac{dz_3}{dt}&= -3 z_1^2 z_3 + 2 (z_2-1) z_2 z_3.
\end{array}
\label{eq29}
\end{equation}
In this local chart $U_3$ we only need to study the infinite equilibria located in its origin because all the other infinite equilibrium points have been studied in the local charts $U_1$ and $U_2$.
After changing the time scale $d\tau_7=z_3dt$ we obtain
\begin{equation}
\begin{array}{rl}\vspace{2mm}
\dfrac{dz_1}{dt}&= -3 z_1 z_3 + \sqrt{6} s \left[-z_1^2 + (z_2-1)^2 + z_3^2\right],\\\vspace{2mm}
\dfrac{dz_2}{dt}&= -2 z_2 z_3,\\\vspace{2mm}
\dfrac{dz_3}{dt}&= -3 z_1^2 + 2 (z_2-1) z_2.
\end{array}
\label{22}
\end{equation}
Obviously the origin $(0,0,0)$ is not an equilibrium point of system (\ref{22}), so we will not continue to investigate other equilibrium points at infinity in the local chart $U_3$.

In summary the equilibrium points filling up the infinity with these different stable and unstable manifolds are summarized in Figures 11, 12 and 13.
  %Figure 13
\begin{figure}[]
\begin{minipage}{130mm}
\centering
\subfigure{
    \includegraphics[width=8cm]{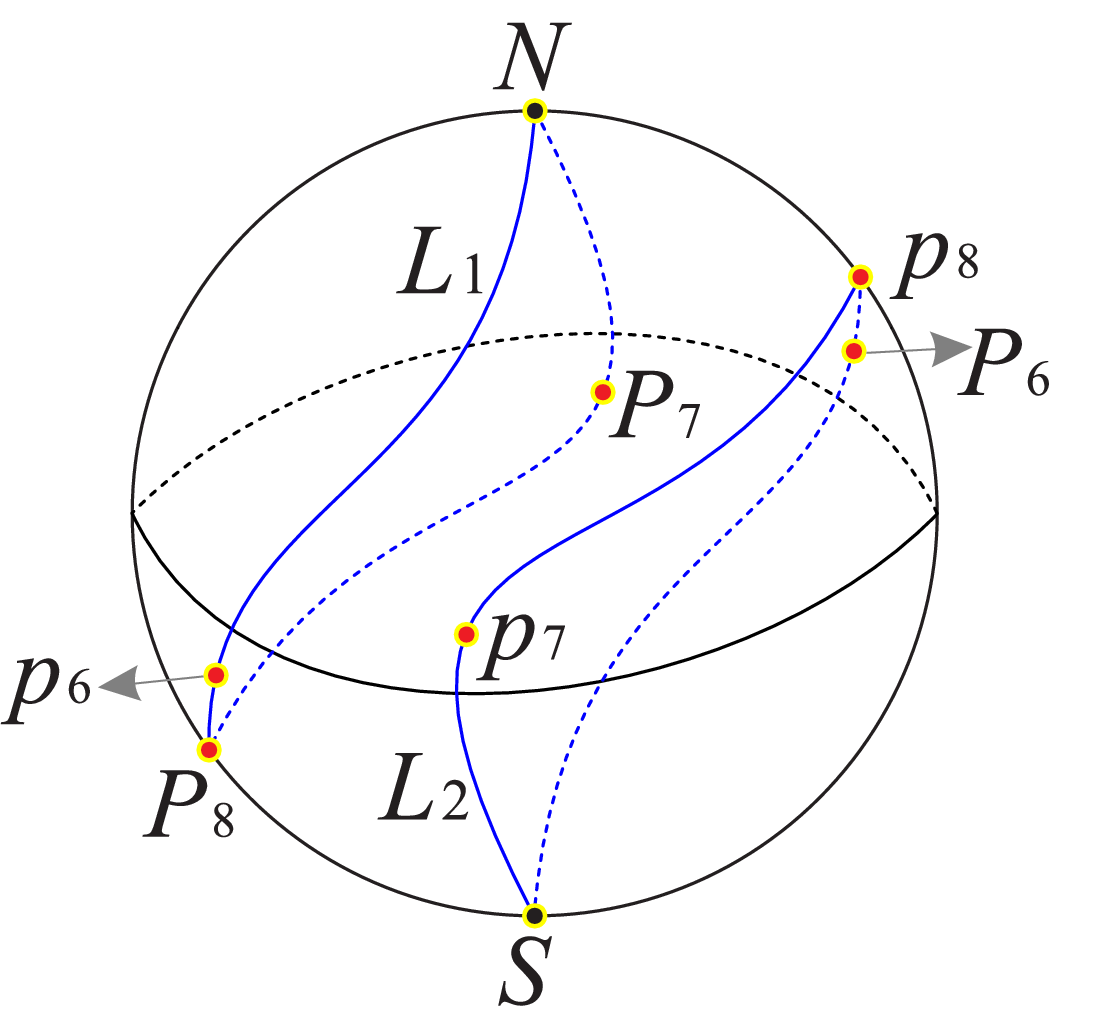}}
\subfigure{\includegraphics[width=2cm]{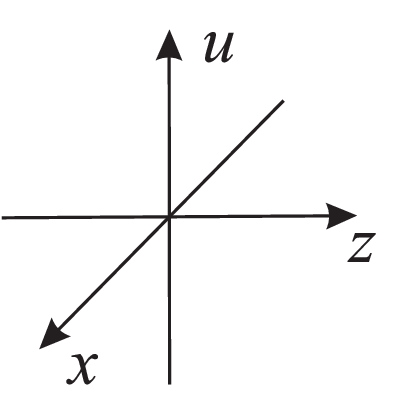}}
\subfigure{\includegraphics[width=12cm]{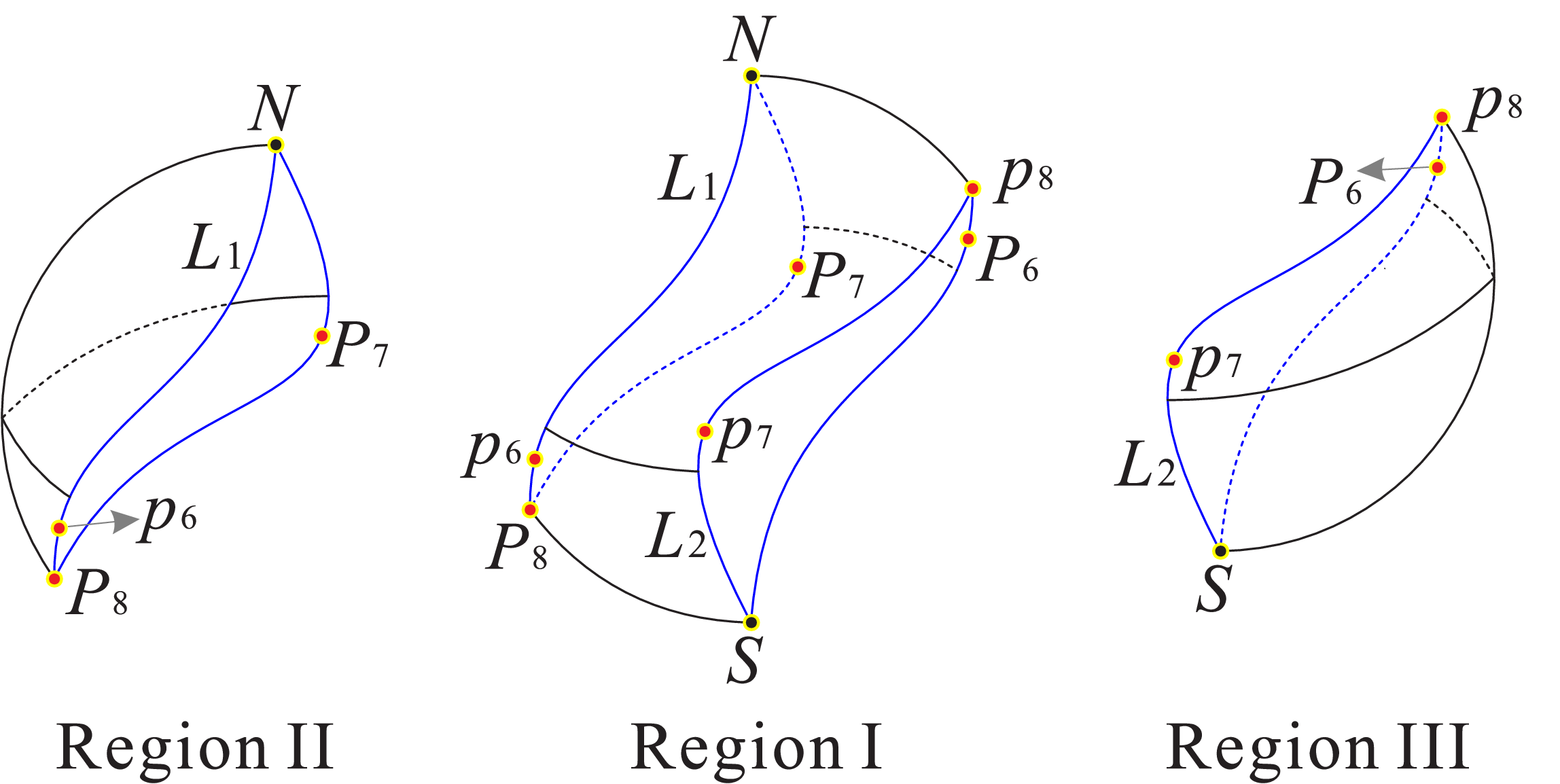}}
    \caption{The sphere $\mathbb{S}^2$ (the infinity of $\mathbb{R}^3$) is filled up with equilibrium points. The stable and unstable manifolds of these equilibrium points are different in the regions I, II, III, lines $L_1$, $L_2$, and points $p_6$, $p_7$, $p_8$, $P_6$, $P_7$, $P_8$ on the sphere defined by the closure of $U_1$ and its symmetric closure of $V_1$ with respect to the origin of $\mathbb{R}^3$. Thus $P_6$, $P_7$ and $P_8$ are the symmetric points with respect to the origin of $p_6$, $p_7$ and $p_8$, respectively. The points $N$ and $S$ denote the north pole and south pole of the Poincar\'e ball, respectively.}
  \label{fig:subfig}
  \end{minipage}
 \end{figure}

\subsection{Phase portrait inside the Poincar\'e ball restricted to the physical region of interest $x^2-(u-z)^2\leq1$}

Note that system (\ref{1}) is invariant with respect to the $x$-axis due to the symmetry $(x,z,u)\mapsto(x,-z,-u)$. Now we divide the Poincar\'e ball restricted to the region $x^2-(u-z)^2-1\leq0$ into four regions as follows
\begin{equation*}
\begin{array}{rl}
R_1:\ z\geq0,\ u\geq0.\ \ \
R_2:\ z\geq0,\ u\leq0.\\
R_3:\ z\leq0,\ u\leq0.\ \ \
R_4:\ z\leq0,\ u\geq0.
\end{array}
\end{equation*}
Due to the symmetry with respect to the $x$-axis, we only need to discuss the phase portrait of system (\ref{1}) in the regions $R_1$ and $R_2$.

Combining the phase portraits on the invariant surface $f_+(x,z,u)=0$, on the invariant planes $z=0$ and $u=0$, and at infinity, we obtain the phase portrait on the boundary of the regions $R_1$ and $R_2$ as shown in Figures 14-17.
 %Figure 14
\begin{figure}[]
  \begin{minipage}{150mm}
\centering
\subfigure[]{\label{fig:subfig:a}
    \includegraphics[width=6.5cm]{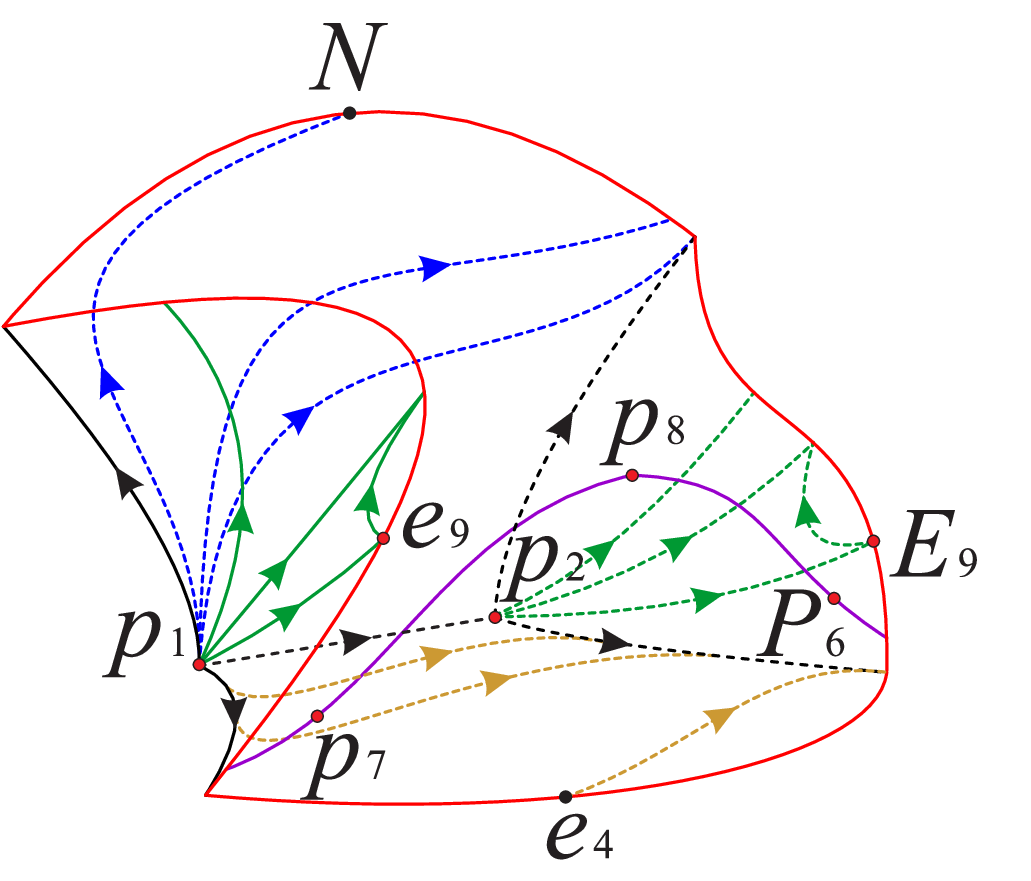}}
\subfigure[]{\label{fig:subfig:b}
    \includegraphics[width=6.5cm]{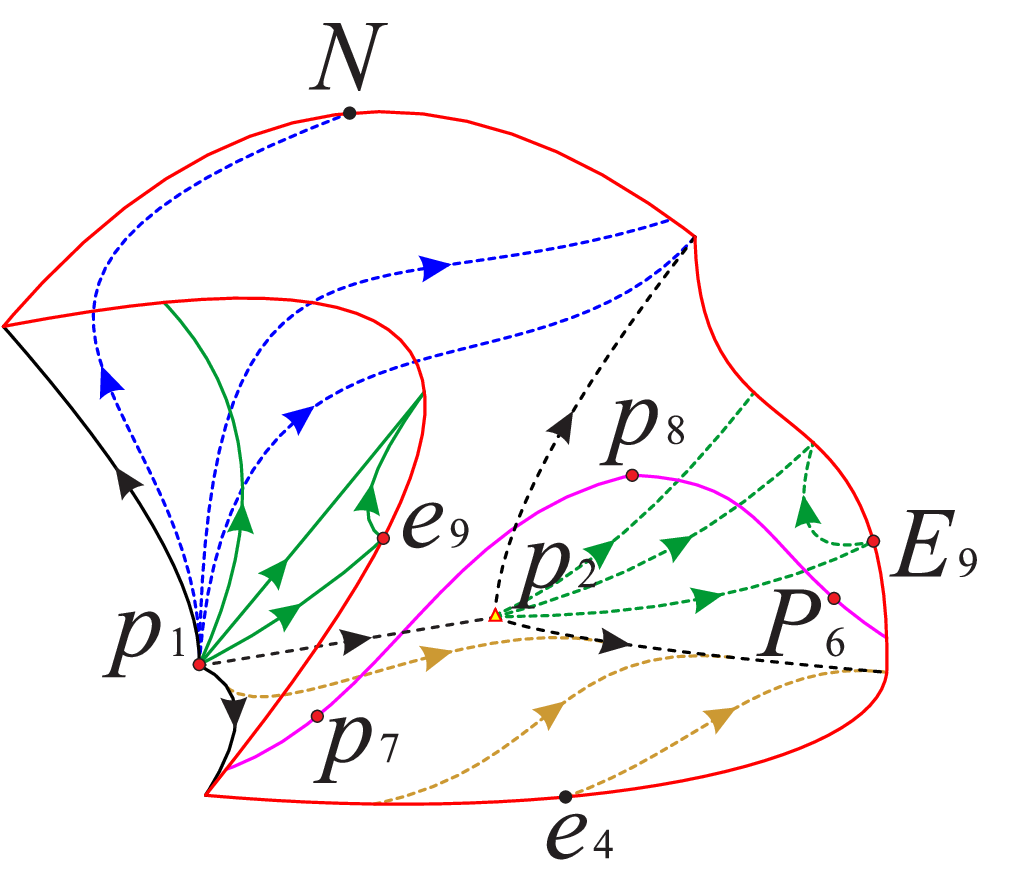}}\\
\subfigure[]{\label{fig:subfig:c}
    \includegraphics[width=6.5cm]{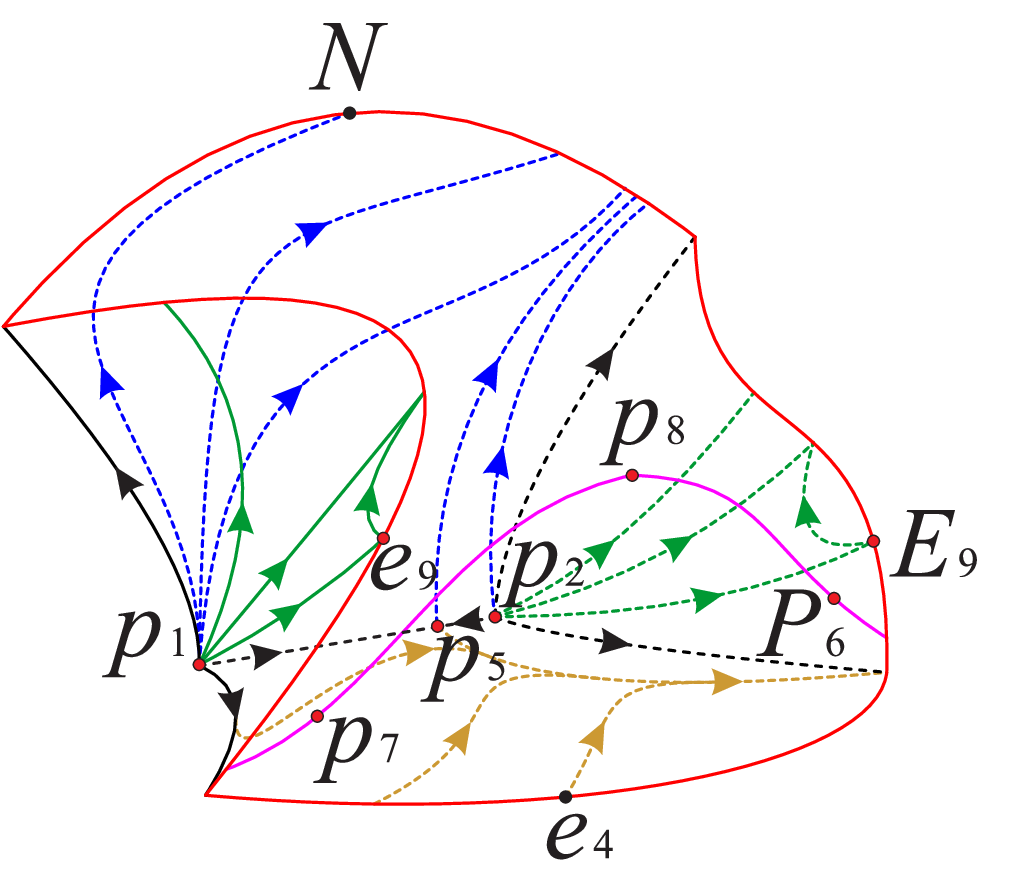}}
\subfigure[]{\label{fig:subfig:d}
    \includegraphics[width=6.5cm]{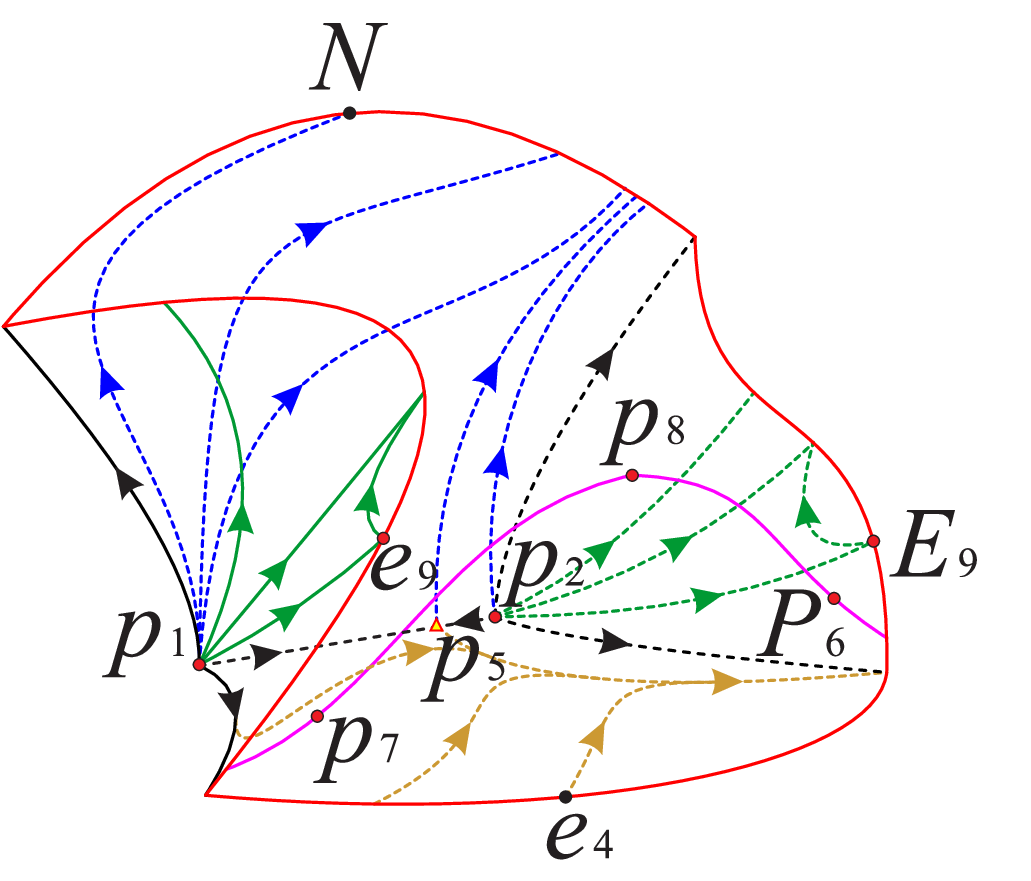}}\\
\subfigure[]{\label{fig:subfig:e}
    \includegraphics[width=6.5cm]{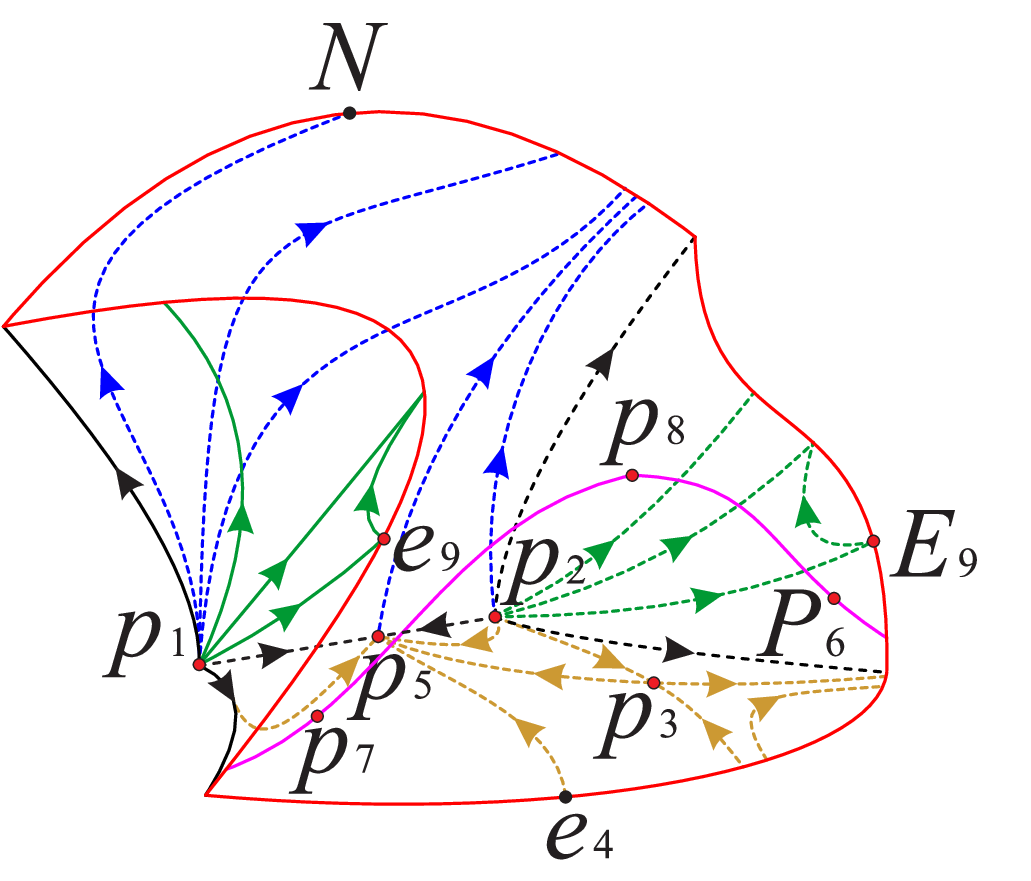}}
\subfigure{\includegraphics[width=2cm]{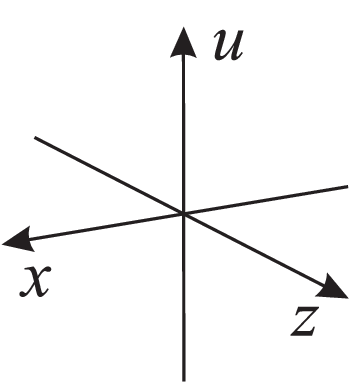}}
    \caption{Phase portrait in the boundary of the region $R_1$ for different values of $s$: (a) $s<-\sqrt{6}/2$, (b) $s=-\sqrt{6}/2$, (c) $-\sqrt{6}/2<s<-1$, (d) $s=-1$, (e) $-1<s<0$. The surfaces above and below the long dashed line represent regions I and III respectively.}
  \label{fig:subfig}
  \end{minipage}
 \end{figure}
 %Figure 15
\begin{figure}[]
  \begin{minipage}{150mm}
\centering
\subfigure[]{\label{fig:subfig:a}
    \includegraphics[width=6.5cm]{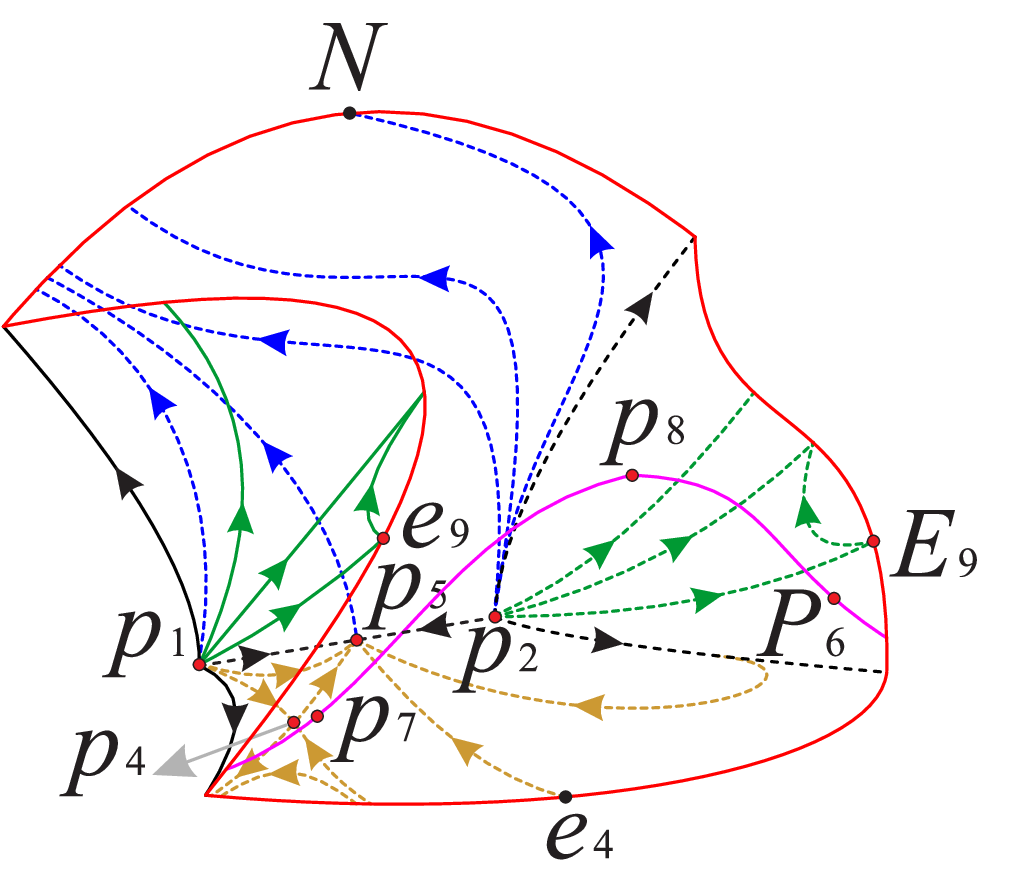}}
\subfigure[]{\label{fig:subfig:b}
    \includegraphics[width=6.5cm]{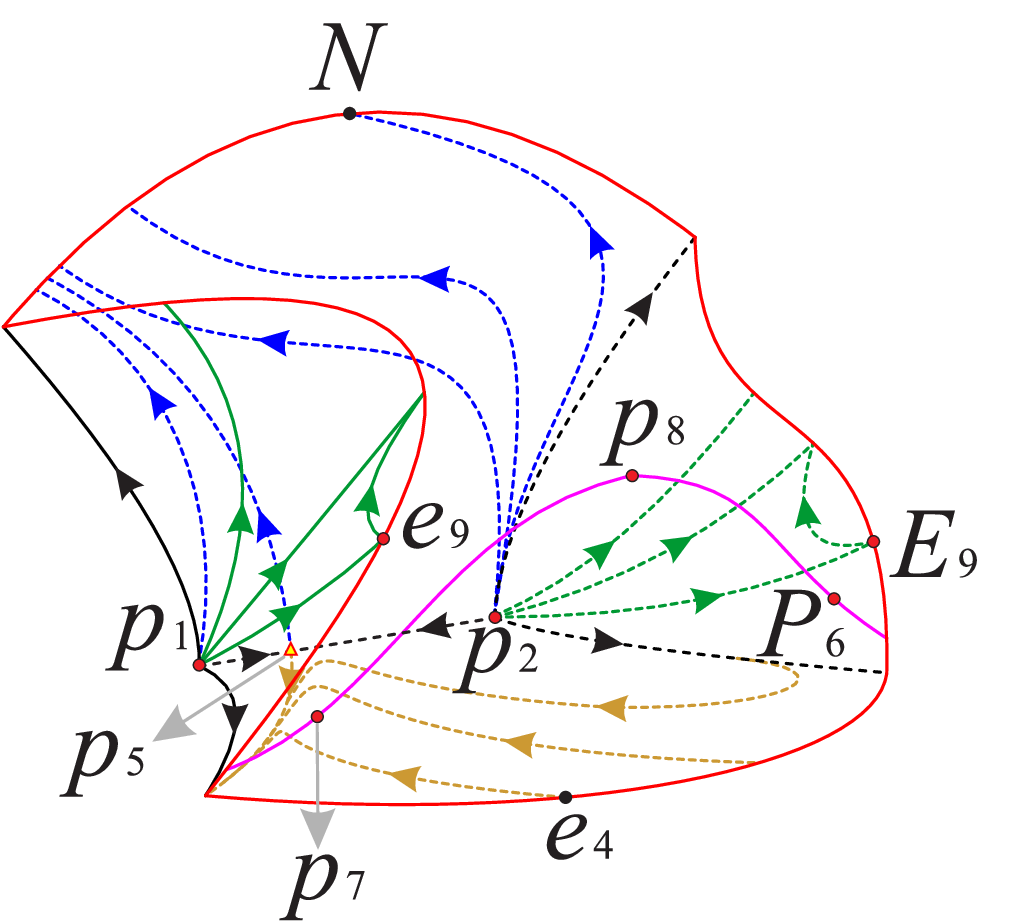}}\\
\subfigure[]{\label{fig:subfig:c}
    \includegraphics[width=6.5cm]{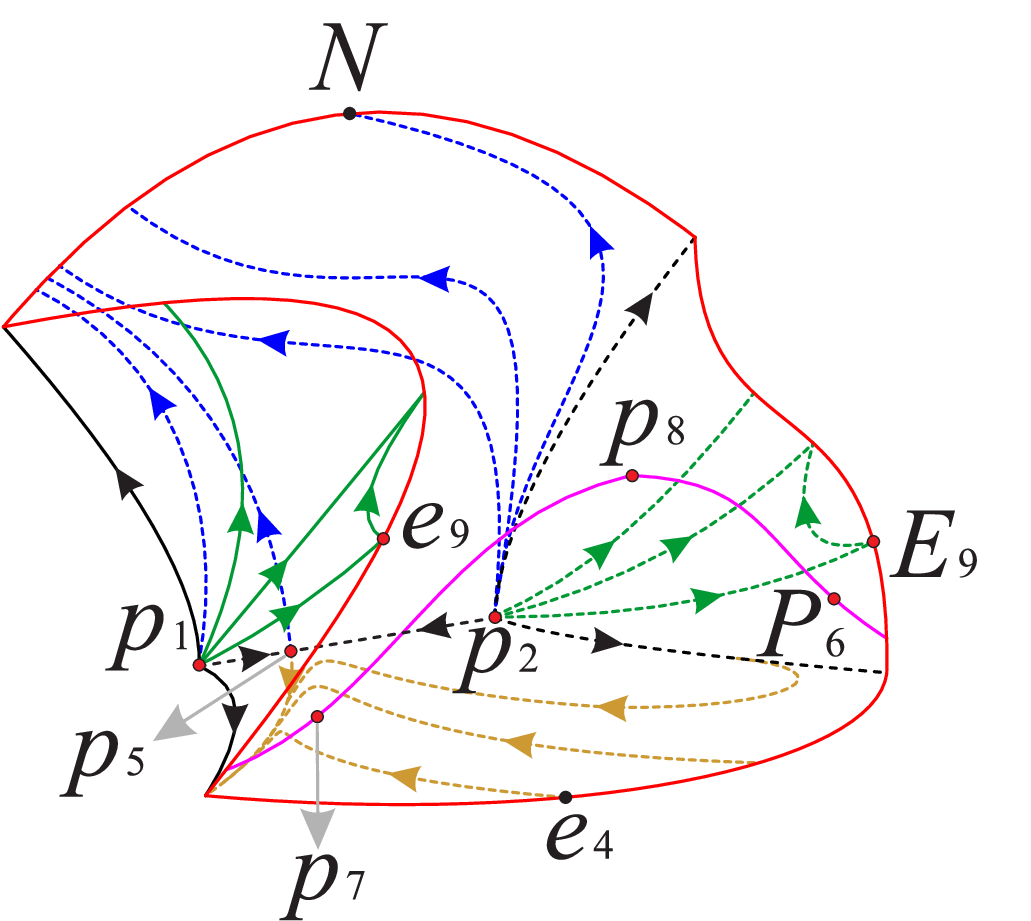}}
\subfigure[]{\label{fig:subfig:d}
    \includegraphics[width=6.5cm]{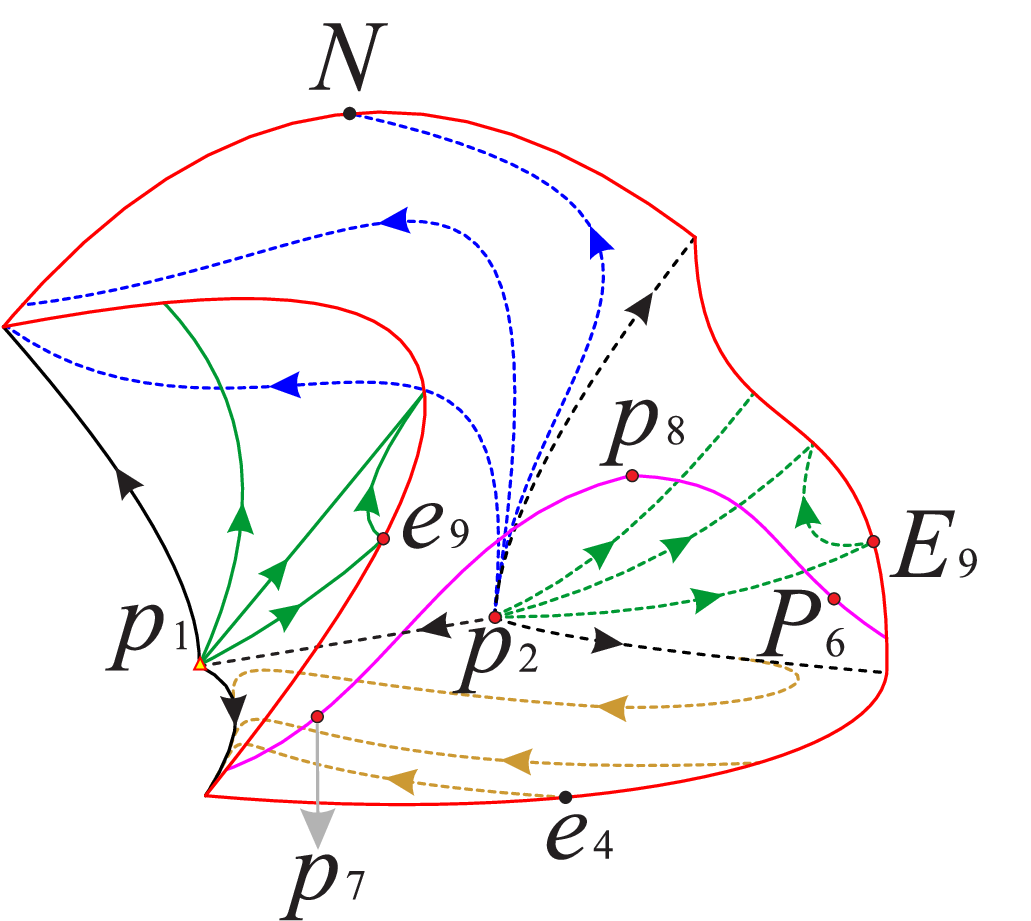}}\\
\subfigure[]{\label{fig:subfig:e}
    \includegraphics[width=6.5cm]{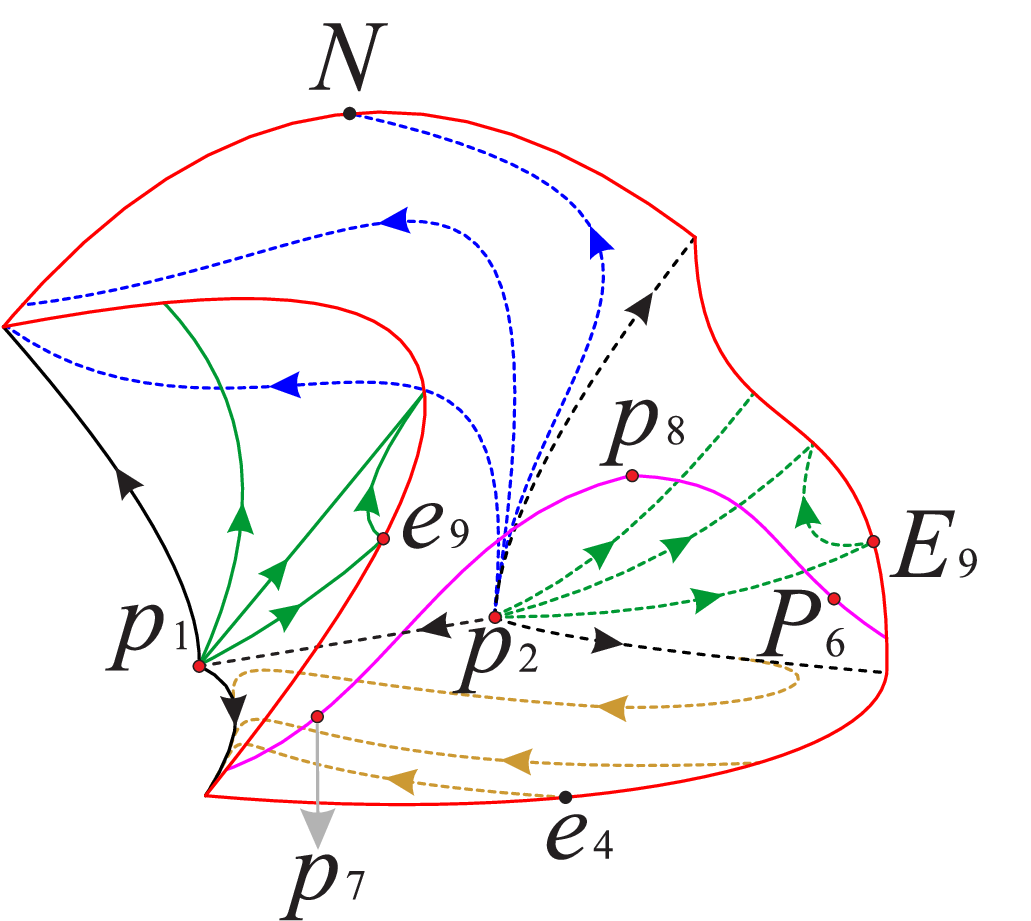}}
\subfigure{\includegraphics[width=2cm]{Fig14f15f16f17f21b23b24c26b30b32b33b34b_frame2_x_z_u.eps}}
    \caption{Phase portrait in the boundary of the region $R_1$ for different values of $s$: (a) $0<s<1$, (b) $s=1$, (c) $1<s<\sqrt{6}/2$, (d) $s=\sqrt{6}/2$, (e) $s>\sqrt{6}/2$. The surfaces above and below the long dashed line represent regions I and III respectively.}
  \label{fig:subfig}
  \end{minipage}
 \end{figure}
  %Figure 16
\begin{figure}[]
  \begin{minipage}{150mm}
\centering
\subfigure[]{\label{fig:subfig:a}
    \includegraphics[width=6.5cm]{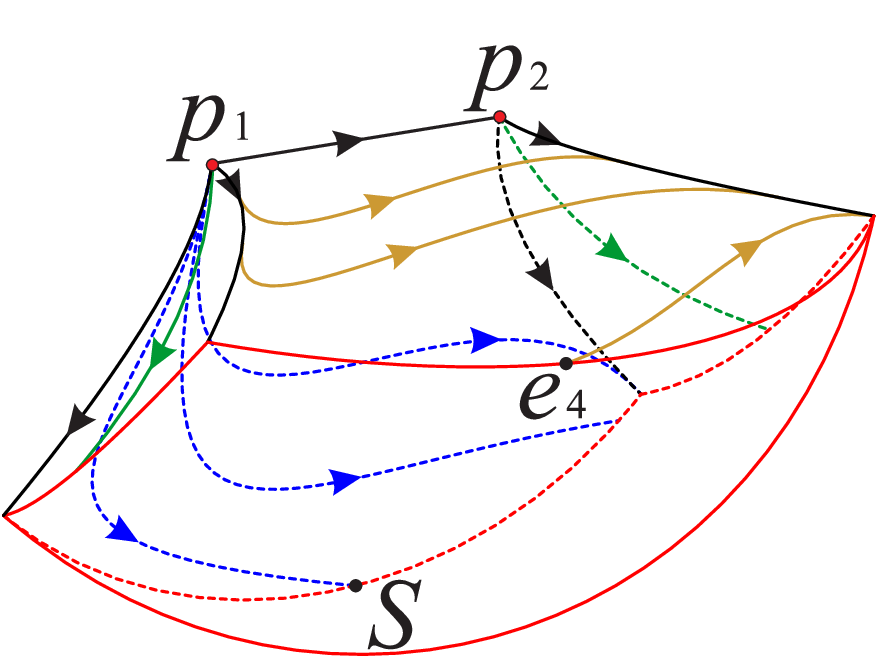}}
\subfigure[]{\label{fig:subfig:b}
    \includegraphics[width=6.5cm]{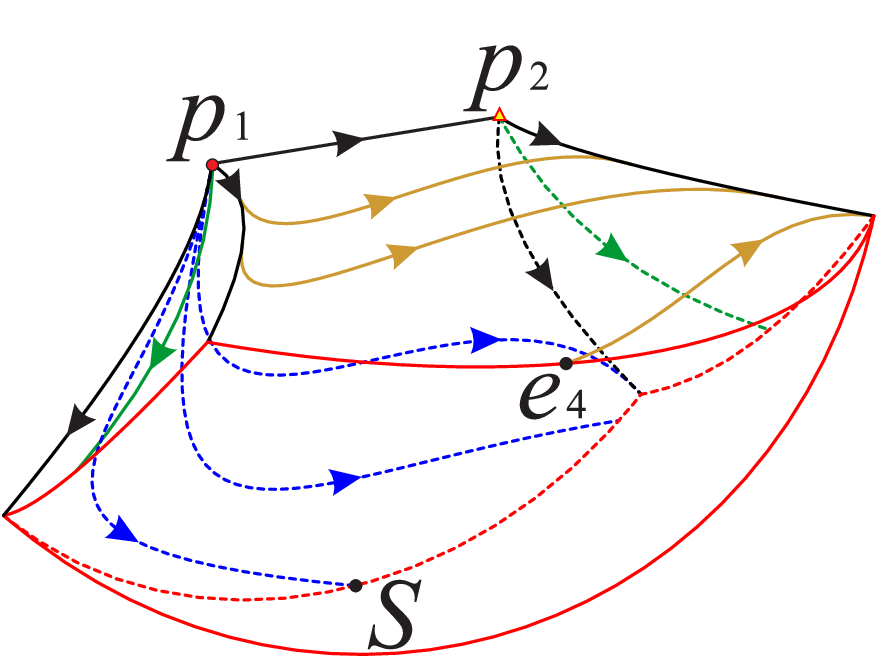}}\\
\subfigure[]{\label{fig:subfig:c}
    \includegraphics[width=6.5cm]{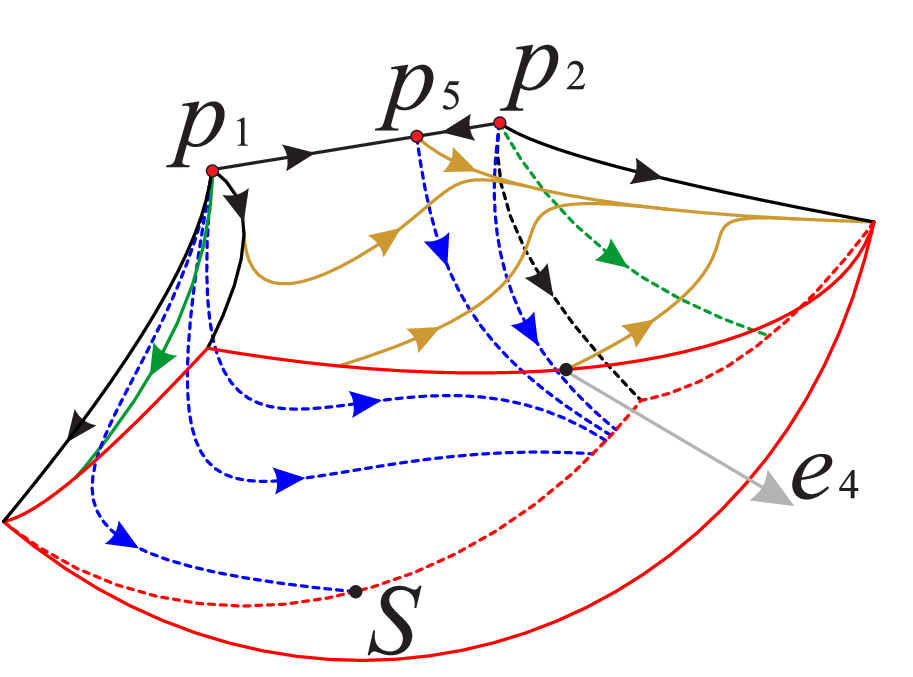}}
\subfigure[]{\label{fig:subfig:d}
    \includegraphics[width=6.5cm]{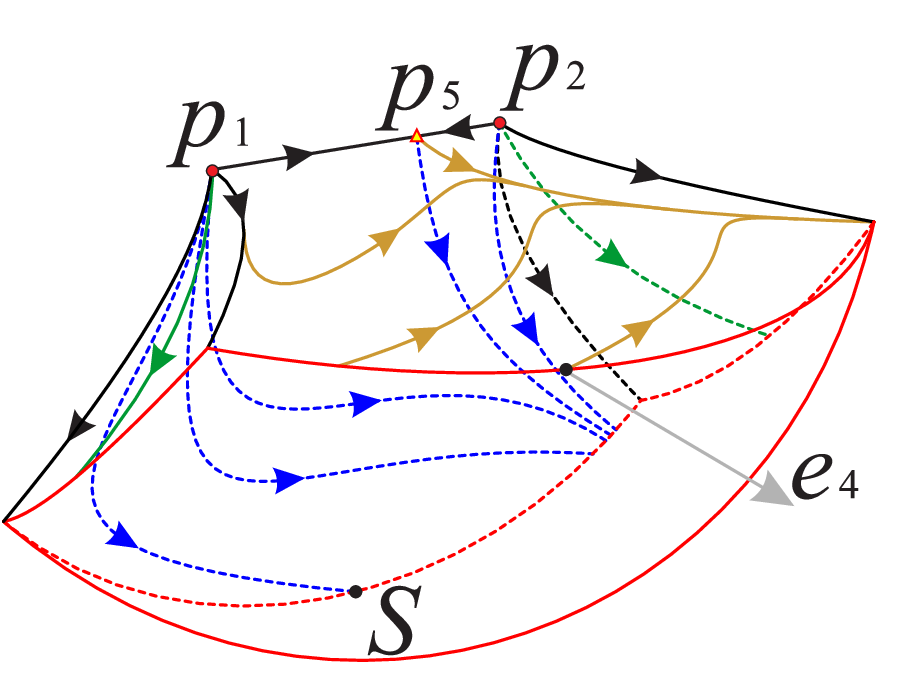}}\\
\subfigure[]{\label{fig:subfig:e}
    \includegraphics[width=6.5cm]{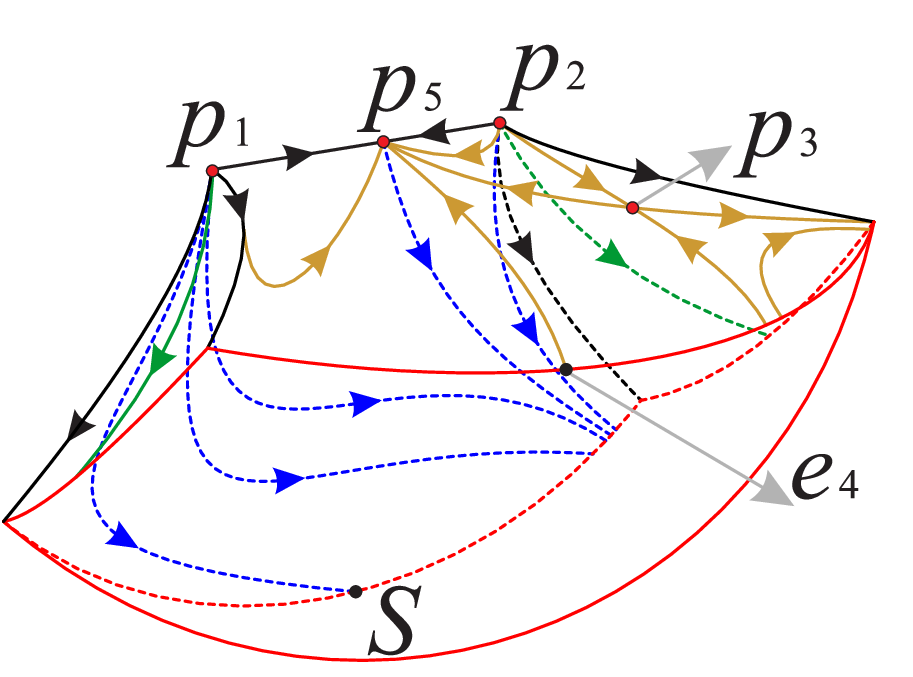}}
\subfigure{\includegraphics[width=2cm]{Fig14f15f16f17f21b23b24c26b30b32b33b34b_frame2_x_z_u.eps}}
    \caption{Phase portrait in the boundary of the region $R_2$ for different values of $s$: (a) $s<-\sqrt{6}/2$, (b) $s=-\sqrt{6}/2$, (c) $-\sqrt{6}/2<s<-1$, (d) $s=-1$, (e) $-1<s<0$. The surfaces above and below the long dashed line represent regions I and III respectively.}
  \label{fig:subfig}
  \end{minipage}
 \end{figure}
 %Figure 17
\begin{figure}[]
  \begin{minipage}{150mm}
\centering
\subfigure[]{\label{fig:subfig:a}
    \includegraphics[width=6.5cm]{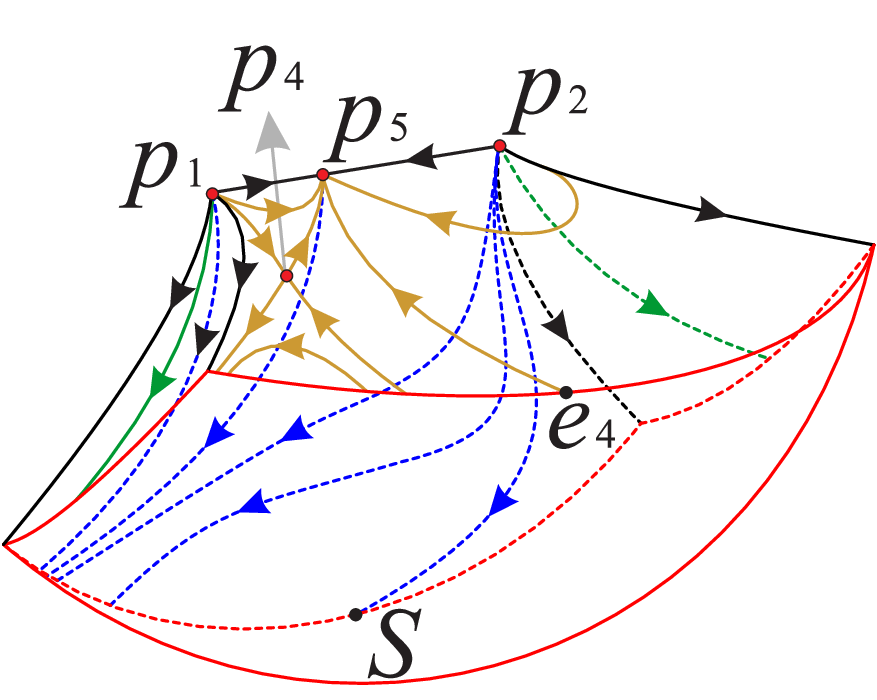}}
\subfigure[]{\label{fig:subfig:b}
    \includegraphics[width=6.5cm]{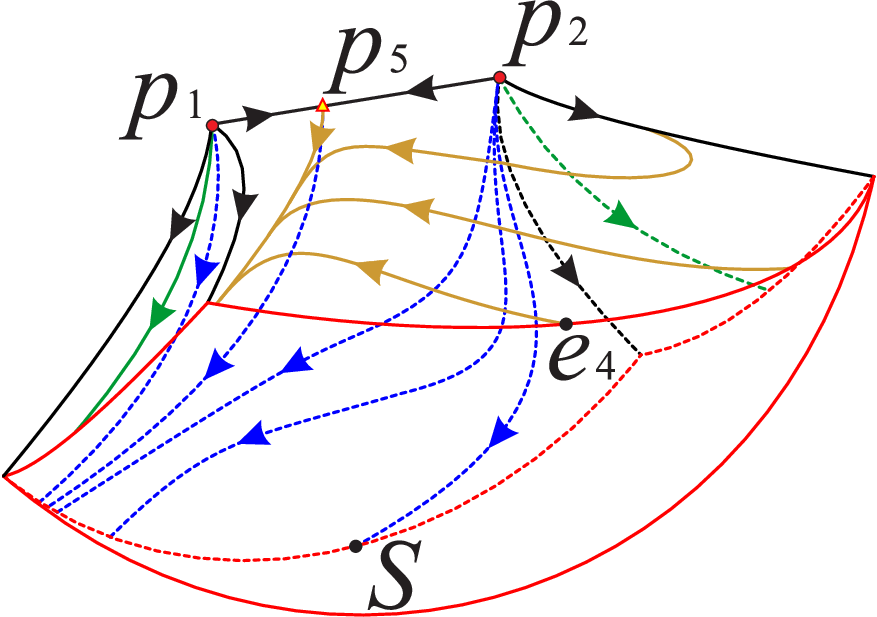}}\\
\subfigure[]{\label{fig:subfig:c}
    \includegraphics[width=6.5cm]{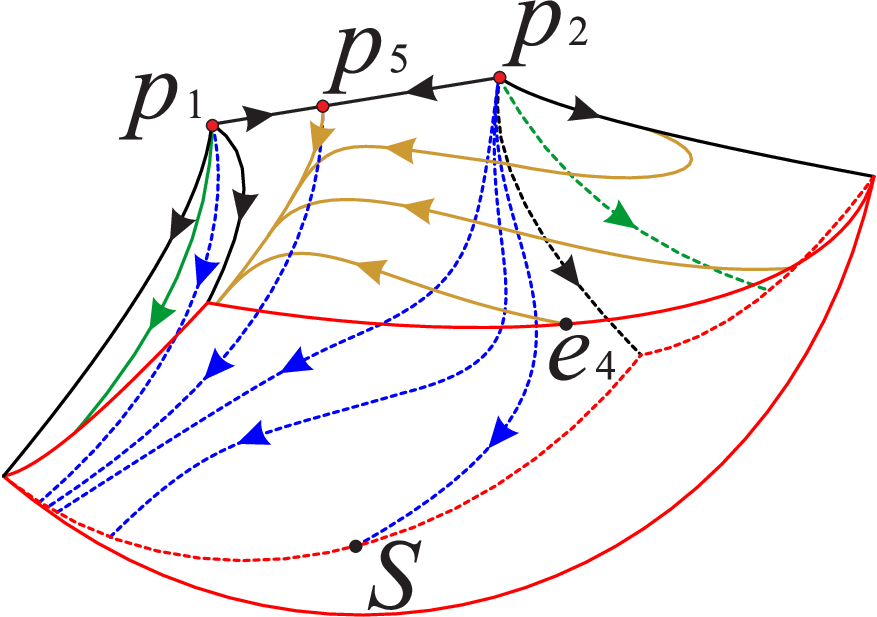}}
\subfigure[]{\label{fig:subfig:d}
    \includegraphics[width=6.5cm]{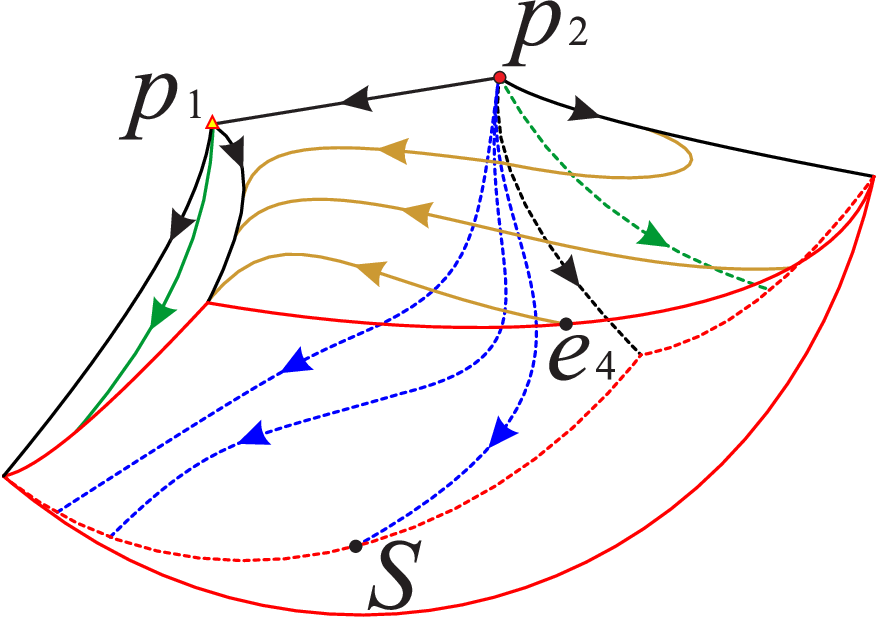}}\\
\subfigure[]{\label{fig:subfig:e}
    \includegraphics[width=6.5cm]{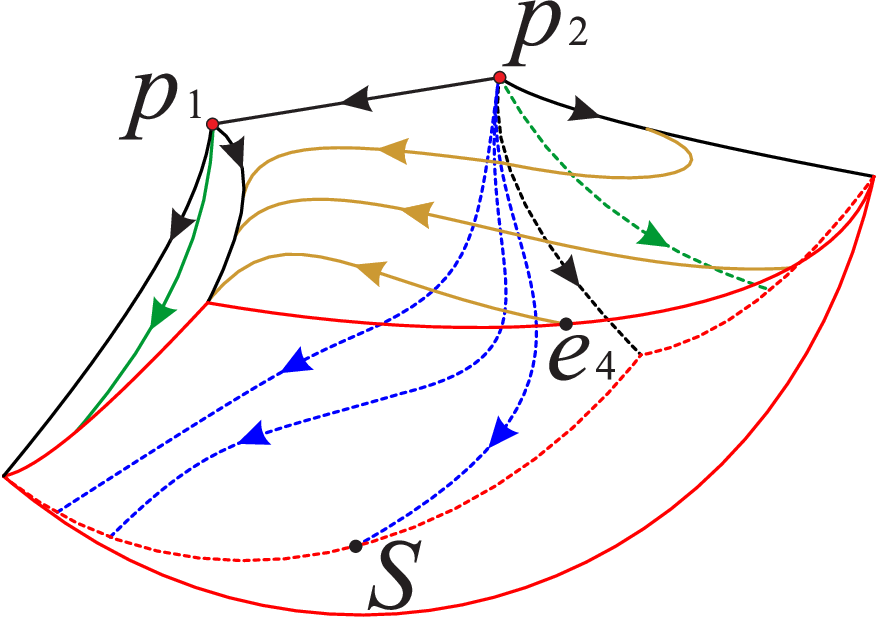}}
\subfigure{\includegraphics[width=2cm]{Fig14f15f16f17f21b23b24c26b30b32b33b34b_frame2_x_z_u.eps}}
    \caption{Phase portrait in the boundary of the region $R_2$ for different values of $s$: (a) $0<s<1$, (b) $s=1$, (c) $1<s<\sqrt{6}/2$, (d) $s=\sqrt{6}/2$, (e) $s>\sqrt{6}/2$. The surfaces above and below the long dashed line represent regions I and III respectively.}
  \label{fig:subfig}
  \end{minipage}
 \end{figure}

\begin{figure}[htbp]
\begin{minipage}[t]{130mm}
\vspace {2mm}
\centering\includegraphics[width=9cm]{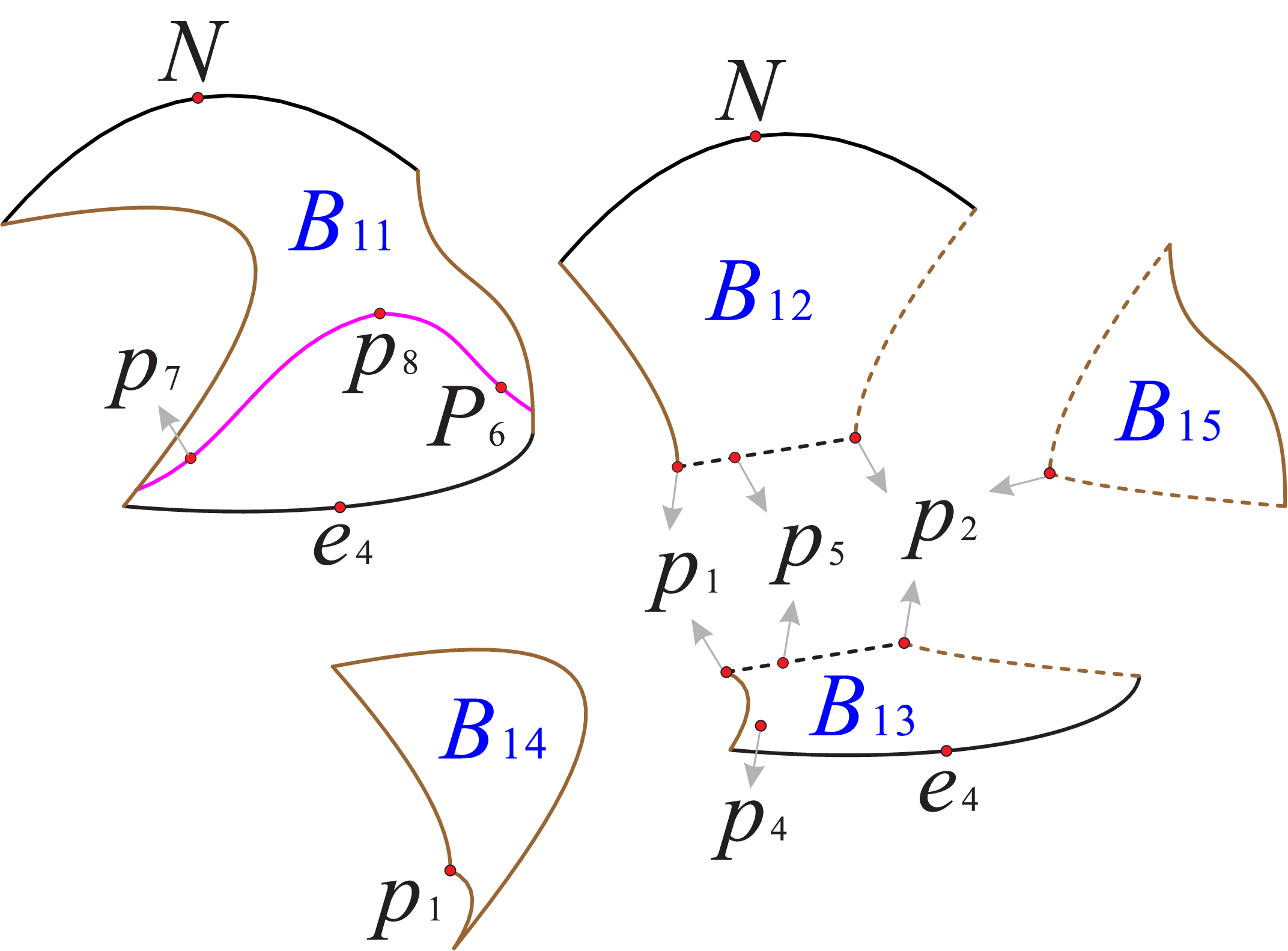}
\caption{The three boundary surfaces and two boundary planes of the region $R_1$.}
\label{fig18}
\end{minipage}
\end{figure}

\begin{figure}[htbp]
\begin{minipage}[t]{130mm}
\vspace {2mm}
\centering\includegraphics[width=9cm]{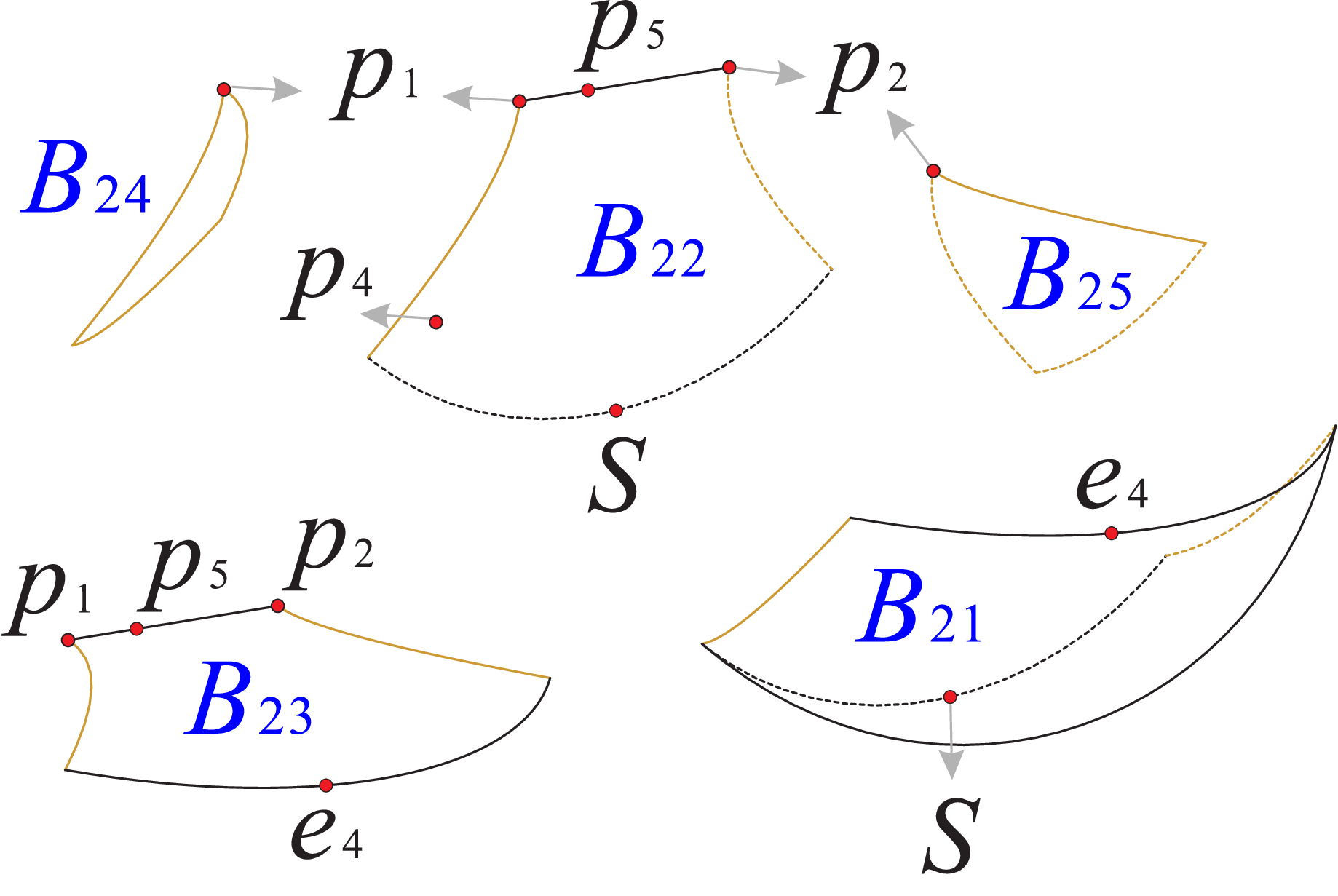}
\caption{The three boundary surfaces and two boundary planes of the region $R_2$.}
\label{fig19}
\end{minipage}
\end{figure}

Now as shown in Figures \ref{fig18} and \ref{fig19}, we divided the boundary of the regions $R_1$ and $R_2$ into sub-surfaces $B_{11}$, $B_{12}$, $\cdots$, $B_{15}$ and $B_{21}$, $B_{22}$, $\cdots$, $B_{25}$ respectively, then the phase portrait on the boundary of $R_1$ will be displayed more clearly. Therefore we can find from Figures 13, 15 and 17 that there is a hyperbolic sector at the north pole $N$ on spherical boundary $B_{11}$ of the Poincar\'e ball, and $N$ is stable on the back boundary plane $B_{12}$. The equilibrium points $p_1$ and $p_2$ are unstable on the boundary surfaces $B_{l_1l_2}\ (l_1=1,2, l_2=2,3,4)$. Moreover the equilibrium point $p_5$ is unstable on the back boundary planes $B_{12}$ and $B_{22}$, and it is stable on the bottom boundary planes $B_{13}$ and $B_{23}$ and on their intersection. In addition the properties of the remaining equilibrium points that are not located at the intersection of these boundary surfaces and planes have been studied in the previous section and will not be repeated here.
%Figure 18
%Figure 19

\subsection{Dynamics in the interior of the regions $R_1$ and $R_2$}

Without loss of generality and considering the physical region of interest, we take $s = \sqrt{6}/4$, and the dynamics of system (\ref{1}) can be studied in the same way when we take other values of $s$. Then the five finite equilibrium points of system (\ref{1}) have the form $p_1=\{1,0,0\}$, $p_2=\{-1,0,0\}$, $p_3=\{4/3,-\sqrt{15}/3,0\}$, $p_4=\{4/3,\sqrt{15}/3,0\}$ and $p_5=\{1/2,0,0\}$. The dynamical behavior of system (\ref{1}) inside the region $R_1$ is determined by the behavior of the flow in the following planes and surfaces
\begin{equation*}
\begin{array}{rl}
&z=0,\ u=0,\ f_+(x,z,u)=0,\\
&h_1(x,z,u)=0,\ h_2(x,z,u)=0,\ h_3(x,z,u)=0,
\end{array}
\end{equation*}
where
\begin{equation*}
\begin{array}{rl}
h_1(x,z,u)&=\sqrt{6} s \left[-x^2 + (u - z)^2 + 1\right] + x \left[3 x^2 + 2 (u - z) z - 3\right],\\
h_2(x,z,u)&=3 x^2 + 2 (u - z) z - 2,\\
h_3(x,z,u)&=3 x^2 + 2 (u - z) z.
\end{array}
\end{equation*}

\begin{figure}[]
\begin{minipage}{130mm}
\centering
\subfigure{
     \includegraphics[width=10cm]{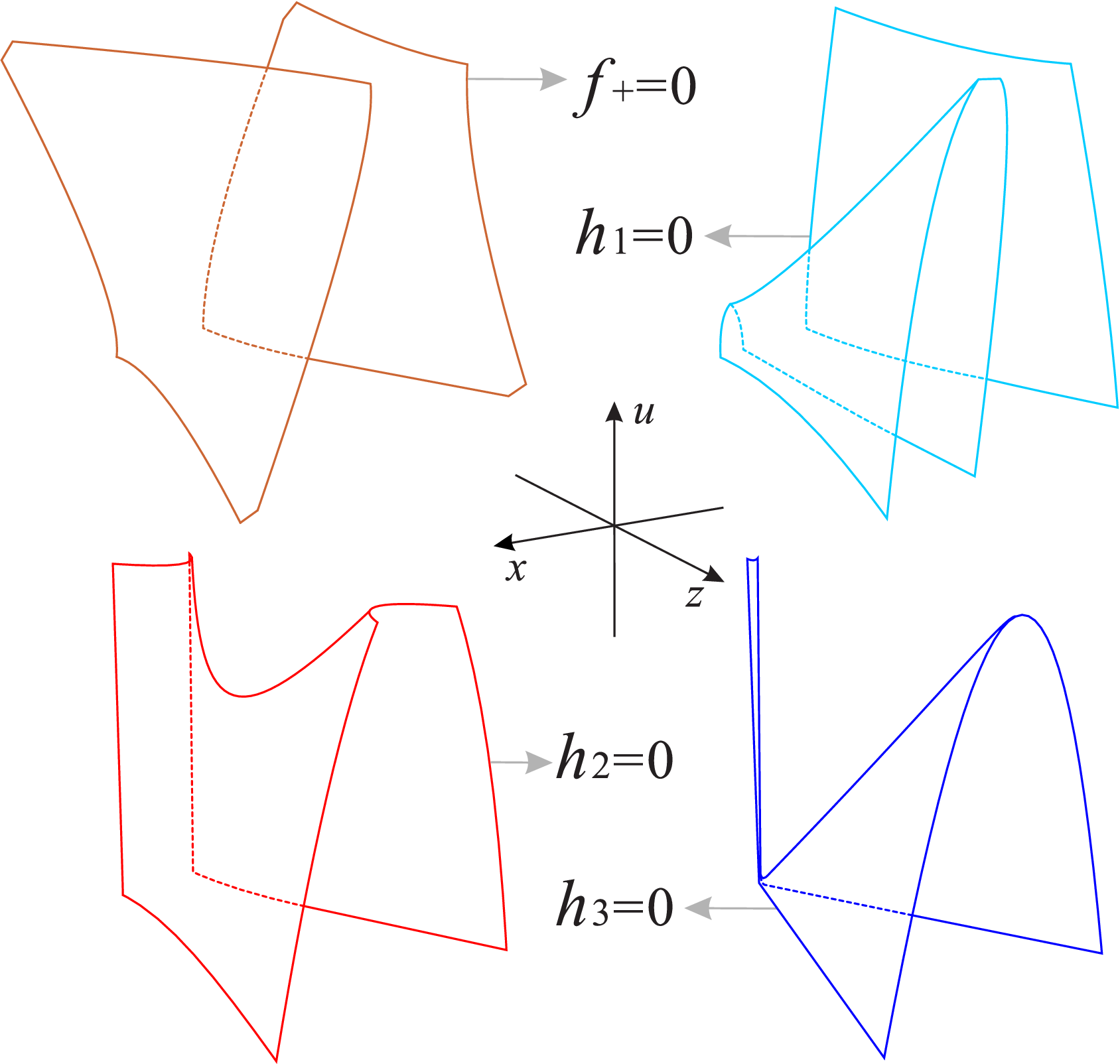}}
    \includegraphics[width=6cm]{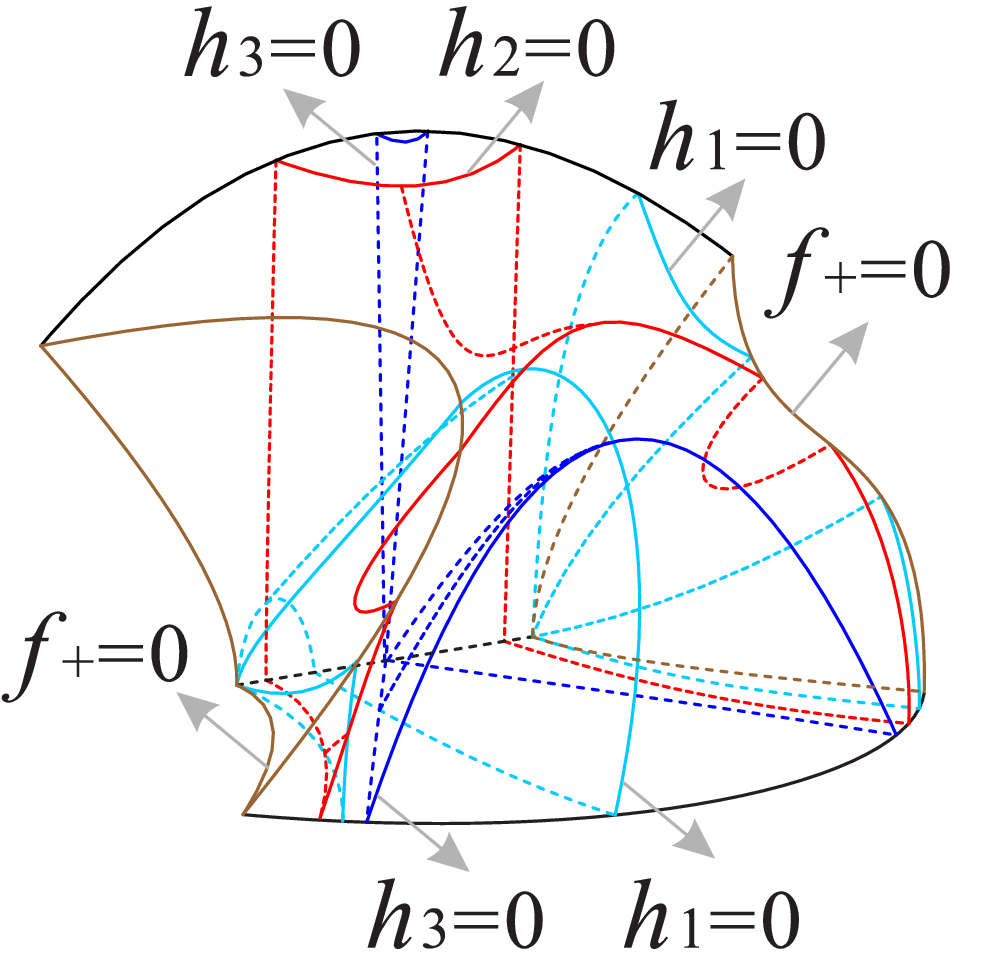}
    \caption{The surfaces $h_1$, $h_2$ and $h_3$ restricted to the invariant surface $f_+(x,z,u)= 0$ and the region $R_1$ of the Poincar\'e ball.}
  \label{fig20}
  \end{minipage}
 \end{figure}

\begin{figure}[htbp]
\begin{minipage}[t]{120mm}
\vspace {2mm}
\centering\includegraphics[width=10cm]{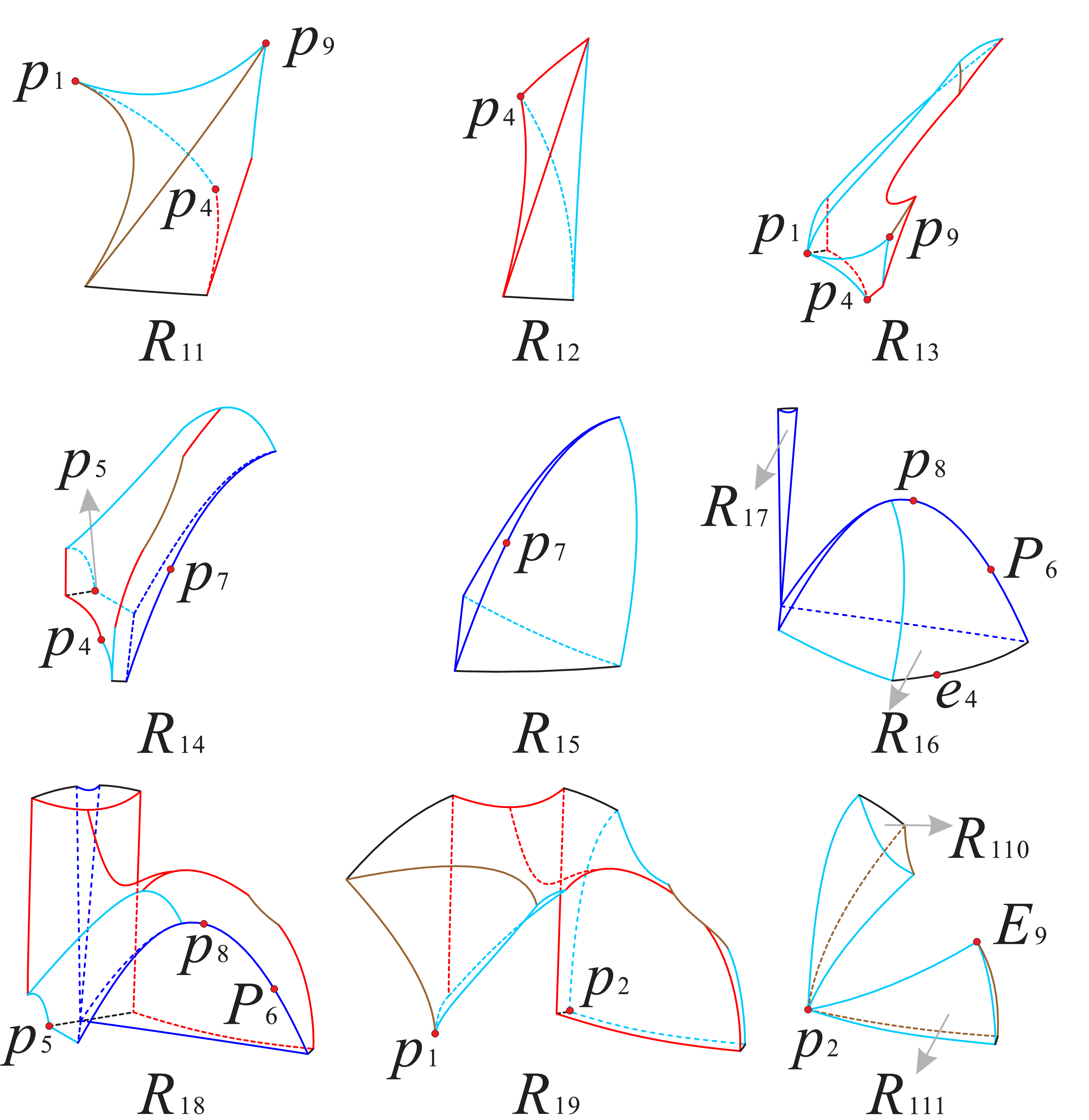}
  \subfigure{\includegraphics[width=2cm]{Fig14f15f16f17f21b23b24c26b30b32b33b34b_frame2_x_z_u.eps}}
\caption{There are the eleven subregions inside the region $R_1$ of the Poincar\'e ball.}
\label{fig21}
\end{minipage}
\end{figure}

\begin{figure}[]
\begin{minipage}{130mm}
\centering
\subfigure{\includegraphics[width=10cm]{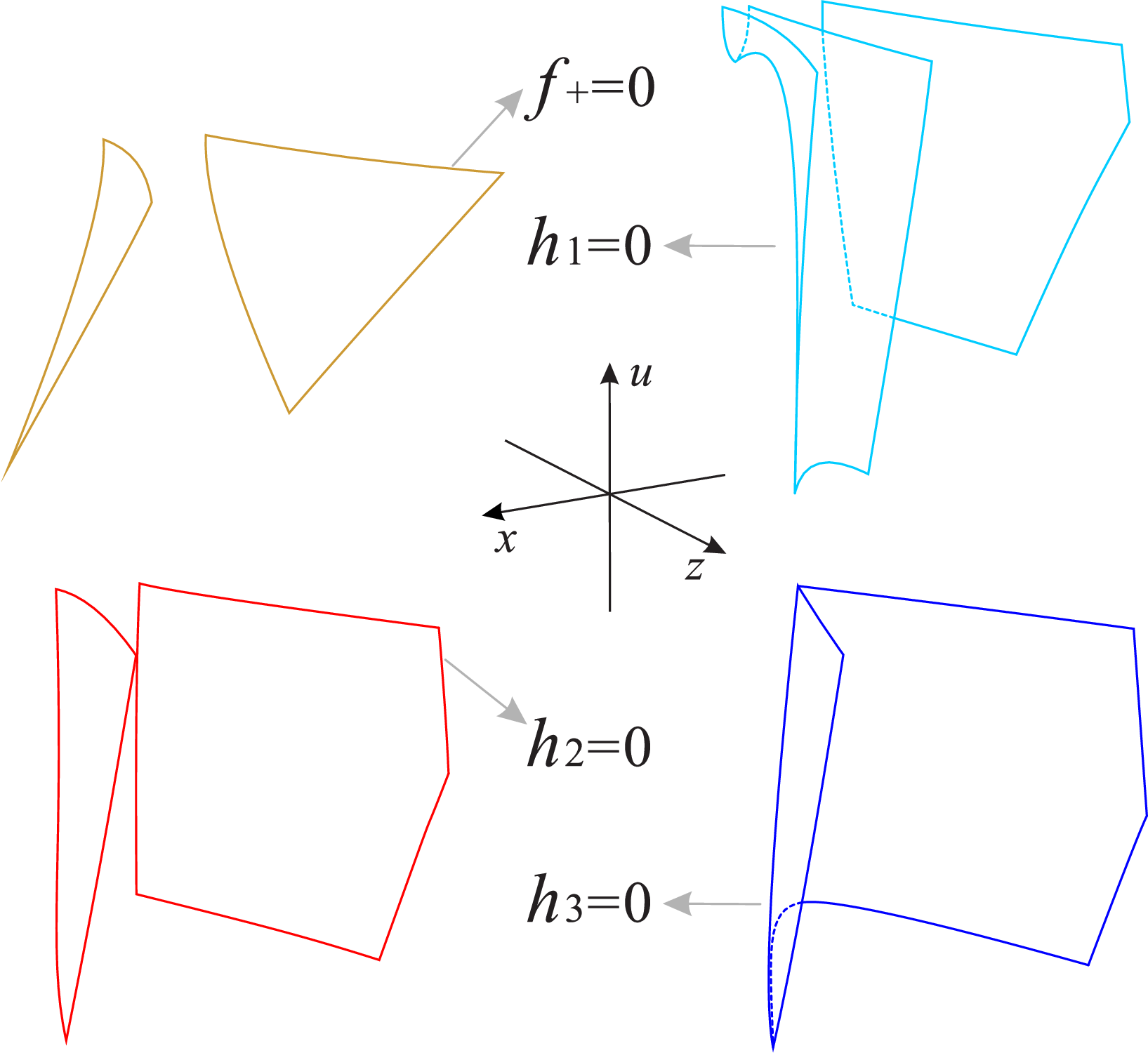}}
     \includegraphics[width=8cm]{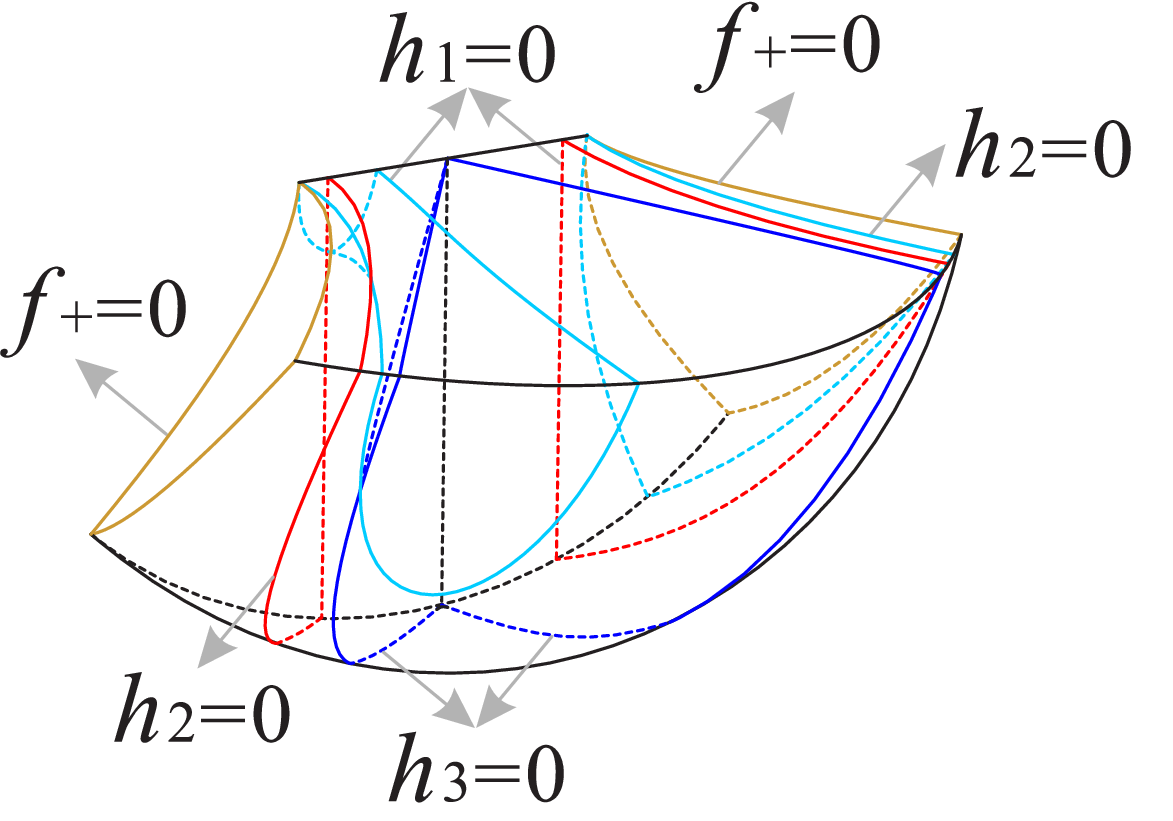}
    \caption{The surfaces $h_1$, $h_2$ and $h_3$ restricted to the invariant surface $f_+(x,z,u)= 0$ and the region $R_2$ of the Poincar\'e ball.}
  \label{fig22}
  \end{minipage}
 \end{figure}

\begin{figure}[htbp]
\begin{minipage}[t]{120mm}
\vspace {2mm}
\centering\includegraphics[width=10cm]{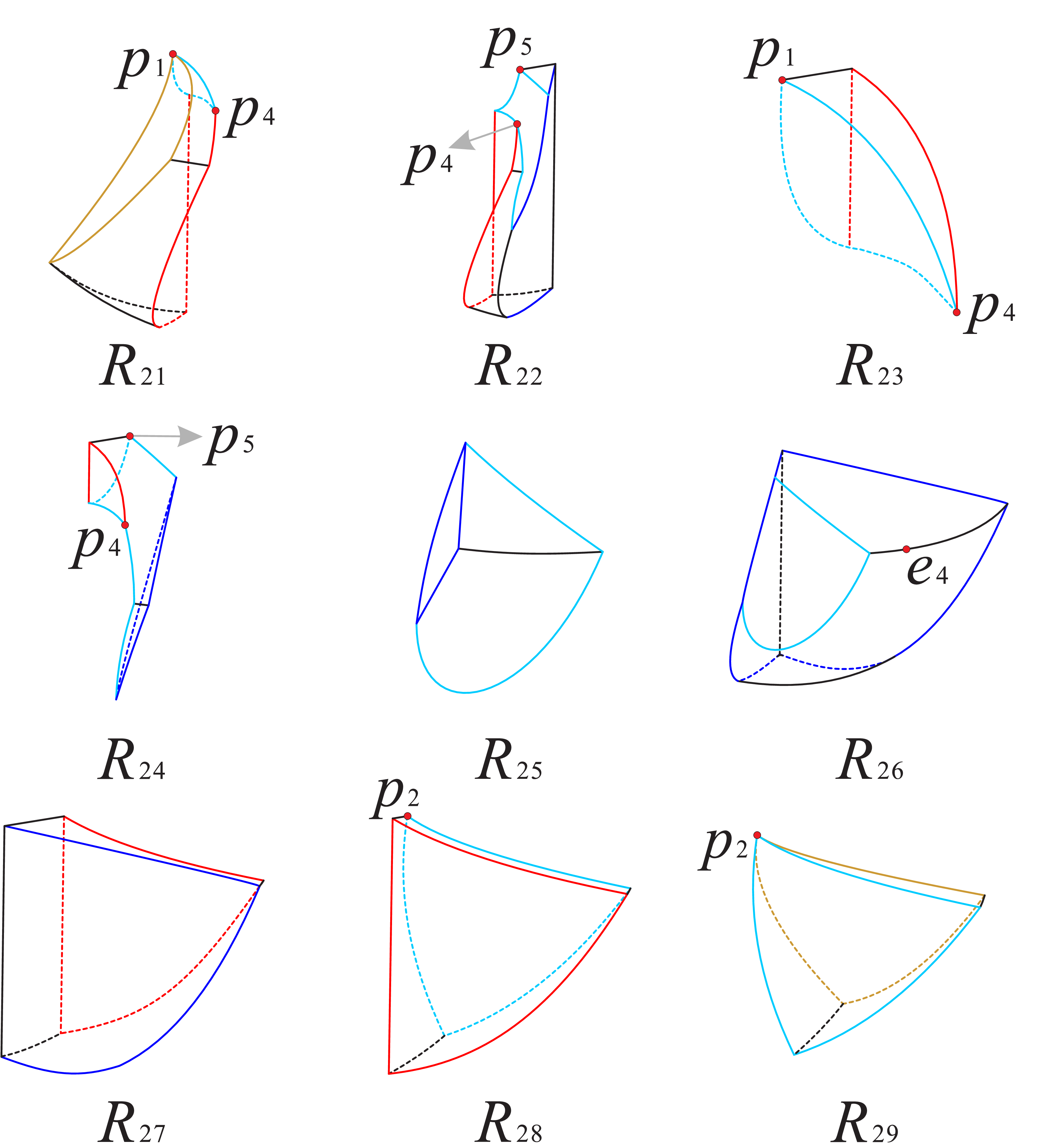}
  \subfigure{\includegraphics[width=2cm]{Fig14f15f16f17f21b23b24c26b30b32b33b34b_frame2_x_z_u.eps}}
\caption{There are the nine subregions inside the region $R_2$ of the Poincar\'e ball.}
\label{fig23}
\end{minipage}
\end{figure}

The above planes and surfaces divide the regions $R_1$ and $R_2$ into eleven subregions $R_{1i},\ i=(1,\dots,11)$ and nine subregions $R_{2j},\ j=(1,\dots,9)$ respectively, see Figures \ref{fig20}-\ref{fig23} for more details. The signs of the functions $h_1$, $h_2$ and $h_3$ in these subregions of $R_1$ and $R_2$ can be found in Tables \ref{table4} and \ref{table5} respectively.
%Figure 20
   %Figure 21
%Figure 22
%Figure 23
%Table 4
\begin{table}[!htb]
\newcommand{\tabincell}[2]{\begin{tabular}{@{}#1@{}}#2\end{tabular}}
\centering
\caption{\label{opt}Signs of functions $h_1$, $h_2$ and $h_3$ in the subregions of $R_1$.}
\footnotesize
\rm
\centering
\begin{tabular}{@{}*{12}{l}}
\specialrule{0em}{2pt}{2pt}
 \toprule
\hspace{2mm}\textbf{Functions}&\textbf{Positive}&\textbf{Negative}\\
\specialrule{0em}{2pt}{2pt}
\toprule
\tabincell{l}{\hspace{8mm}$h_1$}&\tabincell{l}{$R_{11}, R_{12}, R_{16}, R_{17}, R_{18}, R_{19}$}&\tabincell{l}{$R_{13}, R_{14}, R_{15}, R_{110}, R_{111}$}\\
\specialrule{0em}{2pt}{2pt}
\hline
\specialrule{0em}{2pt}{2pt}
\tabincell{l}{\hspace{8mm}$h_2$}&\tabincell{l}{$R_{11}, R_{13}, R_{19}, R_{110}, R_{111}$}&\tabincell{l}{$R_{12}, R_{14}, R_{15}, R_{16}, R_{17}, R_{18}$}\\
\specialrule{0em}{2pt}{2pt}
\hline
\specialrule{0em}{2pt}{2pt}
\tabincell{l}{\hspace{8mm}$h_3$}&\tabincell{l}{$R_{11}, R_{12}, R_{13}, R_{14}, R_{18}, R_{19}, R_{110}, R_{111}$}&\tabincell{l}{$R_{15}, R_{16}, R_{17}$}\\
\specialrule{0em}{2pt}{2pt}
 \toprule
\label{table4}
\end{tabular}
\end{table}
 %Table 5
\begin{table}[!htb]
\newcommand{\tabincell}[2]{\begin{tabular}{@{}#1@{}}#2\end{tabular}}
\centering
\caption{\label{opt}Signs of functions $h_1$, $h_2$ and $h_3$ in the subregions of $R_2$.}
\footnotesize
\rm
\centering
\begin{tabular}{@{}*{12}{l}}
\specialrule{0em}{2pt}{2pt}
 \toprule
\hspace{2mm}\textbf{Functions}&\textbf{Positive}&\textbf{Negative}\\
\specialrule{0em}{2pt}{2pt}
\toprule
\tabincell{l}{\hspace{8mm}$h_1$}&\tabincell{l}{$R_{21}, R_{22}, R_{26}, R_{27}, R_{28}$}&\tabincell{l}{$R_{23}, R_{24}, R_{25}, R_{29}$}\\
\specialrule{0em}{2pt}{2pt}
\hline
\specialrule{0em}{2pt}{2pt}
\tabincell{l}{\hspace{8mm}$h_2$}&\tabincell{l}{$R_{21}, R_{23}, R_{28}, R_{29}$}&\tabincell{l}{$R_{22}, R_{24}, R_{25}, R_{26}, R_{27}$}\\
\specialrule{0em}{2pt}{2pt}
\hline
\specialrule{0em}{2pt}{2pt}
\tabincell{l}{\hspace{8mm}$h_3$}&\tabincell{l}{$R_{21}, R_{22}, R_{23}, R_{24}, R_{27}, R_{28}, R_{29}$}&\tabincell{l}{$R_{25}, R_{26}$}\\
\specialrule{0em}{2pt}{2pt}
 \toprule
\label{table5}
\end{tabular}
\end{table}

In addition it should also be noted that in order to avoid visual confusion, we change the dashed lines and solid lines in Figures \ref{fig21} and \ref{fig23} to the normal perspective, instead of corresponding to the dashed lines and solid lines in Figures \ref{fig20} and \ref{fig22} respectively. What we need to pay attention here is that for any value of $s$, the finite equilibrium point $p_4$ is located at the intersection of the surfaces $h_1$ and $h_2$ on the invariant plane $u = 0$, and the finite equilibrium point $p_5$ is located at the intersection of the surface $h_1$ and $x$-axis (see Figures 21 and 23). For the infinite equilibrium point $p_7 = (3/2,1/2,0)$ of equations (25), according to the three-dimensional Poincar\'e transformation in the local chart $U_1$, we know that $p_7 = (1/z_3, 3/(2z_3), 1/(2z_3))\ (z_3\to 0)$ is on the intersection of the surface $h_3$ and the Poincar\'e sphere. Similarly in the $xzu$ coordinate system we have $p_6 = (1/z_3, -3/(2z_3), -1/(2z_3))\ (z_3\to 0)$, since the infinite equilibrium points $P_6$ and $p_6$ are symmetric about the origin, then we obtain $P_6 = (-1/z_3, 3/(2z_3), 1/(2z_3))\ (z_3\to 0)$, which is symmetric with $p_7$ with respect to the plane $x = 0$. For the infinite equilibrium point $e_9=(-1/2,0)$ of equations (20) when $u\geq z$, we combine the relationship $x = \pm\sqrt{1+y^2}$ in equations (18) and the invariant surface $f_+(x,z,u)= 0$ to know that the coordinate of $e_9$ in the three-dimensional coordinate system $xzu$ can be denoted as $p_9 = (\sqrt{1+V^2}/V, 1/(2V), -1/(2V))\ (V\to 0)$, obviously the infinite equilibrium point $p_9$ is not in the region $R_1$. But when $u \leq z$ we have $p_9 = (\sqrt{1+V^2}/V, 3/(2V), 1/(2V))\ (V\to 0)$, and it is exactly located on the surface $h_1$, on the invariant surface $f_p = 0$ and in their intersection with the Poincar\'e sphere (see Figure 21).
	
As it is shown in the subregion $R_{11}$ (see Figure \ref{fig21}) the left side surface is contained in the invariant surface $f_+(x,z,u)= 0$, the bottom plane is contained in the invariant plane $u = 0$, the right-back segment surface is contained in $h_1=0$, and the right-front segment surface is contained in $h_2=0$. From Table \ref{table6} we find that the orbits of system (\ref{1}) increase monotonically along the positive directions of the three coordinate axes, which means that the orbits in $R_{11}$ come from the finite equilibrium points $p_1$, $p_4$, or from the subregion $R_{12}$, and then go to the boundary of Poincar\'e sphere restricted to this subregion.
 %Table 6
\begin{table}[!htb]
\newcommand{\tabincell}[2]{\begin{tabular}{@{}#1@{}}#2\end{tabular}}
\centering
\caption{\label{opt}Dynamical behavior in the twenty subregions.}
\footnotesize
\rm
\centering
\begin{tabular}{@{}*{12}{l}}
\specialrule{0em}{2pt}{2pt}
 \toprule
\hspace{2mm}\textbf{Subregions}&\textbf{Corresponding Region}&\textbf{Increase or decrease}\\
\specialrule{0em}{2pt}{2pt}
\toprule
\tabincell{l}{\hspace{8mm}$R_{11}$}&\tabincell{l}{$h_1>0,\ h_2>0,\ h_3>0,\ z>0,\ u>0$}&\tabincell{l}{$\dot{x}>0,\ \dot{z}>0,\ \dot{u}>0$}\\
\specialrule{0em}{2pt}{2pt}
\hline
\specialrule{0em}{2pt}{2pt}
\tabincell{l}{\hspace{8mm}$R_{12}$}&\tabincell{l}{$h_1>0,\ h_2<0,\ h_3>0,\ z>0,\ u>0$}&\tabincell{l}{$\dot{x}>0,\ \dot{z}<0,\ \dot{u}>0$}\\
\specialrule{0em}{2pt}{2pt}
\hline
\specialrule{0em}{2pt}{2pt}
\tabincell{l}{\hspace{8mm}$R_{13}$}&\tabincell{l}{$h_1<0,\ h_2>0,\ h_3>0,\ z>0,\ u>0$}&\tabincell{l}{$\dot{x}<0,\ \dot{z}>0,\ \dot{u}>0$}\\
\specialrule{0em}{2pt}{2pt}
\hline
\specialrule{0em}{2pt}{2pt}
\tabincell{l}{\hspace{8mm}$R_{14}$}&\tabincell{l}{$h_1<0,\ h_2<0,\ h_3>0,\ z>0,\ u>0$}&\tabincell{l}{$\dot{x}<0,\ \dot{z}<0,\ \dot{u}>0$}\\
\specialrule{0em}{2pt}{2pt}
\hline
\specialrule{0em}{2pt}{2pt}
\tabincell{l}{\hspace{8mm}$R_{15}$}&\tabincell{l}{$h_1<0,\ h_2<0,\ h_3<0,\ z>0,\ u>0$}&\tabincell{l}{$\dot{x}<0,\ \dot{z}<0,\ \dot{u}<0$}\\
\specialrule{0em}{2pt}{2pt}
\hline
\specialrule{0em}{2pt}{2pt}
\tabincell{l}{\hspace{8mm}$R_{16}$}&\tabincell{l}{$h_1>0,\ h_2<0,\ h_3<0,\ z>0,\ u>0$}&\tabincell{l}{$\dot{x}>0,\ \dot{z}<0,\ \dot{u}<0$}\\
\specialrule{0em}{2pt}{2pt}
\hline
\specialrule{0em}{2pt}{2pt}
\tabincell{l}{\hspace{8mm}$R_{17}$}&\tabincell{l}{$h_1>0,\ h_2<0,\ h_3<0,\ z>0,\ u>0$}&\tabincell{l}{$\dot{x} >0,\ \dot{z}<0,\ \dot{u}<0$}\\
\specialrule{0em}{2pt}{2pt}
\hline
\specialrule{0em}{2pt}{2pt}
\tabincell{l}{\hspace{8mm}$R_{18}$}&\tabincell{l}{$h_1>0,\ h_2<0,\ h_3>0,\ z>0,\ u>0$}&\tabincell{l}{$\dot{x} >0,\ \dot{z}<0,\ \dot{u}>0$}\\
\specialrule{0em}{2pt}{2pt}
\hline
\specialrule{0em}{2pt}{2pt}
\tabincell{l}{\hspace{8mm}$R_{19}$}&\tabincell{l}{$h_1>0,\ h_2>0,\ h_3>0,\ z>0,\ u>0$}&\tabincell{l}{$\dot{x}>0,\ \dot{z}>0,\ \dot{u}>0$}\\
\specialrule{0em}{2pt}{2pt}
\hline
\specialrule{0em}{2pt}{2pt}
\tabincell{l}{\hspace{8mm}$R_{110}$}&\tabincell{l}{$h_1<0,\ h_2>0,\ h_3>0,\ z>0,\ u>0$}&\tabincell{l}{$\dot{x}<0,\ \dot{z}>0,\ \dot{u}>0$}\\
\specialrule{0em}{2pt}{2pt}
\hline
\specialrule{0em}{2pt}{2pt}
\tabincell{l}{\hspace{8mm}$R_{111}$}&\tabincell{l}{$h_1<0,\ h_2>0,\ h_3>0,\ z>0,\ u>0$}&\tabincell{l}{$\dot{x}<0,\ \dot{z}>0,\ \dot{u}>0$}\\
\specialrule{0em}{2pt}{2pt}
\hline
\specialrule{0em}{2pt}{2pt}
\tabincell{l}{\hspace{8mm}$R_{21}$}&\tabincell{l}{$h_1>0,\ h_2>0,\ h_3>0,\ z>0,\ u<0$}&\tabincell{l}{$\dot{x}>0,\ \dot{z}>0,\ \dot{u}<0$}\\
\specialrule{0em}{2pt}{2pt}
\hline
\specialrule{0em}{2pt}{2pt}
\tabincell{l}{\hspace{8mm}$R_{22}$}&\tabincell{l}{$h_1>0,\ h_2<0,\ h_3>0,\ z>0,\ u<0$}&\tabincell{l}{$\dot{x}>0,\ \dot{z}<0,\ \dot{u}<0$}\\
\specialrule{0em}{2pt}{2pt}
\hline
\specialrule{0em}{2pt}{2pt}
\tabincell{l}{\hspace{8mm}$R_{23}$}&\tabincell{l}{$h_1<0,\ h_2>0,\ h_3>0,\ z>0,\ u<0$}&\tabincell{l}{$\dot{x}<0,\ \dot{z}>0,\ \dot{u}<0$}\\
\specialrule{0em}{2pt}{2pt}
\hline
\specialrule{0em}{2pt}{2pt}
\tabincell{l}{\hspace{8mm}$R_{24}$}&\tabincell{l}{$h_1<0,\ h_2<0,\ h_3>0,\ z>0,\ u<0$}&\tabincell{l}{$\dot{x}<0,\ \dot{z}<0,\ \dot{u}<0$}\\
\specialrule{0em}{2pt}{2pt}
\hline
\specialrule{0em}{2pt}{2pt}
\tabincell{l}{\hspace{8mm}$R_{25}$}&\tabincell{l}{$h_1<0,\ h_2<0,\ h_3<0,\ z>0,\ u<0$}&\tabincell{l}{$\dot{x}<0,\ \dot{z}<0,\ \dot{u}>0$}\\
\specialrule{0em}{2pt}{2pt}
\hline
\specialrule{0em}{2pt}{2pt}
\tabincell{l}{\hspace{8mm}$R_{26}$}&\tabincell{l}{$h_1>0,\ h_2<0,\ h_3<0,\ z>0,\ u<0$}&\tabincell{l}{$\dot{x}>0,\ \dot{z}<0,\ \dot{u}>0$}\\
\specialrule{0em}{2pt}{2pt}
\hline
\specialrule{0em}{2pt}{2pt}
\tabincell{l}{\hspace{8mm}$R_{27}$}&\tabincell{l}{$h_1>0,\ h_2<0,\ h_3>0,\ z>0,\ u<0$}&\tabincell{l}{$\dot{x}>0,\ \dot{z}<0,\ \dot{u}<0$}\\
\specialrule{0em}{2pt}{2pt}
\hline
\specialrule{0em}{2pt}{2pt}
\tabincell{l}{\hspace{8mm}$R_{28}$}&\tabincell{l}{$h_1>0,\ h_2>0,\ h_3>0,\ z>0,\ u<0$}&\tabincell{l}{$\dot{x}>0,\ \dot{z} >0,\ \dot{u}<0$}\\
\specialrule{0em}{2pt}{2pt}
\hline
\specialrule{0em}{2pt}{2pt}
\tabincell{l}{\hspace{8mm}$R_{29}$}&\tabincell{l}{$h_1<0,\ h_2>0,\ h_3>0,\ z>0,\ u<0$}&\tabincell{l}{$\dot{x}<0,\ \dot{z}>0,\ \dot{u}<0$}\\
\specialrule{0em}{2pt}{2pt}
 \toprule
\label{table6}
\end{tabular}
\end{table}

In the subregion $R_{12}$, its bottom plane is contained in the invariant plane $u = 0$, the left side surface is contained in the surface $h_2=0$, and the right side surface is contained in the surface $h_1=0$. Then we can see that the orbits of system (\ref{1}) in this subregion monotonically decrease along the positive direction of the $z$-axis, but increase monotonically along the positive direction of the $x$-axis and $u$-axis, so the orbits start at the infinite equilibrium points on the Poincar\'e sphere in this subregion and then enter into the subregion $R_{11}$.

The bottom plane in the subregion $R_{13}$ is contained in the invariant plane $u = 0$, the left side surface is contained in the surface $h_1=0$, the right side surface is contained in the surface $h_2=0$, and the backplane is contained in the invariant plane $z = 0$, The upper and lower surfaces of the front are composed of the intersection of the surfaces $h_1=0$, $h_2=0$ and the invariant surface $f_+(x,z,u)= 0$ on the Poincar\'e sphere. The orbits in this subregion increase monotonically along the positive $z$-axis and $u$-axis, but decrease monotonically along the positive $x$-axis, so the orbits start at the finite point $p_1$, and then cross the side surfaces of this subregion and eventually go to the subregion $R_{14}$ respectively.

The bottom plane of the subregion $R_{14}$ is contained in the invariant plane $u = 0$, the left surface is contained in the surface composed of the surfaces $h_1=0$, $h_2=0$ and the invariant plane $f_+(x,z,u)= 0$, and the right surface is contained in the surface $h_3=0$, the back surface is contained in the surface $h_1=0$, the backplane is contained in the invariant plane $z = 0$, the front surface is contained in the surface enclosed by the intersection lines of the surfaces $h_1=0$, $h_2=0$, $h_3=0$ and the invariant surface $f_+(x,z,u)= 0$ on the Poincar\'e sphere. The orbits in this subregion increase monotonically along the positive $u$-axis, and decrease monotonically along the positive two other coordinate axes, which indicates that the orbits of this subregion start from the finite point $p_4$ or the equilibrium points on the Poincar\'e sphere at infinity in this subregion, or come from the subregion $R_{13}$ and finally enter into subregions $R_{15}$ or $R_{18}$.

The bottom plane of the subregion $R_{15}$ is contained in the invariant plane $u = 0$, the front surface is contained in the Poincar\'e sphere, the left surface is contained in the surface $h_3=0$, and the right surface is contained in the surface $h_1=0$. The dynamic behavior of the orbits in this subregion are the same as that in the subregion $R_{14}$, and they decrease monotonically along the positive direction of the three coordinate axes, which means that the orbits of start from subregion $R_{14}$ or the infinite equilibrium points on the Poincar\'e sphere in this subregion, and then enter into subregion $R_{16}$.

The bottom plane of the subregion $R_{16}$ is contained in the invariant plane $u = 0$, the left and right surfaces are contained in the surface $h_3 = 0$, the front left surface is contained in the surface $h_1 = 0$ and the front right surface is contained in Poincar\'e sphere. The orbits in this subregion increase monotonically along the positive $x$-axis, and decrease monotonically along the positive two other coordinate axes, thereby the orbits originate from the infinite equilibrium points on the Poincar\'e sphere in this subregion, and then enter into the subregion $R_{18}$.

The front surface of the subregion $R_{17}$ is contained in the surface $h_3 = 0$, and the back surface is contained in the invariant plane $z = 0$. The dynamic behavior of the orbits in this subregion is the same as in the subregion $R_{16}$, they come from the subregion $R_{18}$, after crossing the right part boundary surface of the subregion $R_{17}$ and then from its left boundary surface back to the subregion $R_{18}$.

The bottom plane of the subregion $R_{18}$ is contained in the invariant plane $u = 0$, the left surface is contained in the surface $h_1 = 0$, and the upper surface and the right surface are contained in the surface $h_2 = 0$ and the invariant surface $f_+ =0$, the front surface is contained in the Poincar\'e sphere, the surface below the front surface is contained in the surface $h_3 = 0$, and the back surface is composed of the invariant plane $z = 0$ and the surface $h_3 = 0$. The orbits in this subregion decrease monotonically in the positive direction along the $z$-axis, and decrease monotonically in the positive direction along the other two coordinate axes, for this reason the orbits start at the equilibrium points on the Poincar\'e sphere at infinity or come from the subregions $R_{14}$, $R_{16}$, the left side part surface of $R_{17}$, and then tend to the subregion $R_{19}$.

The backplane of the subregion $R_{19}$ is divided into the left and right parts by the surface $h_2 = 0$, but both of them are contained in the invariant plane $z = 0$. The left and right side surfaces of $R_{19}$ are contained in the surface composed of the surface $h_1 = 0$ and the invariant surface $f_+(x,z,u)= 0$, the top surface is contained in the Poincar\'e sphere, the lower-right surface is contained in the surface $h_2 = 0$, and the lower-left surface is contained in the surface $h_1 = 0$. The orbits in this subregion are monotonically increasing along the positive direction of the three coordinate axes, so the orbits come from the finite equilibrium points $p_1$ and $p_2$, or from the subregion $R_{18}$, and finally approach the equilibrium points on the Poincar\'e sphere at infinity.

The subregions $R_{110}$ and $R_{111}$ are connected by the finite equilibrium point $p_2$, their front surfaces are contained in the surface $h_1 = 0$, the right surface of $R_{110}$ and the back surface of $R_{111}$ are contained in the invariant surface $f_+(x,z,u)= 0$, the back surface of $R_{110}$ is contained in the invariant plane $z = 0$, the top surface of $R_{110}$ and the surface on the right side of $R_{111}$ are contained in the Poincar\'e sphere. The orbits in these two subregions decrease monotonically along the positive direction of the $x$-axis, and monotonically increase along the positive direction along the other two coordinate axes. Thus the orbits in these two subregions start at the equilibrium point $p_2$, and then eventually go to the corresponding infinite equilibrium point on the Poincar\'e sphere.

Therefore the dynamic process of the orbits inside the eleven subregions of $R_1$ discussed above can be summarized as
$$\begin{tikzpicture}[->, thick]
 \node (R12) at (-3.77, 2.1) [] {$R_{12}$};
  \node (R11) at (-2.37, 2.1) [] {$R_{11}$};
    \node (p4) at (-2.37, 1.0) [] {$p_4$};
      \node (PSR14) at (-0.26, 1.0) [] {PS in $R_{14}$};
  \node (PSR11) at (-0.26, 2.1) [] {PS in $R_{11}$};
  \node (PSR12) at (-5.75, 2.1) [] {PS in $R_{12}$ };
 \node (p2) at (-5.08, 0) [] {$p_2$};
  \node (R111) at (-7.01, 0) [] {PS in $R_{111}$};
  \node (R19) at (-3.8, -1.17) [] {$R_{19}$};
  \node (p1) at (-3.8, 1) [] {$p_1$};
  \node (PSR19) at (-5.78, -1.17) [] {PS in $R_{19}$};
  \node (PSR110) at (-5.9, 1) [] {PS in $R_{110}$};
  \node (R15) at (-0.08, 0) [] {$R_{15}$};
    \node (PSR15) at (1.9, 0) [] {PS in $R_{15}$};
  \node (R14) at (-1.5, 0) [] {$R_{14}$};
  \node (R13) at (-2.91, 0) [] {$R_{13}$};
  \node (R18) at (-1.5, -1.17) [] {$R_{18}$};
  \node (R16) at (-0.08, -1.17) [] {$R_{16}$};
      \node (PSR16) at (1.9, -1.17) [] {PS in $R_{16}$};
  \node (lR17) at (-0.11, -2.17) [] {LPS of $R_{17}$.};
  \node (rR17) at (-2.93, -2.17) [] {RPS of $R_{17}$};
\draw (PSR12) -- (R12) ;
\draw (R12) -- (R11) ;
\draw (R11) -- (PSR11) ;
\draw (p4) -- (R11) ;
\draw (p4) -- (R14) ;
\draw (PSR14) -- (R14) ;
\draw (p2) -- (R111) ;
\draw (p2) -- (PSR110) ;
\draw (p2) -- (R19) ;
\draw (p1) -- (R19) ;
\draw (p1) -- (R11) ;
\draw (p1) -- (R13) ;
\draw (R19) -- (PSR19) ;
\draw (R18) -- (R19) ;
\draw (R13) -- (R14) ;
\draw (R14) -- (R18) ;
\draw (R14) -- (R15) ;
\draw (R15) -- (R16) ;
\draw (PSR15) -- (R15) ;
\draw (PSR16) -- (R16) ;
\draw (R18) -- (rR17) ;
\draw (rR17) -- (lR17) ;
\draw (lR17) -- (R18) ;
\draw (R16) -- (R18) ;
\end{tikzpicture}$$
{\footnotesize{Note: PS represents Poincar\'e sphere.\\
\indent\hspace{4.4mm} RPS denotes the right part of surface.\\
\indent\hspace{4.4mm} LPS stands for the left part of surface.}}

The flow chart discussed above shows that the orbits of system (\ref{1}) contained in the region $R_1$ have an $\omega$-limit in the subregions $R_{11}$, $R_{19}$, $R_{110}$ and $R_{111}$ when these four subregions are restricted to the Poincar\'e sphere at infinity. Furthermore the orbits have an $\alpha$-limit at the finite equilibrium points $p_1$, $p_2$ and $p_4$. In addition the orbits also have an $\omega$-limit in the subregions $R_{12}$, $R_{14}$, $R_{15}$ and $R_{16}$ when they are confined to the Poincar\'e sphere at infinity.

For the subregions in $R_2$ (see Figure 23), the front and bottom surfaces of the subregion $R_{21}$ are contained in the Poincar\'e sphere, the back surface is contained in the invariant plane $z = 0$, the top surface is contained in the invariant plane $u = 0$, and the left surface is contained in the invariant plane $f_+(x,z,u)= 0$, the lower right surface is contained in the surface $h_2 = 0$, and the upper right surface is contained in the surface $h_1 = 0$. The orbits in this subregion increase monotonically along the positive direction of the $x$-axis and $z$-axis, and the monotonically decrease along the positive direction of the $u$-axis, so the orbits in this subregion start from the finite equilibrium points $p_1$, $p_4$ and the right surface $h_2 = 0$, and finally approaching the point of infinity at Poincar\'e sphere.

The front and bottom surfaces of the subregion $R_{22}$ are contained in the Poincar\'e sphere, the top plane is divided into two parts by the surface $h_1 = 0$, but these two parts are contained in the same invariant plane $u = 0$, the left side surface is contained in the surface $h_2 = 0$, the right side surface is contained in the surface $h_3 = 0$, and the back surface is contained in the invariant plane $z = 0$. The orbits in this subregion increase monotonically along the positive direction of the $x$-axis, and decrease along the positive direction of the other two axes. Therefore the orbits in this subregion start from the connected subregions $R_{24}$ and $R_{26}$, they enter into this subregion through the surfaces $h_1 = 0$ and $h_3 = 0$, and then go through the surface $h_2 = 0$ into the subregion $R_{21}$.

The top and back planes of the subregion $R_{23}$ are contained in the invariant plane $u = 0$ and the invariant plane $z = 0$ respectively, the left surface is contained in the surface $h_1 = 0$, and the right surface is contained in the surface $h_2 = 0$. The orbits in this subregion increase monotonically along the positive direction of the $z$-axis and decrease monotonically along the positive direction of the other two axes, which indicates that the orbits in this subregion start at the finite equilibrium point $p_1$ and cross the surface $h_2 = 0$, and then enter into the subregion $R_{24}$.

The top and back planes of the subregion $R_{24}$ are also contained in the invariant plane $u = 0$ and the invariant plane $z = 0$ respectively, the bottom surface is contained in the surface $h_1 = 0$, and the left surface is contained in the surface $h_2 = 0$, the right surface is contained in the surface $h_3 = 0$, and the front surface is contained in the Poincar\'e sphere. Then the orbits in this subregion are monotonically decreasing along the positive direction of the three axes, which means that the orbits in this subregion start at the finite equilibrium point $p_4$, or come from the subregion $R_{23}$ and the infinite equilibrium points on the Poincar\'e sphere, and then directly enter into the subregions $R_{22}$ and $R_{25}$ through the surfaces $h_1 = 0$ and $h_3 = 0$ respectively.

The top plane of the subregion $R_{25}$ is contained in the invariant plane $u = 0$, the bottom surface is contained in the surface $h_1 = 0$, the left surface is contained in the surface $h_3 = 0$, and the front surface is contained in the Poincar\'e sphere. The orbits in this subregion are monotonically decreasing along the positive direction of the $x$-axis and $z$-axis, but monotonically increasing along the positive direction of the $u$-axis, which indicates that the orbits in this subregion start from the infinity equilibrium point on the Poincar\'e sphere, then cross the surface $h_1 = 0$ into the subregion $R_{26}$.

The top plane of the subregion $R_{26}$ is also contained in the invariant plane $u = 0$, the right and bottom surfaces are contained in the Poincar\'e sphere, the front surface is contained in the surface $h_1 = 0$, and the back-left and back-right surfaces are contained in the surface $h_3 = 0$. The orbits in this subregion are monotonically decreasing along the positive direction of the $z$-axis, but are increasing monotonically along the positive directions of the remaining two axes, then the orbits in this subregion start from the subregion $R_{27}$ and cross the back-right surface $h_3 = 0$ or from the infinite equilibrium points on the Poincar\'e sphere, and then cross the back-left surface $h_3 = 0$ into the subregion $R_{22}$.

The subregion $R_{27}$ consists of the top plane contained in the invariant plane $u = 0$, the back plane contained in the invariant plane $z = 0$, the back surface contained in the surface $h_2 = 0$, the front surface contained in the surface $h_3 = 0$, and the bottom surface contained in the Poincar\'e sphere. The orbits in this subregion are monotonically increasing along the positive $x$-axis, but are monotonically decreasing along the positive directions of the remaining two axes, so they start from the subregion $R_{28}$ or the equilibrium points on the Poincar\'e sphere at infinity, then go through the left surface $h_3 = 0$ to the subregion $R_{26}$.

The subregion $R_{28}$ consists of the top plane contained in the invariant plane $u = 0$, the left plane contained in the invariant plane $z = 0$, the front surface contained in the surface $h_2= 0$, the back surface contained in $h_1 = 0$, and the right surface contained in the Poincar\'e sphere. All the orbits in this subregion are monotonically decreasing along the positive direction of the $u$-axis, and monotonically increasing along the positive direction of the other two axes, so they start from the finite equilibrium point $p_2$ and then enter into the subregion $R_{27}$ through the front surface $h_2 = 0$.

The top plane of the subregion $R_{29}$ is contained in the invariant plane $u = 0$, the left plane is contained in the invariant plane $z = 0$, the right surface is contained in the Poincar\'e sphere, and the front surface is contained in the surface $h_1 = 0$, the back surface is contained in the constant surface $f_+(x,z,u)= 0$. The orbits in this subregion are monotonically increasing along the positive direction of the $z$-axis, and monotonically decreasing along the positive direction of the other two axes, which means that the orbits start at the finite equilibrium point $p_2$ and finally tend to the equilibrium points on Poincar\'e sphere at infinity.

Therefore the dynamic behavior of the orbits inside the nine subregions of $R_2$ discussed above can be represented as
$$\begin{tikzpicture}[->, thick]
 \node (R26) at (0, 0) [] {$R_{26}$};
  \node (R22) at (1.4, 0) [] {$R_{22}$};
  \node (R21) at (2.8, 0) [] {$R_{21}$};
  \node (p1) at (4.15, 0) [] {$p_1$.};
  \node (R24) at (1.4, 2.24) [] {$R_{24}$};
  \node (p4) at (2.8, 1.12) [] {$p_4$};
  \node (R23) at (4.15, 2.24) [] {$R_{23}$};
  \node (R25) at (0, 1.12) [] {$R_{25}$};
    \node (PSR21) at (2.8, -1.12) [] {PS in $R_{21}$};
  \node (PSR25) at (-2, 1.12) [] {PS in $R_{25}$};
   \node (PSR26) at (0, -1.12) [] {PS in $R_{26}$};
  \node (R27) at (-1.4, 0) [] {$R_{27}$};
  \node (PSR27) at (-3.38, 0) [] {PS in $R_{27}$ };
  \node (R28) at (-1.4, -1.12) [] {$R_{28}$};
  \node (R29) at (-2.7, -2.24) [] {$R_{29}$};
  \node (p2) at (-2.7, -1.12) [] {$p_2$};
  \node (PSR29) at (-0.71, -2.24) [] {PS in $R_{29}$};
\draw (R26) -- (R22) ;
\draw (R22) -- (R21) ;
\draw (p1) -- (R21) ;
\draw (R24) -- (R22) ;
\draw (p4) -- (R21) ;
\draw (p4) -- (R24) ;
\draw (R23) -- (R24) ;
\draw (R24) -- (R25) ;
\draw (p1) -- (R23) ;
\draw (R25) -- (R26) ;
\draw (PSR25) -- (R25) ;
\draw (PSR26) -- (R26) ;
\draw (R27) -- (R26) ;
\draw (PSR27) -- (R27) ;
\draw (R28) -- (R27) ;
\draw (p2) -- (R29) ;
\draw (p2) -- (R28) ;
\draw (R29) -- (PSR29) ;
\draw (R21) -- (PSR21) ;
\end{tikzpicture}$$

The flow chart in the region $R_2$ discussed above shows that the orbits of system (\ref{1}) contained in this region have an $\alpha$-limit at the finite equilibrium points $p_1$, $p_2$ and $p_4$, as well as on the Poincar\'e sphere restricted to the subregions $R_{25}$, $R_{26}$ and $R_{27}$. Moreover the orbits have an $\omega$-limit on the Poincar\'e sphere restricted to the subregions $R_{21}$ and $R_{29}$. Therefore we fully describe all qualitative global dynamic behaviors of system (\ref{1}).

\section{Case II: $s \neq 0, k=-1$}
	
For an open universe the phase portraits of system (\ref{1}) on the invariant planes $z = 0$ and $u = 0$ are the same as those in sections 3.1.1 and 3.1.2. Moreover the local phase portraits of the finite and infinite equilibrium points on the Poincar\'e sphere are consistent with sections 3.1.4 and 3.1.5 respectively. However the physical region of interest $f_-(x,z,u) = 0$ is no more an invariant surface.

Taking the same approach as in section 3.2, we also divide the Poincar\'e ball restricted to the region $x^2-(u+z)^2\leq1$ into four regions as below
\begin{equation*}
\begin{array}{rl}
S_1:\ z\geq0,\ u\geq0.\ \ \
S_2:\ z\geq0,\ u\leq0.\\
S_3:\ z\leq0,\ u\leq0.\ \ \
S_4:\ z\leq0,\ u\geq0.
\end{array}
\end{equation*}
and we shall only study the phase portrait of system (\ref{1}) in the regions $S_1$ and $S_2$ due to the symmetry as mentioned before. The phase portrait of system (\ref{1}) on the boundaries of these two regions are the same as the closed universe except for the boundary surfaces $B_{14}$, $B_{15}$, $B_{24}$, $B_{25}$ contained on the surface $f_-(x,z,u) = 0$ (see Figures 14-19).

\subsection{Dynamics in the interior of the regions $S_1$ and $S_2$}

\begin{figure}[]
\begin{minipage}{130mm}
\centering
\subfigure[]{\label{fig:subfig:a}
    \includegraphics[width=7cm]{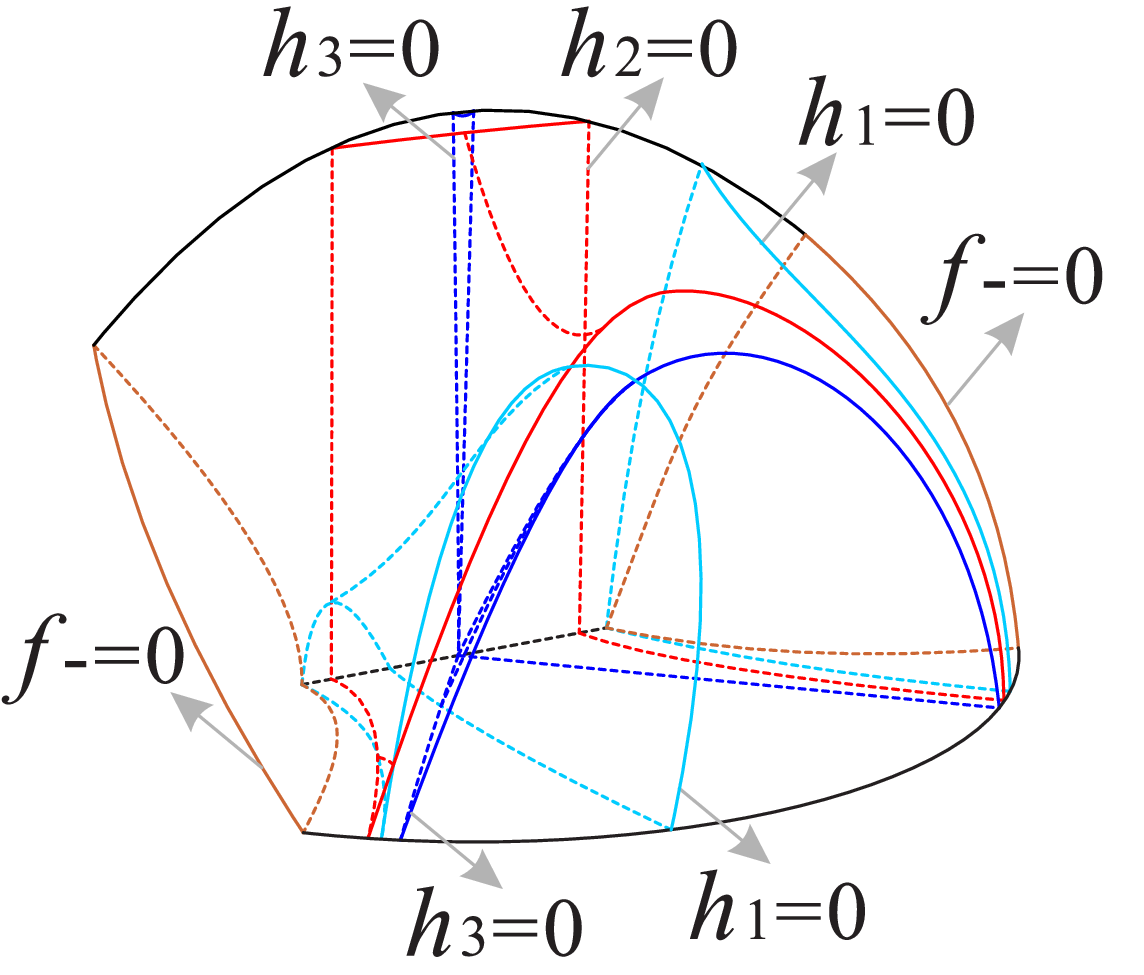}}
\subfigure[]{\label{fig:subfig:b}
    \includegraphics[width=7cm]{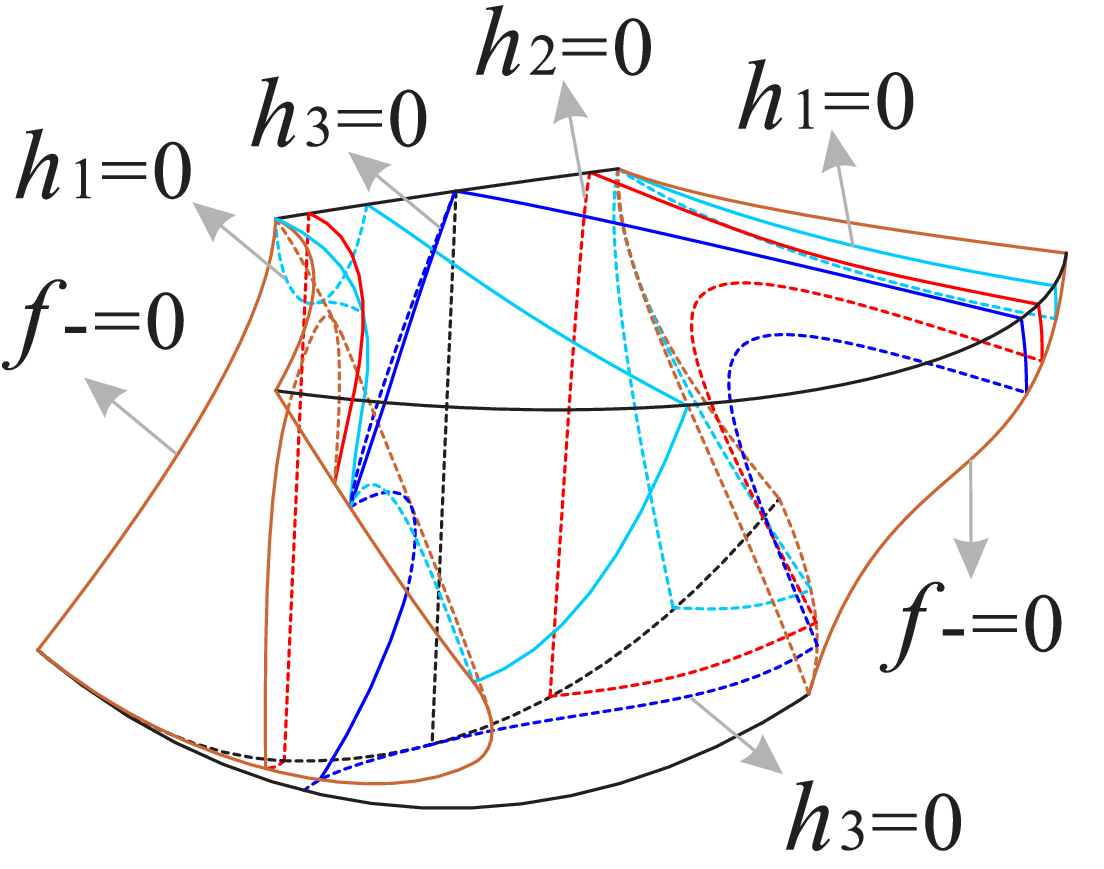}}\\
\subfigure{\includegraphics[width=2cm]{Fig14f15f16f17f21b23b24c26b30b32b33b34b_frame2_x_z_u.eps}}
    \caption{The surfaces $h_1$, $h_2$ and $h_3$ restricted to the surface $f_-(x,z,u)= 0$ and the regions $S_1$ and $S_2$ of the Poincar\'e ball, respectively.}
  \label{fig24}
  \end{minipage}
 \end{figure}

\begin{figure}[htbp]
\begin{minipage}[t]{120mm}
\vspace {2mm}
\centering\includegraphics[width=10cm]{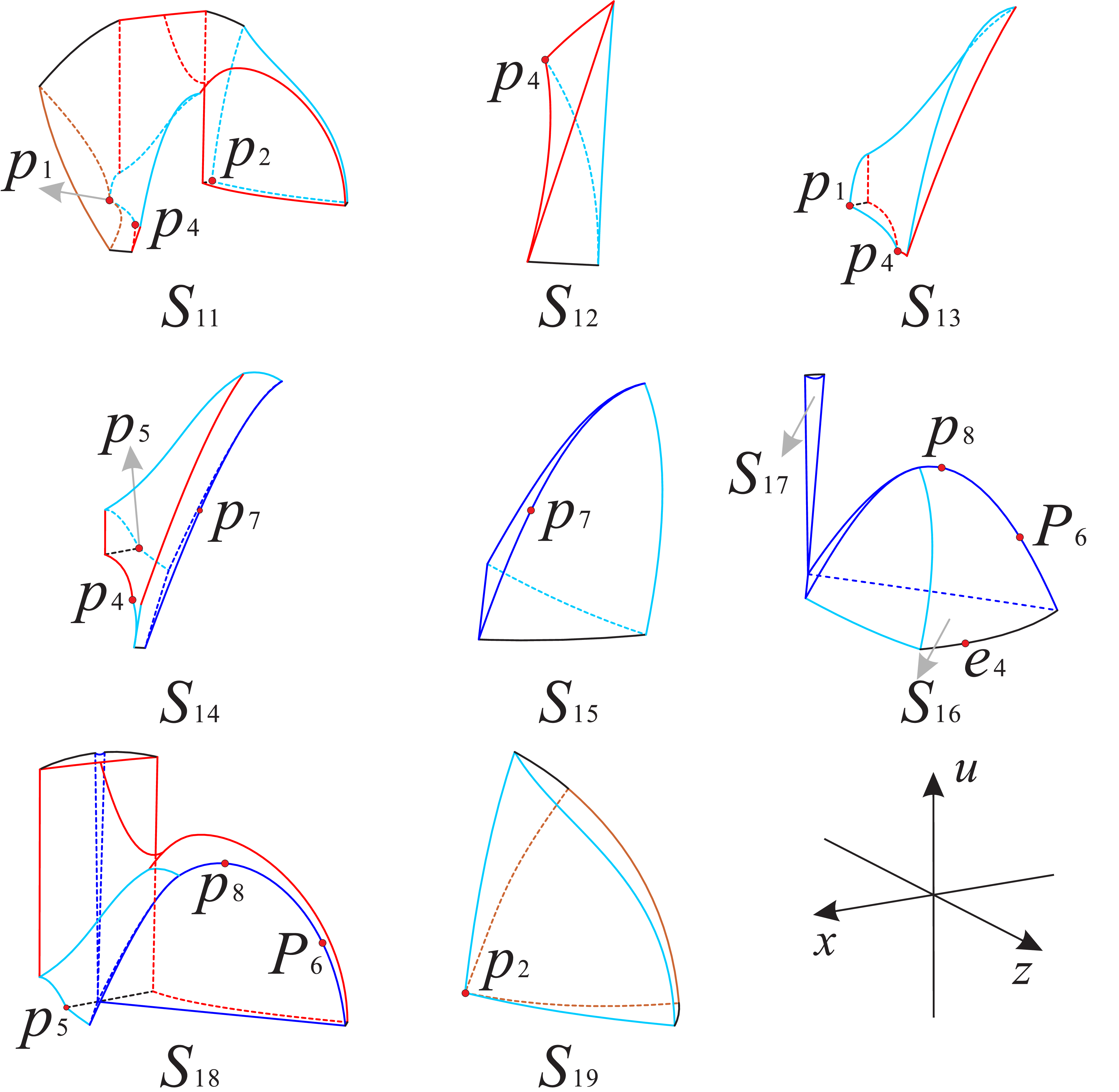}
\caption{The nine subregions inside the region $S_1$ of the Poincar\'e ball.}
\label{fig25}
\end{minipage}
\end{figure}

\begin{figure}[htbp]
\begin{minipage}[t]{120mm}
\vspace {2mm}
\centering\includegraphics[width=10cm]{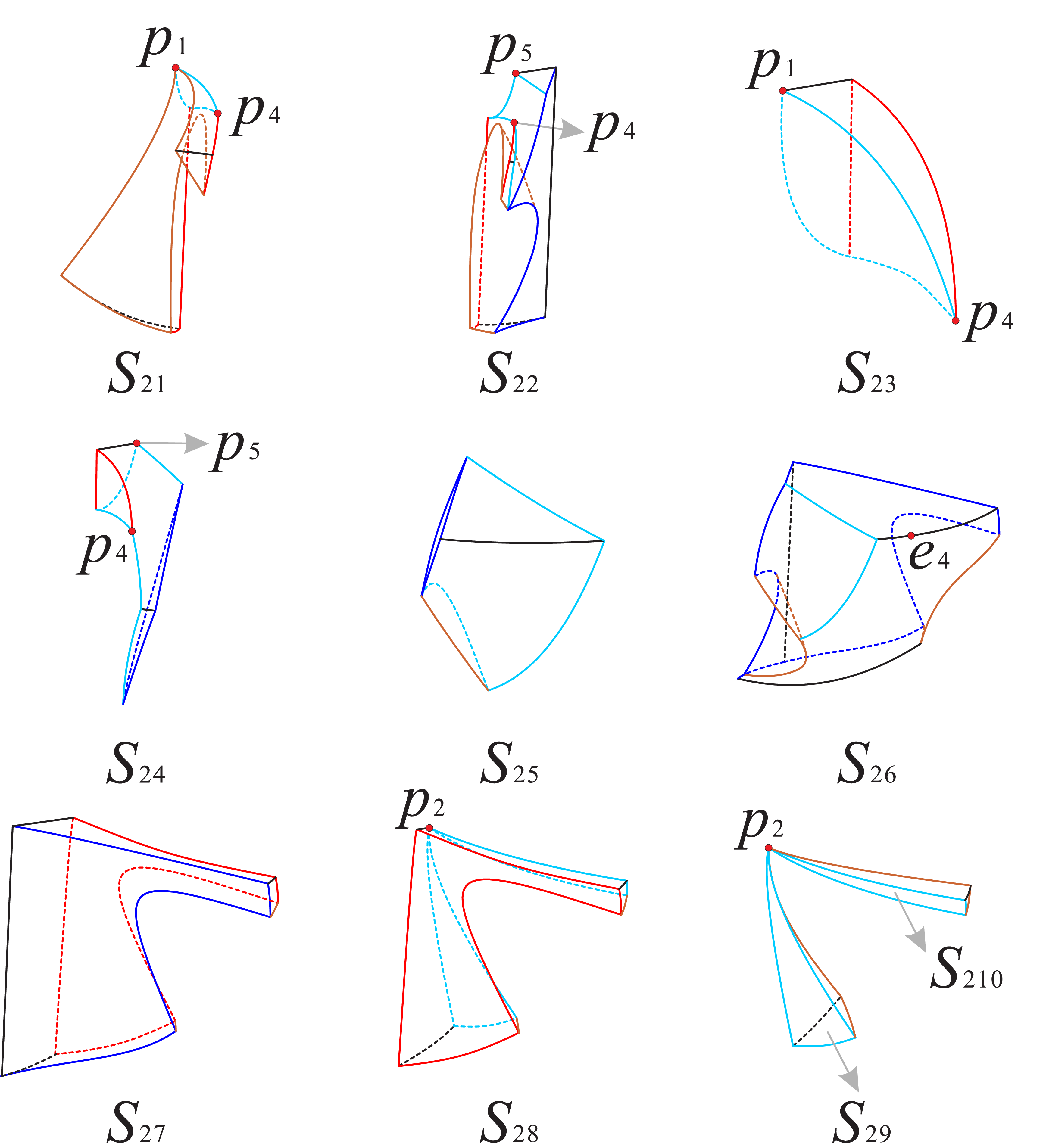}
  \subfigure{\includegraphics[width=2cm]{Fig14f15f16f17f21b23b24c26b30b32b33b34b_frame2_x_z_u.eps}}
\caption{The ten subregions inside the region $S_2$ of the Poincar\'e ball.}
\label{fig26}
\end{minipage}
\end{figure}

For convenience we continue the discussion of the case $s = \sqrt{6}/4$. The invariant planes $z=0$, $u=0$, and the surfaces $h_1$, $h_2$, $h_3$ as well as $f_-(x,z,u)=0$ divide the regions $S_1$ and $S_2$ into nine subregions $S_{11},\ S_{12},\ \cdots, S_{19}$ and ten subregions $S_{21},\ S_{22},\ \cdots, S_{210}$ respectively, see Figures \ref{fig24}-\ref{fig26} for more details. The signs of the functions $h_1$, $h_2$ and $h_3$ in these subregions of $S_1$ and $S_2$ can be found in Tables \ref{table7} and \ref{table8} respectively.
 %Figure 24
 %Figure 25
%Figure 26
 %Table 7
\begin{table}[!htb]
\newcommand{\tabincell}[2]{\begin{tabular}{@{}#1@{}}#2\end{tabular}}
\centering
\caption{\label{opt}Signs of functions $h_1$, $h_2$ and $h_3$ in the subregions of $S_1$.}
\footnotesize
\rm
\centering
\begin{tabular}{@{}*{12}{l}}
\specialrule{0em}{2pt}{2pt}
 \toprule
\hspace{2mm}\textbf{Functions}&\textbf{Positive}&\textbf{Negative}\\
\specialrule{0em}{2pt}{2pt}
\toprule
\tabincell{l}{\hspace{8mm}$h_1$}&\tabincell{l}{$S_{11}, S_{12}, S_{16}, S_{17}, S_{18}$}&\tabincell{l}{$S_{13}, S_{14}, S_{15}, S_{19}$}\\
\specialrule{0em}{2pt}{2pt}
\hline
\specialrule{0em}{2pt}{2pt}
\tabincell{l}{\hspace{8mm}$h_2$}&\tabincell{l}{$S_{11}, S_{13}, S_{19}$}&\tabincell{l}{$S_{12}, S_{14}, S_{15}, S_{16}, S_{17}, S_{18}$}\\
\specialrule{0em}{2pt}{2pt}
\hline
\specialrule{0em}{2pt}{2pt}
\tabincell{l}{\hspace{8mm}$h_3$}&\tabincell{l}{$S_{11}, S_{12}, S_{13}, S_{14}, S_{18}, S_{19}$}&\tabincell{l}{$S_{15}, S_{16}, S_{17}$}\\
\specialrule{0em}{2pt}{2pt}
 \toprule
\label{table7}
\end{tabular}
\end{table}
 %Table 8
\begin{table}[!htb]
\newcommand{\tabincell}[2]{\begin{tabular}{@{}#1@{}}#2\end{tabular}}
\centering
\caption{\label{opt}Signs of functions $h_1$, $h_2$ and $h_3$ in the subregions of $S_2$.}
\footnotesize
\rm
\centering
\begin{tabular}{@{}*{12}{l}}
\specialrule{0em}{2pt}{2pt}
 \toprule
\hspace{2mm}\textbf{Functions}&\textbf{Positive}&\textbf{Negative}\\
\specialrule{0em}{2pt}{2pt}
\toprule
\tabincell{l}{\hspace{8mm}$h_1$}&\tabincell{l}{$S_{21}, S_{22}, S_{26}, S_{27}, S_{28}$}&\tabincell{l}{$S_{23}, S_{24}, S_{25}, S_{29}, S_{210}$}\\
\specialrule{0em}{2pt}{2pt}
\hline
\specialrule{0em}{2pt}{2pt}
\tabincell{l}{\hspace{8mm}$h_2$}&\tabincell{l}{$S_{21}, S_{23}, S_{28}, S_{29}, S_{210}$}&\tabincell{l}{$S_{22}, S_{24}, S_{25}, S_{26}, S_{27}$}\\
\specialrule{0em}{2pt}{2pt}
\hline
\specialrule{0em}{2pt}{2pt}
\tabincell{l}{\hspace{8mm}$h_3$}&\tabincell{l}{$S_{21}, S_{22}, S_{23}, S_{24}, S_{27}, S_{28}, S_{29}, S_{210}$}&\tabincell{l}{$S_{25}, S_{26}$}\\
\specialrule{0em}{2pt}{2pt}
 \toprule
\label{table8}
\end{tabular}
\end{table}

As shown in Figure \ref{fig25} the top surface of the subregion $S_{11}$ is contained in the Poincar\'e sphere, the bottom surface is contained in the surfaces $h_1=0$ and $h_2=0$, the bottom plane is contained in the invariant plane $u = 0$, and the left and right sides are contained in the surfaces $f_-= 0$ and $h_1=0$, the back plane is contained in the invariant plane $z = 0$. Noting the dynamic behavior of the orbits in Table \ref{table9}, we find that all the orbits in this subregion increase monotonically, which means that the orbits in this subregion start at the finite equilibrium points $p_1$, $p_2$ and $p_4$, or come from the subregion $S_{12}$, eventually approaching the infinite equilibrium points on the Poincar\'e sphere.
\begin{table}[!htb]
\newcommand{\tabincell}[2]{\begin{tabular}{@{}#1@{}}#2\end{tabular}}
\centering
\caption{\label{opt}Dynamical behavior in the nineteen subregions.}
\footnotesize
\rm
\centering
\begin{tabular}{@{}*{12}{l}}
\specialrule{0em}{2pt}{2pt}
 \toprule
\hspace{2mm}\textbf{Subregions}&\textbf{Corresponding Region}&\textbf{Increase or decrease}\\
\specialrule{0em}{2pt}{2pt}
\toprule
\tabincell{l}{\hspace{8mm}$S_{11}$}&\tabincell{l}{$h_1>0,\ h_2>0,\ h_3>0,\ z>0,\ u>0$}&\tabincell{l}{$\dot{x}>0,\ \dot{z}>0,\ \dot{u}>0$}\\
\specialrule{0em}{2pt}{2pt}
\hline
\specialrule{0em}{2pt}{2pt}
\tabincell{l}{\hspace{8mm}$S_{12}$}&\tabincell{l}{$h_1>0,\ h_2<0,\ h_3>0,\ z>0,\ u>0$}&\tabincell{l}{$\dot{x}>0,\ \dot{z}<0,\ \dot{u}>0$}\\
\specialrule{0em}{2pt}{2pt}
\hline
\specialrule{0em}{2pt}{2pt}
\tabincell{l}{\hspace{8mm}$S_{13}$}&\tabincell{l}{$h_1<0,\ h_2>0,\ h_3>0,\ z>0,\ u>0$}&\tabincell{l}{$\dot{x}<0,\ \dot{z}>0,\ \dot{u}>0$}\\
\specialrule{0em}{2pt}{2pt}
\hline
\specialrule{0em}{2pt}{2pt}
\tabincell{l}{\hspace{8mm}$S_{14}$}&\tabincell{l}{$h_1<0,\ h_2<0,\ h_3>0,\ z>0,\ u>0$}&\tabincell{l}{$\dot{x}<0,\ \dot{z}<0,\ \dot{u}>0$}\\
\specialrule{0em}{2pt}{2pt}
\hline
\specialrule{0em}{2pt}{2pt}
\tabincell{l}{\hspace{8mm}$S_{15}$}&\tabincell{l}{$h_1<0,\ h_2<0,\ h_3<0,\ z>0,\ u>0$}&\tabincell{l}{$\dot{x}<0,\ \dot{z}<0,\ \dot{u}<0$}\\
\specialrule{0em}{2pt}{2pt}
\hline
\specialrule{0em}{2pt}{2pt}
\tabincell{l}{\hspace{8mm}$S_{16}$}&\tabincell{l}{$h_1>0,\ h_2<0,\ h_3<0,\ z>0,\ u>0$}&\tabincell{l}{$\dot{x}>0,\ \dot{z}<0,\ \dot{u}<0$}\\
\specialrule{0em}{2pt}{2pt}
\hline
\specialrule{0em}{2pt}{2pt}
\tabincell{l}{\hspace{8mm}$S_{17}$}&\tabincell{l}{$h_1>0,\ h_2<0,\ h_3<0,\ z>0,\ u>0$}&\tabincell{l}{$\dot{x} >0,\ \dot{z}<0,\ \dot{u}<0$}\\
\specialrule{0em}{2pt}{2pt}
\hline
\specialrule{0em}{2pt}{2pt}
\tabincell{l}{\hspace{8mm}$S_{18}$}&\tabincell{l}{$h_1>0,\ h_2<0,\ h_3>0,\ z>0,\ u>0$}&\tabincell{l}{$\dot{x} >0,\ \dot{z}<0,\ \dot{u}>0$}\\
\specialrule{0em}{2pt}{2pt}
\hline
\specialrule{0em}{2pt}{2pt}
\tabincell{l}{\hspace{8mm}$S_{19}$}&\tabincell{l}{$h_1<0,\ h_2>0,\ h_3>0,\ z>0,\ u>0$}&\tabincell{l}{$\dot{x}<0,\ \dot{z}>0,\ \dot{u}>0$}\\
\specialrule{0em}{2pt}{2pt}
\hline
\specialrule{0em}{2pt}{2pt}
\tabincell{l}{\hspace{8mm}$S_{21}$}&\tabincell{l}{$h_1>0,\ h_2>0,\ h_3>0,\ z>0,\ u<0$}&\tabincell{l}{$\dot{x}>0,\ \dot{z}>0,\ \dot{u}<0$}\\
\specialrule{0em}{2pt}{2pt}
\hline
\specialrule{0em}{2pt}{2pt}
\tabincell{l}{\hspace{8mm}$S_{22}$}&\tabincell{l}{$h_1>0,\ h_2<0,\ h_3>0,\ z>0,\ u<0$}&\tabincell{l}{$\dot{x}>0,\ \dot{z}<0,\ \dot{u}<0$}\\
\specialrule{0em}{2pt}{2pt}
\hline
\specialrule{0em}{2pt}{2pt}
\tabincell{l}{\hspace{8mm}$S_{23}$}&\tabincell{l}{$h_1<0,\ h_2>0,\ h_3>0,\ z>0,\ u<0$}&\tabincell{l}{$\dot{x}<0,\ \dot{z}>0,\ \dot{u}<0$}\\
\specialrule{0em}{2pt}{2pt}
\hline
\specialrule{0em}{2pt}{2pt}
\tabincell{l}{\hspace{8mm}$S_{24}$}&\tabincell{l}{$h_1<0,\ h_2<0,\ h_3>0,\ z>0,\ u<0$}&\tabincell{l}{$\dot{x}<0,\ \dot{z}<0,\ \dot{u}<0$}\\
\specialrule{0em}{2pt}{2pt}
\hline
\specialrule{0em}{2pt}{2pt}
\tabincell{l}{\hspace{8mm}$S_{25}$}&\tabincell{l}{$h_1<0,\ h_2<0,\ h_3<0,\ z>0,\ u<0$}&\tabincell{l}{$\dot{x}<0,\ \dot{z}<0,\ \dot{u}>0$}\\
\specialrule{0em}{2pt}{2pt}
\hline
\specialrule{0em}{2pt}{2pt}
\tabincell{l}{\hspace{8mm}$S_{26}$}&\tabincell{l}{$h_1>0,\ h_2<0,\ h_3<0,\ z>0,\ u<0$}&\tabincell{l}{$\dot{x}>0,\ \dot{z}<0,\ \dot{u}>0$}\\
\specialrule{0em}{2pt}{2pt}
\hline
\specialrule{0em}{2pt}{2pt}
\tabincell{l}{\hspace{8mm}$S_{27}$}&\tabincell{l}{$h_1>0,\ h_2<0,\ h_3>0,\ z>0,\ u<0$}&\tabincell{l}{$\dot{x}>0,\ \dot{z}<0,\ \dot{u}<0$}\\
\specialrule{0em}{2pt}{2pt}
\hline
\specialrule{0em}{2pt}{2pt}
\tabincell{l}{\hspace{8mm}$S_{28}$}&\tabincell{l}{$h_1>0,\ h_2>0,\ h_3>0,\ z>0,\ u<0$}&\tabincell{l}{$\dot{x}>0,\ \dot{z} >0,\ \dot{u}<0$}\\
\specialrule{0em}{2pt}{2pt}
\hline
\specialrule{0em}{2pt}{2pt}
\tabincell{l}{\hspace{8mm}$S_{29}$}&\tabincell{l}{$h_1<0,\ h_2>0,\ h_3>0,\ z>0,\ u<0$}&\tabincell{l}{$\dot{x}<0,\ \dot{z}>0,\ \dot{u}<0$}\\
\specialrule{0em}{2pt}{2pt}
\hline
\specialrule{0em}{2pt}{2pt}
\tabincell{l}{\hspace{8mm}$S_{210}$}&\tabincell{l}{$h_1<0,\ h_2>0,\ h_3>0,\ z>0,\ u<0$}&\tabincell{l}{$\dot{x}<0,\ \dot{z}>0,\ \dot{u}<0$}\\
\specialrule{0em}{2pt}{2pt}
 \toprule
\label{table9}
\end{tabular}
\end{table}

For the subregion $S_{12}$ the left and right surfaces are contained in the surfaces $h_2 = 0$ and $h_1 = 0$ respectively, the bottom plane is contained in the invariant plane $u = 0$, and the front plane is contained in the Poincar\'e sphere. Table \ref{table9} shows that the orbits in this subregion increase monotonically along the positive direction of the $x$-axis and $u$-axis, and decrease monotonically along the positive direction of the $z$-axis. Therefore the orbits in this subregion start from the infinite equilibrium points on the Poincar\'e sphere, and then cross the left surface $h_2 = 0$ into the subregion $S_{11}$.

In the subregion $S_{13}$ the left and right surfaces are contained in the surfaces $h_1 = 0$ and $h_2 = 0$ respectively, the bottom plane is contained in the invariant plane $u = 0$, the back plane is contained in the invariant plane $z = 0$, and the front plane is contained in the Poincar\'e sphere. Table \ref{table9} means that the orbits in this subregion decrease monotonically along the positive direction of the $x$-axis, while the orbits increase monotonically along the other two axes. It means that the orbits in this subregion start at the finite equilibrium point $p_1$ and cross the surface $h_2 = 0$ into the neighboring subregion $S_{14}$.

In the subregion $S_{14}$ the left side surface is contained in the surfaces $h_1 = 0$ and $h_2 = 0$, the right side surface is contained in the surface $h_3 = 0$, the back plane is contained in the invariant plane $z = 0$, and the back surface is contained in the surface $h_1 = 0$, the bottom plane is contained in the invariant plane $u = 0$, and the front surface is contained in the Poincar\'e sphere. From Table \ref{table9} we know that the orbits in this subregion are monotonically decreasing along the positive direction of the $x$-axis and $z$-axis, while the orbits are monotonically increasing along the $u$-axis in the positive direction, then the orbits start at the finite equilibrium point $p_4$, or the equilibrium points on the Poincar\'e sphere at infinity in this subregion, and then cross the surfaces $h_1 = 0$ and $h_3=0$, and finally enter the adjacent subregions $S_{15}$ and $S_{18}$ respectively.

The subregion $S_{15}$ is composed of the left side surface contained in the surface $h_3 = 0$, the right side surface contained in the surface $h_1 = 0$, the bottom surface contained in the invariant plane $u = 0$, and the front surface contained in the Poincar\'e sphere. From Table \ref{table9} we find that the orbits are monotonically decreasing along the three directions in the positive direction in this subregion, so the orbits originate from the subregion $S_{14}$ or from the infinite equilibrium points on the Poincar\'e sphere in the subregion $S_{15}$, and then cross the surface $h_1 = 0$ into the connecting subregion $S_{16}$.

The top surface of the subregion $S_{16}$ is contained in the surface $h_3 = 0$, the front left surface is contained in the surface $h_1 = 0$, the front right surface is contained in the Poincar\'e sphere, and the bottom plane is contained in the invariant plane $u = 0$. The subregion $S_{17}$ is composed of the front surface contained in $h_3 = 0$ and the back plane contained in the invariant plane $z = 0$. Table \ref{table9} indicates that the orbits have the same dynamic behavior in these two subregions, they increase monotonically along the positive direction of the $x$-axis and decrease monotonically along the other two axes, so the orbits in the subregion $S_{16}$ start from the subregion $S_{15}$ or from the infinite equilibrium points on Poincar\'e sphere, then cross the surface $h_3 = 0$ into the subregion $S_{18}$. The orbits in the subregion $S_{17}$ come from the subregion $S_{18}$ and finally return to this subregion, that is the orbits in the subregion $S_{18}$ can cross the subregion $S_{17}$ from the right to left.

The top surface of the subregion $S_{18}$ is contained in the surface $h_2 = 0$, the bottom surface is contained in the surface $h_3 = 0$, the bottom plane is contained in the invariant plane $u = 0$, the left side surface is contained in the surface $h_1 = 0$, and the front surface is contained in the Poincar\'e sphere, the back surface is contained in the invariant plane $z = 0$ and the surface $h_3 = 0$. Table \ref{table9} implies that the orbits decrease monotonically along the $z$-axis and increase monotonically along the other two axes in this subregion, so some orbits may start from infinite equilibrium points on Poincar\'e sphere, some orbits start from the finite equilibrium point $p_2$, and then cross the surface $h_2 = 0$ into the subregion $S_{18}$ and eventually into the subregion $S_{11}$.

The front surface of the subregion $S_{19}$ is contained in the surface $h_1 = 0$, the bottom plane is contained in the invariant plane $u = 0$, the back surface part is contained in the invariant plane $z = 0$ and the surface $f_-(x,z,u)= 0$. Table \ref{table9} shows that the orbits in $S_{19}$ decrease monotonically along the positive direction of the $x$-axis, and increase monotonically along the positive direction of the other two axes, so the orbits in this subregion start at the finite equilibrium point $p_2$ and then tend to equilibrium points at infinity on the Poincar\'e sphere or go through the right part of the surface $f_-(x,z,u)= 0$ into outer space. Therefore the dynamic behavior of the orbits of system (\ref{1}) in the region $S_1$ can be concluded into the following flow chart
$$\begin{tikzpicture}[->, thick]
  \node (S18) at (0, 0) [] {$S_{18}$};
  \node (S14) at (-1.36, 0) [] {$S_{14}$};
  \node (S13) at (-2.72, 0) [] {$S_{13}$};
  \node (S16) at (0, -1.15) [] {$S_{16}$};
  \node (S15) at (-1.36, -1.15) [] {$S_{15}$};
  \node (PSS15) at (-1.36, -2.3) [] {PS in $S_{15}$};
     \node (PSS14) at (-2.72, -1.15) [] {PS in $S_{14}$};
     \node (lS17) at (2.05, 0) [] {LPS of $S_{17}$};
     \node (rS17) at (2.05, -1.15) [] {RPS of $S_{17}$};
      \node (PSS18) at (1.36, 1.1) [] {PS in $S_{18}$};
 \node (p4) at (-1.36, 1.1) [] {$p_4$};
 \node (p1) at (-2.72, 1.1) [] {$p_1$};
 \node (PSS16) at (1.36, -2.3) [] {PS in $S_{16}$.};
    \node (S11) at (-1.36, 2.22) [] {$S_{11}$};
    \node (p2) at (0, 2.2) [] {$p_2$};
    \node (PSS19) at (1.82, 2.2) [] {PS in $S_{19}$};
        \node (OSn) at (1.45, 3.37) [] {OS ($x<0$)};
    \node (S12) at (-2.72, 2.2) [] {$S_{12}$};
    \node (PSS12) at (-4.65, 2.2) [] {PS in $S_{12}$};
 \node (PSS11) at (-1.36, 3.37) [] {PS in $S_{11}$};
\draw (PSS15) -- (S15);
\draw (S15) -- (S16);
\draw (S16) -- (S18);
\draw (S14) -- (S18);
\draw (S14) -- (S15);
\draw (S13) -- (S14);
\draw (p1) -- (S13);
\draw (p1) -- (S11);
\draw (p4) -- (S11);
\draw (p4) -- (S14);
\draw (PSS18) -- (S18);
\draw (PSS16) -- (S16);
\draw (PSS12) -- (S12);
\draw (S12) -- (S11);
\draw (S11) -- (PSS11);
\draw (p2) -- (S11);
\draw (p2) -- (S18);
\draw (p2) -- (PSS19);
\draw (p2) -- (PSS11);
\draw (p2) -- (OSn);
\draw (PSS14) -- (S14);
\draw (S18) -- (S11);
\draw (S18) -- (rS17);
\draw (rS17) -- (lS17);
\draw (lS17) -- (S18);
\end{tikzpicture}$$
{\footnotesize{Note: OS represents the outer space.}}

The above flow chart in the region $S_1$ indicates that the orbits of system (\ref{1}) have an $\alpha$-limit at the finite equilibrium points $p_1$, $p_2$ and $p_4$, as well as on the Poincar\'e sphere restricted to the subregions $S_{12}$, $S_{14}$, $S_{15}$, $S_{16}$ and $S_{18}$. In addition the orbits have an $\omega$-limit in the subregions $S_{11}$ and $S_{19}$, which are restricted to Poincar\'e sphere.

For the region $S_2$ the left surface of the subregion $S_{21}$ is contained in the surface $f_-(x,z,u)= 0$, the right surface is contained in the surfaces $h_1 = 0$ and $h_2 = 0$, the top plane is contained in the invariant plane $u = 0$, the bottom surface and the front triangle-shaped surface are contained in the Poincar\'e sphere, and the back plane is contained in the invariant plane $z = 0$. Table \ref{table9} shows that the orbits in this subregion decrease monotonically along the positive direction of $u$-axis, and monotonically increase along the positive directions of the other two axes, so the orbits in this subregion start at finite equilibrium points $p_1$ and $p_4$, or from the neighboring subregion $S_{22}$, they eventually tend to the infinite equilibrium points of the Poincar\'e sphere that is restricted to the subregion $S_{21}$ or go through the left part of the surface $f_-(x,z,u)= 0$ into the outer space.

The top plane and surface of the subregion $S_{22}$ are contained in the invariant plane $u = 0$ and the surface $h_1 = 0$ respectively, the back plane is contained in the invariant plane $z = 0$, and the left side surface is contained in the surfaces $h_2 = 0$ and $f_-(x,z,u)= 0$, the right side surface is contained in the surface $h_3 = 0$, and the front and bottom surfaces are contained in the Poincar\'e sphere. Table \ref{table9} implies that the orbits in this subregion increase monotonically along the positive $x$-axis direction and decrease monotonically along the positive $z$-axis and $u$-axis directions, which means that these orbits may come from the subregions $S_{24}$ and $S_{26}$, and then enter in the subregion $S_{21}$ or tend to the infinite equilibrium points on the Poincar\'e sphere in the subregion $S_{22}$, or cross the surface $f_-(x,z,u)= 0$ into the outer space.

The structure of subregions $S_{23}$ and $S_{24}$ and the dynamic behavior of the orbits in them are the same as those of subregions $R_{23}$ and $R_{24}$ (see Figures \ref{fig26} and \ref{fig23}), then the orbits in the subregion $S_{23}$ start at the finite equilibrium point $p_1$ and cross the surface $h_2 = 0$, and then enter into the subregion $S_{24}$, and the orbits in the subregion $S_{24}$ start at the finite equilibrium point $p_4$, the connecting subregion $S_{23}$, as well as the infinite equilibrium points on the Poincar\'e sphere, and then cross the surfaces $h_1 = 0$ and $h_3 = 0$, and eventually tend to the subregions $S_{22}$ and $S_{25}$ respectively.

For the subregion $S_{25}$ except that the surface on the lower left side is contained in the surface $f_-(x,z,u)= 0$, the remaining structure is the same as that of $R_{25}$. According to the Tables \ref{table6} and \ref{table9} the monotonicity of the orbits in these two subregions are also the same, so the orbits in the subregion $S_{25}$ start from the infinity equilibrium points on the Poincar\'e sphere, or from the outer space through the left side surface $f_-(x,z,u)= 0$, but all eventually cross the surface $h_1 = 0$ into the subregion $S_ {26}$.

The top plane of the subregion $S_{26}$ is contained in the invariant plane $u = 0$, the left and right side surfaces are contained in the surfaces $h_3 = 0$ and $f_-(x,z,u)= 0$, and the front surface located at the upper left corner is contained in the surface $h_1 = 0$. The front surface containing the infinite equilibrium point $e_4$ is contained in the Poincar\'e sphere. Table \ref{table9} shows that the orbits decrease monotonically along the positive direction of the $z$-axis in the subregion $S_{26}$, while the orbits along the positive direction of the remaining two axes is just the opposite. Therefore the orbits in this subregion start at the infinite equilibrium points on the Poincar\'e sphere or from the subregion $S_{27}$ and the outer space passing through the right side surface $f_-(x,z,u)= 0$, and finally enter the subregion $S_{22}$ through the left side surface $h_3 = 0$.

The front and back surfaces of the subregion $S_{27}$ are contained in the surfaces $h_3 = 0$ and $h_2 = 0$ respectively, the back plane is contained in the invariant plane $z = 0$, the top plane is contained in the invariant plane $u = 0$, and the right side surface is contained in the Poincar\'e sphere and the right part of the surface $f_-(x,z,u)= 0$. Table \ref{table9} shows that the orbits decrease monotonically along the positive direction of the $x$-axis and increase monotonically along the positive direction of the $z$-axis and $u$-axis in this subregion. In this way the orbits in $S_{27}$ start from the adjacent subregion $S_{28}$ or come from the outer space passing through the right side surface $f_-(x,z,u)= 0$ and the infinite equilibrium points on the Poincar\'e sphere, and finally pass through the front surface $h_3 = 0$ and enter the subregion $S_{26}$.

The structure of the subregion $S_{28}$ is similar to the subregion $S_{27}$ discussed above, except that the back surface of the subregion $S_{27}$ happens to be the front surface of the subregion $S_{28}$, and the back surface of the subregion $S_{28}$ is contained in the surface $h_1 = 0$. Note that Table \ref{table9} implies that the orbits monotonically decrease along the positive direction of the $u$-axis in this subregion, and increase monotonically along the positive direction of the other two axes. This means that the orbits start at the finite equilibrium point $p_2$, and then cross the surface $h_2 = 0$ into the subregion $S_{27}$.

The subregions $S_{29}$ and $S_{210}$ are connected by the finite equilibrium point $p_2$. Their front surfaces are contained in the surface $h_1 = 0$, the bottom surface of $S_{29}$ and the right surface of $S_{210}$ are contained in the Poincar\'e sphere, the right side surface of $S_{29}$ and the back surface of $S_{210}$ are contained in the right part of the surface $f_-(x,z,u)= 0$, and the back surface plane of $S_{29}$ is contained in the invariant plane $z = 0$. From Table \ref{table9} we find that the dynamic behavior of the orbits in these two subregions is the same, both decrease monotonically along the positive direction of the $x$-axis and $u$-axis, and increase monotonically along the positive direction of the $z$-axis. Thus the orbits in these two subregions start at the finite equilibrium point $p_2$, and then tend to the Poincar\'e sphere or go through the surface $f_-(x,z,u)= 0$ into the outer space.

Accordingly the dynamic behavior of system (\ref {1}) in the region $S_2$ can be shown by the following flow chart
$$\begin{tikzpicture}[->, thick]
  \node (S25) at (0, -1.15) [] {$S_{25}$};
   \node (S24) at (0, 0) [] {$S_{24}$};
      \node (S22) at (-3, 0) [] {$S_{22}$};
    \node (OSp) at (-4.78, 1.5) [] {OS ($x>0$)};
    \node (PSS22) at (-4.93, 0) [] {PS in $S_{22}$};
      \node (S21) at (-3, 3) [] {$S_{21}$};
    \node (PSS21) at (-4.93, 3) [] {PS in $S_{21}$};
  \node (S23) at (-1, 1) [] {$S_{23}$};
  \node (p1) at (-2, 2) [] {$p_1$};
       \node (p4) at (0, 3) [] {$p_4$};
  \node (S26) at (-3, -1.15) [] {$S_{26}$};
     \node (PSS24) at (1.94, 0) [] {PS in $S_{24}$};
     \node (OSn) at (0, -2.3) [] {OS ($x<0$)};
   \node (S27) at (-3, -2.3) [] {$S_{27}$};
     \node (PSS27) at (-4.93, -2.3) [] {PS in $S_{27}$};
  \node (p2) at (0, -3.45) [] {$p_2$};
    \node (S28) at (-3, -3.45) [] {$S_{28}$};
    \node (S29) at (1.27, -3.45) [] {$S_{29}$};
      \node (PSS29) at (3.2, -3.45) [] {PS in $S_{29}$};
     \node (S210) at (0, -4.55) [] {$S_{210}$};
       \node (PSS210) at (2.08, -4.55) [] {PS in $S_{210}$.};
\draw (S22) -- (OSp);
\draw (S22) -- (PSS22);
\draw (S22) -- (S21);
\draw (S21) -- (OSp);
\draw (S21) -- (PSS21);
\draw (p1) -- (S21);
\draw (p1) -- (S23);
\draw (p4) -- (S21);
\draw (p4) -- (S24);
\draw (S23) -- (S24);
\draw (S24) -- (S22);
\draw (PSS24) -- (S24);
\draw (S24) -- (S25);
\draw (S25) -- (S26);
\draw (S26) -- (S22);
\draw (OSn) -- (S25);
\draw (OSn) -- (S26);
\draw (OSn) -- (S27);
\draw (p2) -- (OSn);
\draw (S27) -- (S26);
\draw (S27) -- (PSS27);
\draw (S28) -- (S27);
\draw (p2) -- (S28);
\draw (p2) -- (S29);
\draw (S29) -- (PSS29);
\draw (p2) -- (S210);
\draw (S210) -- (PSS210);
\end{tikzpicture}$$

This flow chart in the region $S_2$ implies that the orbits of system (\ref{1}) have an $\alpha$-limit at the finite equilibrium points $p_1$, $p_2$ and $p_4$, and at the infinite equilibrium points on the Poincar\'e sphere restricted to the subregion $S_{24}$. Besides the orbits have an $\omega$-limit in the subregions $S_{21}$, $S_{22}$, $S_{27}$, $S_{29}$ and $S_{210}$, which are restricted to Poincar\'e sphere.

When the field potential $V(\phi)$ takes the form of constant, i.e. $s=0$. Then system (\ref{1}) is reduced to 
\begin{equation}
\begin{array}{rl} \vspace{2mm}
\dfrac{dx}{dt}&=x \left[3 x^2 + 2 (u - z) z - 3\right],\\ \vspace{2mm}
\dfrac{dz}{dt}&=z \left[3 x^2 + 2 (u - z) z - 2\right] ,\\ \vspace{2mm}
\dfrac{du}{dt}&=u \left[3 x^2 + 2 (u - z) z\right].
\end{array}
\label{eq31}
\end{equation}
Now we continue to describe the global dynamics of system (\ref{eq31}) in the closed and open universe in sections 5 and 6 respectively.

%section 5
\section{Case III: $s=0, k=1$}

\subsection{Phase portraits on the invariant planes and surface}

In this section we will investigate the local and global phase portraits of the finite and infinite equilibrium points of system (\ref{eq31}). The phase portraits on the invariant planes $x = 0$, $z=0$ and $u = 0$, and on the invariant surface $f_+(x,z,u)=0$ are studied in what follows.

%section 5.1.1
\subsubsection{The invariant plane $x=0$}
On this plane system (\ref{eq31}) writes
\begin{equation}
\begin{array}{rl} \vspace{2mm}
\dfrac{dz}{dt}=2z(uz - z^2 - 1),\ \ \ 
\dfrac{du}{dt}=2u z (u - z).
\end{array}
\label{eq32}
\end{equation}
It is easy to check that this system has no other finite equilibrium points except the straight line $z = 0$ which is full of the equilibrium points of system (\ref{eq32}).

By using the Poincar\'e compactification $z = 1 / V$ and $u = U / V$, we obtain that system (\ref{eq32}) on the local chart $U_1$ has the  form
\begin{equation}
\begin{array}{rl}\vspace{2mm}
\dfrac{dU}{dt}=2 U V^2,\ \ \ 
\dfrac{dV}{dt}=2 V (1 - U + V^2).
\end{array}
\label{eq4}
\end{equation}
Rescaling the time via $d\tau_8=2Vdt$ the previous system becomes
\begin{equation}
\begin{array}{rl}\vspace{2mm}
\dfrac{dU}{d\tau_8}=U V,\ \ \ 
\dfrac{dV}{d\tau_8}=1 - U + V^2.
\end{array}
\label{eq5}
\end{equation}
It admits an infinite equilibrium point $e_{11}=(1,0)$ with eigenvalues $\pm i$, which indicates that $e_{11}$ may be either a center or a weak focus of this system. Note that $(1-2U+V^2)/U^2=C$ is a first integral of system (\ref{eq5}), then $e_{11}$ is a center.

On the local chart $U_2$ we have $z = U / V$ and $u = 1 / V$, then system (\ref{eq32}) can be rewritten as
\begin{equation}
\begin{array}{rl}\vspace{2mm}
\dfrac{dU}{dt}=-2 U V^2,\ \ \ 
\dfrac{dV}{dt}=2 (U-1) U V.
\end{array}
\label{eq6}
\end{equation}
On the local chart $U_2$ we only need to examine the origin $(0,0)$ of system (\ref{eq6}). Obviously $e_{12}$ is an equilibrium point with eigenvalues zero, the conventional eigenvalue method cannot be used to determine the type of $e_{12}$ and its local phase portrait. We apply horizontal blow-up by introducing the transformation $W = U/V$ (see \cite{Alvarez} for more details), and then we obtain 
\begin{equation}
\begin{array}{rl}\vspace{2mm}
\dfrac{dV}{dt}=2 V^2 W (V W-1),\ \ \ 
\dfrac{dW}{dt}=-2 V W (V - W + V W^2).
\end{array}
\label{eq7}
\end{equation}
Doing the time transformation $d\tau_9 = 2 V W dt$ we eliminate the common factor, and we have
\begin{equation}
\begin{array}{rl}\vspace{2mm}
\dfrac{dV}{d\tau_9}=V (V W-1),\ \ \ 
\dfrac{dW}{d\tau_9}=-V + W - V W^2.
\end{array}
\label{eq8}
\end{equation}
Then the equilibrium point $e_{12}=(0,0)$ of system (\ref{eq8}) with eigenvalues $\pm1$ is a hyperbolic saddle.

\begin{figure}[]
  \begin{minipage}{130mm}
\centering\subfigure[]{\label{fig:subfig:a}
    \includegraphics[width=5cm]{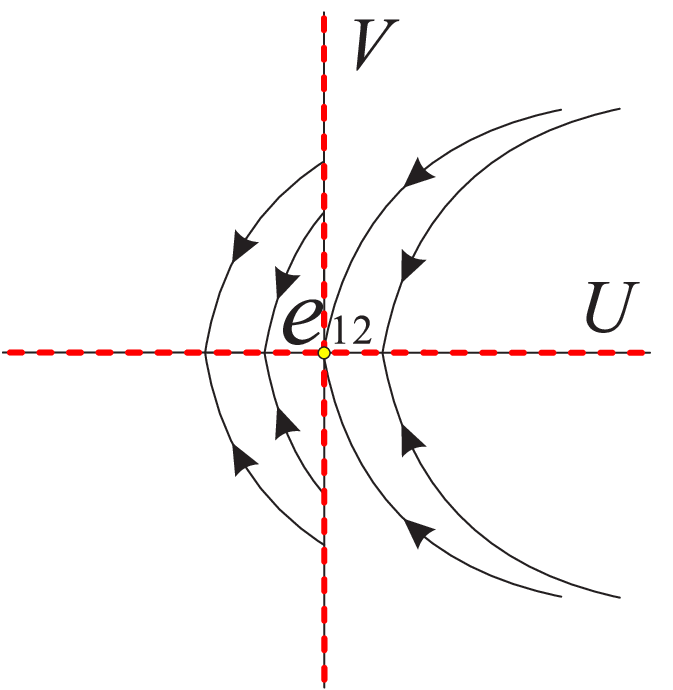}}
  \subfigure[]{\label{fig:subfig:b}
    \includegraphics[width=4cm]{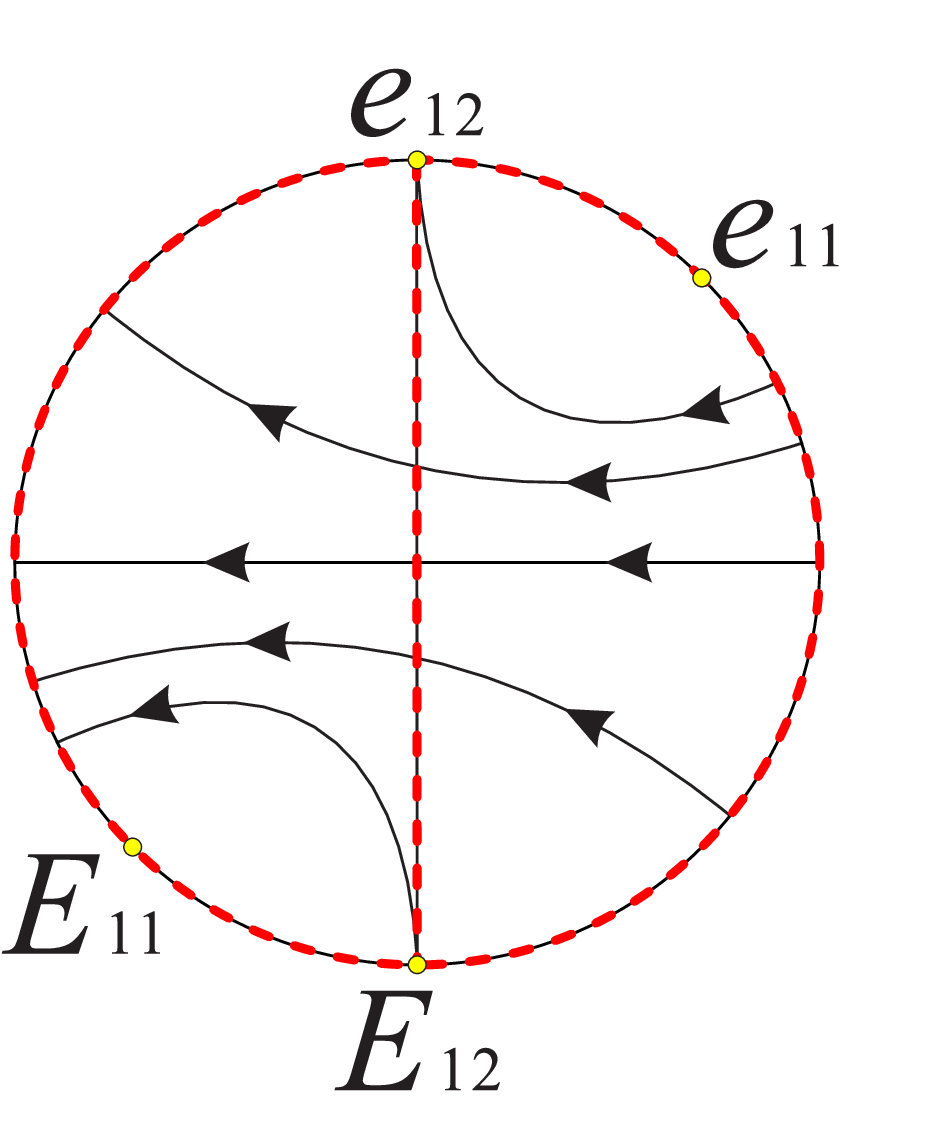}}
     \subfigure{\includegraphics[width=2cm]{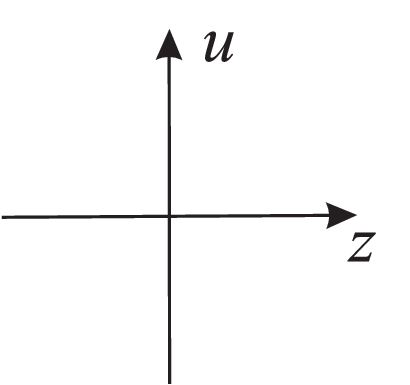}}    
    \caption{(a) The local phase portrait at the infinite equilibrium point $e_{12}$. (b) The global phase portrait of system (\ref{eq32}). $E_{11}$ and $E_{12}$ are the diametrally opposite equilibrium points of $e_{11}$ and $e_{12}$ in the Poincar\'e disc, respectively.}
  \label{fig27}
  \end{minipage}
 \end{figure}

Eliminating the common factor $2 U V$ of system (\ref{eq6}) by taking $d\tau_{10} = 2 U V dt$ yields $dU/d\tau_{10}=-V$ and $dV/d\tau_{10}=U-1$, which implies that the orbits of the local phase portrait of the infinite equilibrium point $e_{12}$ decreases monotonically along the $V$-axis, and increases monotonically along the $U$-axis when $V <0$, or decreases monotonically along the $U$-axis when $V> 0$. Therefore the local phase portrait of $e_{12}$ and the global phase portrait of system (\ref{eq32}) are shown in Figure \ref{fig27}.

%Figure 27

%section 5.1.2
\subsubsection{The invariant plane $z=0$}

\begin{figure}[]
  \begin{minipage}{130mm}
\centering\subfigure{\label{fig:subfig:a}
    \includegraphics[width=4cm]{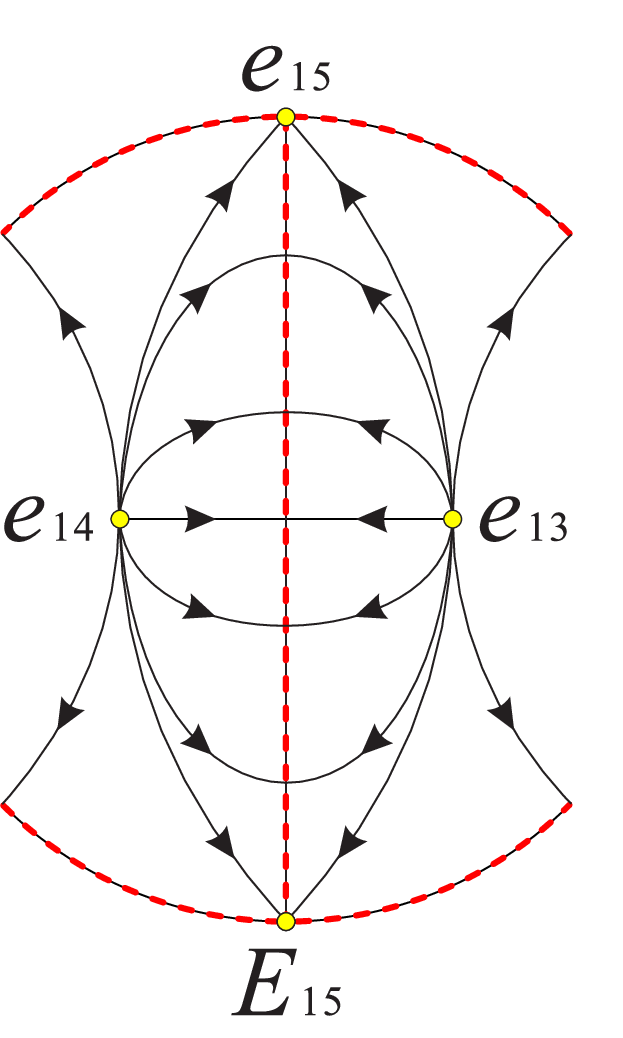}}
     \subfigure{\includegraphics[width=2cm]{Fig2g9b10b28b_frame_x_u.eps}}    
    \caption{The global phase portrait of system (\ref{eq9}) on the invariant plane $z=0$ restricted to the region $x^2-u^2\leq1$. $E_{15}$ is the diametrically opposite point of $e_{15}$ on Poincar\'e disc.}
  \label{fig28}
  \end{minipage}
 \end{figure}

On this plane system (\ref{eq31}) becomes
\begin{equation}
\begin{array}{rl} \vspace{2mm}
\dfrac{dx}{dt}=3x(x^2 - 1),\ \ \ 
\dfrac{du}{dt}=3u x^2.
\end{array}
\label{eq9}
\end{equation}
This system is exactly the same as system (9) in \cite{Gao20201}, so the global phase portrait of system (\ref{eq9}) is shown in Figure \ref{fig28}. The finite equilibrium points $e_{13}=(1,0)$ and $e_{14}=(-1,0)$ are hyperbolic unstable nodes. Besides the line $x = 0$ and the infinity of the local chart $U_1$ are filled with equilibrium points (see \cite{Gao20201} for more details). 

%Figure 28

%section 5.1.3 
\subsubsection{The invariant plane $u=0$}
\begin{figure}[]
  \begin{minipage}{130mm}
\centering\subfigure{\label{fig:subfig:a}
    \includegraphics[width=5.5cm]{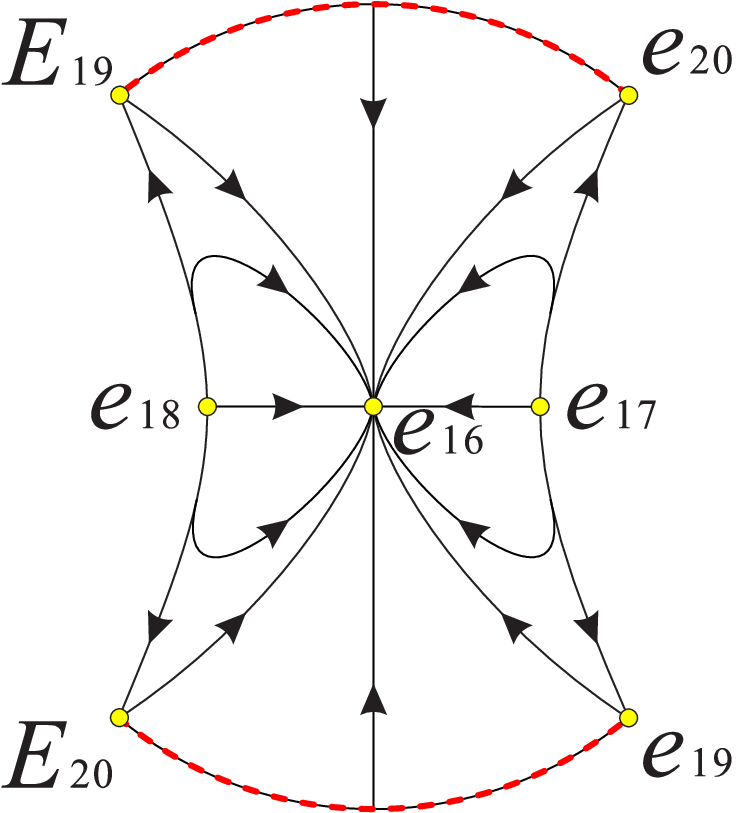}}
     \subfigure{\includegraphics[width=2cm]{Fig4f5f29b_frame_x_z.eps}}    
    \caption{The global phase portrait of system (\ref{eq10}) on the invariant plane $u=0$ restricted to the region $x^2-z^2\leq1$. $E_{19}$ and $E_{20}$ are the diametrically opposite points of $e_{19}$ and $e_{20}$ on Poincar\'e disc respectively.}
  \label{fig29}
  \end{minipage}
 \end{figure}

On this plane system (\ref{eq31}) is reduced to
\begin{equation}
\begin{array}{rl} \vspace{2mm}
\dfrac{dx}{dt}=x (3 x^2 - 2 z^2 - 3),\ \ \ 
\dfrac{dz}{dt}=z (3 x^2 - 2 z^2-2).
\end{array}
\label{eq10}
\end{equation}
Note that system (\ref{eq10}) is the same as system (9) in \cite{Gao20201}, so the global phase portrait of system  (\ref{eq10}) is illustrated in Figure \ref{fig29}. The finite equilibrium point $e_{16}=(0,0)$ is a hyperbolic stable node, $e_{17}=(1,0)$ and $e_{18}=(-1,0)$ are hyperbolic unstable nodes. In addition the infinity of system (\ref{eq10}) is full of the equilibrium points (see \cite{Gao20201} for more details). 

%Figure 29

%section 5.1.4
\subsubsection{The invariant surface $f_+(x,z,u)=0$}

On this surface system (\ref{eq31}) is rewritten in the same form as systems \eqref{eq16} and \eqref{eq17} in section 3.1.3, so the global phase portrait on the invariant surface $f_+(x,z,u)=0$ can be found in Figures \ref{fig9} and \ref{fig10} respectively.

%section 5.1.5
\subsubsection{The finite equilibrium points}
It is easy to find that there are two finite equilibrium points $q_1=(1,0,0)$ and $q_2=(-1,0,0)$ of system (\ref{eq31}), $q_1$ and $q_2$ with the same eigenvalues 6, 3, and 1 are hyperbolic unstable nodes. Besides $u$-axis is full of equilibrium points of system (\ref{eq31}).

%section 5.1.6
\subsubsection{Phase portrait on the Poincar\'{e} sphere at infinity}
According to the three-dimensional Poincar\'{e} compactification, system (\ref{eq31}) can be rewritten as follows.
On the local chart $U_1$
\begin{equation}
\begin{array}{rl}\vspace{2mm}
\dfrac{dz_1}{dt}=z_1 z_3^2,\ 
\dfrac{dz_2}{dt}=3z_2 z_3^2,\ 
\dfrac{dz_3}{dt}=z_3(2 z_1^2 - 2 z_1 z_2 + 3 z_3^2-3).
\end{array}
\label{eq40}
\end{equation}
On the local chart $U_2$
\begin{equation}
\begin{array}{rl}\vspace{2mm}
\dfrac{dz_1}{dt}= -z_1 z_3^2,\ 
\dfrac{dz_2}{dt}= 2 z_2 z_3^2,\ 
\dfrac{dz_3}{dt}= z_3 \left(2 - 3 z_1^2 - 2 z_2 + 2 z_3^2\right).
\end{array}
\label{eq41}
\end{equation}
On the local chart $U_3$
\begin{equation}
\begin{array}{rl}\vspace{2mm}
\dfrac{dz_1}{dt}= -3 z_1 z_3^2,\ 
\dfrac{dz_2}{dt}= -2 z_2 z_3^2,\ 
\dfrac{dz_3}{dt}= -3 z_1^2 z_3 + 2 (z_2-1) z_2 z_3.
\end{array}
\label{eq42}
\end{equation}
It is noted that equations (\ref{eq40}), (\ref{eq41}), (\ref{eq42}) are the same as systems (\ref{eq23}), (\ref{eq27}) and (\ref{eq29}) when $s = 0$ in section 3.1.5 respectively, so the Poincar\'e sphere is filled with equilibrium points of system (\ref{eq31}) at infinity.

%section 5.2
\subsection{Phase portrait inside the Poincar\'e ball restricted to the physical region of interest $x^2-(u-z)^2\leq1$}

As mentioned in section 2 system (\ref{eq31}) is invariant under the three symmetries $(x,z,u)\mapsto(-x,-z,-u)$, $(x,z,u)\mapsto(x,-z,-u)$ and $(x,z,u)\mapsto(-x,z,u)$. Here we divide the Poincar\'e ball into four regions as follows
\begin{equation*}
\begin{array}{rl}
Q_1:\ x\geq0,\ z\geq0.\ \ \
Q_2:\ x\leq0,\ z\geq0.\\
Q_3:\ x\leq0,\ z\leq0.\ \ \
Q_4:\ x\geq0,\ z\leq0.
\end{array}
\end{equation*}
Due to the above symmetries with respect to the origin, the $x$-axis, and the invariant plane $x=0$, we only need to discuss the phase portrait of system (\ref{eq31}) in the region $Q_1$ restricted to the region $x^2-(u-z)^2\leq1$.

\begin{figure}[]
  \begin{minipage}{130mm}
\centering
\subfigure{\includegraphics[width=4cm]{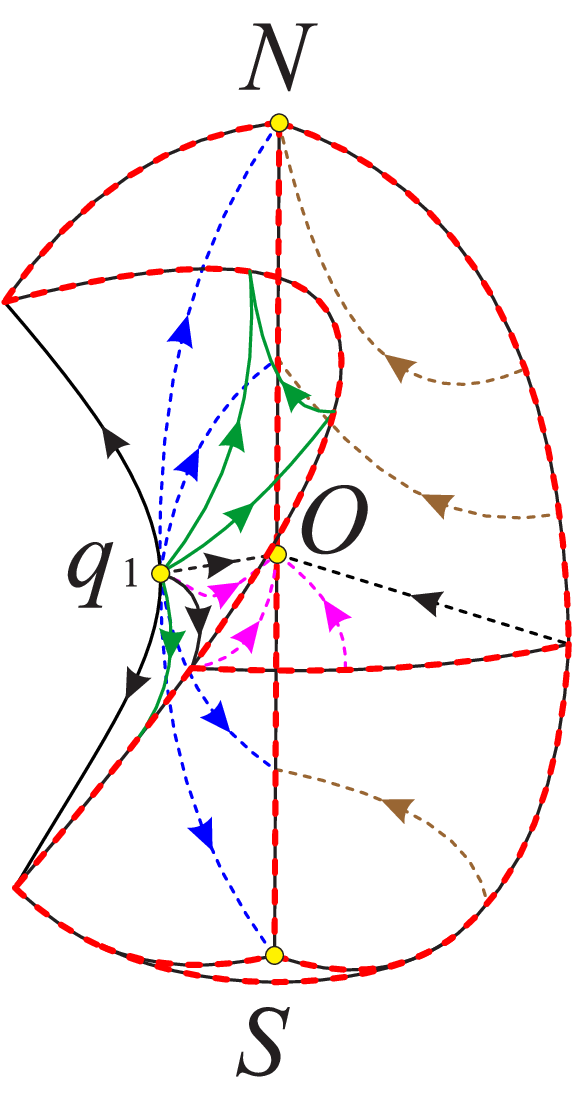}}
\subfigure{\includegraphics[width=2cm]{Fig14f15f16f17f21b23b24c26b30b32b33b34b_frame2_x_z_u.eps}}
    \caption{Phase portrait on the boundary of the region $Q_1$, and $O$ denotes the center of the Poincar\'e ball.}
  \label{fig30}
  \end{minipage}
 \end{figure}

Joining the phase portraits on the invariant planes $x=0$, $z=0$ and $u=0$, as well as on the invariant surface $f_+(x,z,u)=0$, and on the Poincar\'e sphere at infinity, the phase portrait on the boundary of the region $Q_1$ is displayed in Figure \ref{fig30}. It is noted that all the equilibrium points on the $u$-axis are stable along the two intersecting boundary planes $x = 0$ and $z = 0$, and the finite equilibrium point $q_1$ is unstable on the invariant boundary plane $z = 0$ and on the invariant boundary surface $f_+(x,z,u) = 0$.

 %Figure 30

\subsection{Dynamics in the interior of the region $Q_1$}

\begin{figure}[]
  \begin{minipage}{130mm}
\centering
\subfigure{\includegraphics[width=10cm]{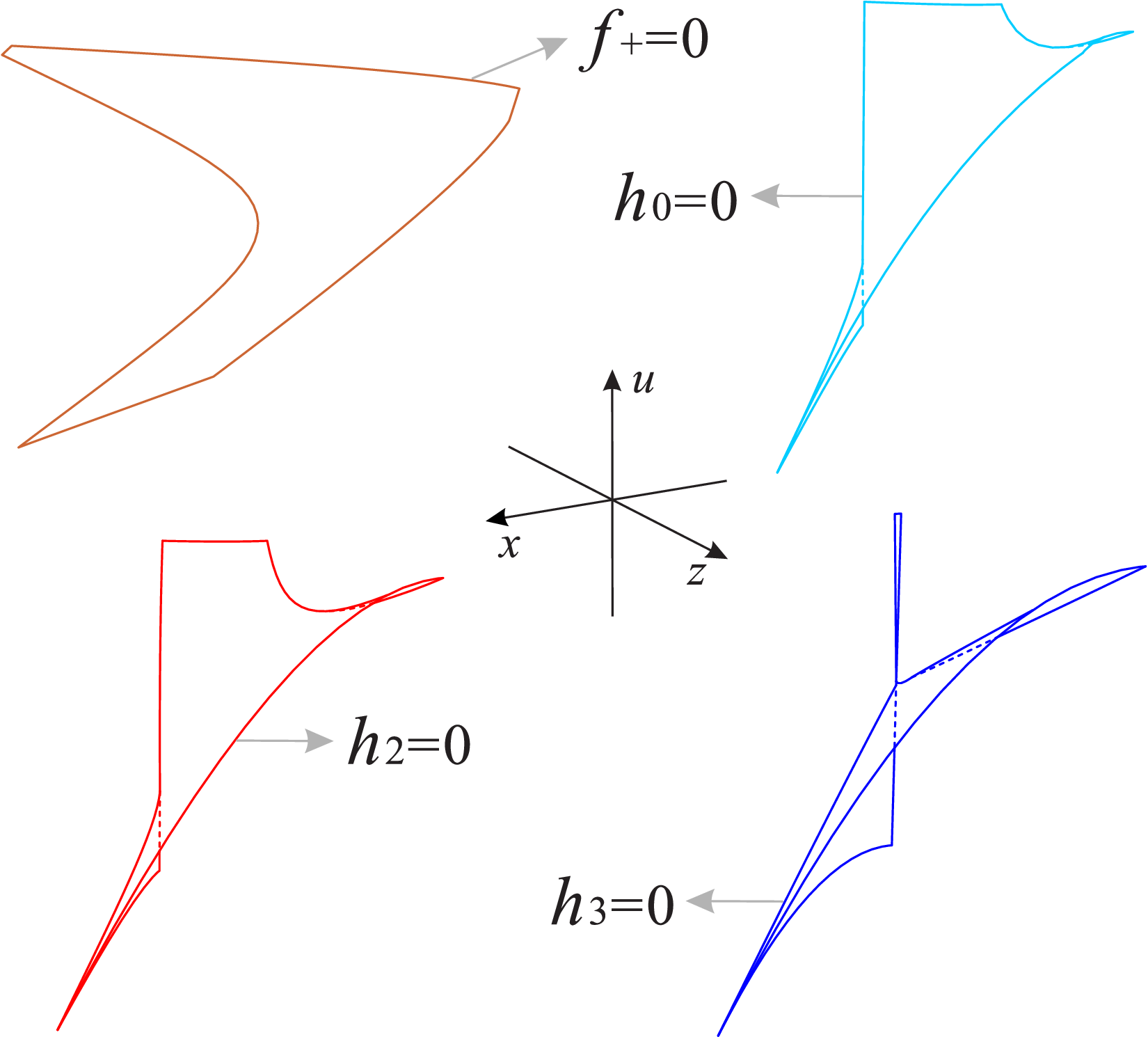}}
\subfigure{\includegraphics[width=5.5cm]{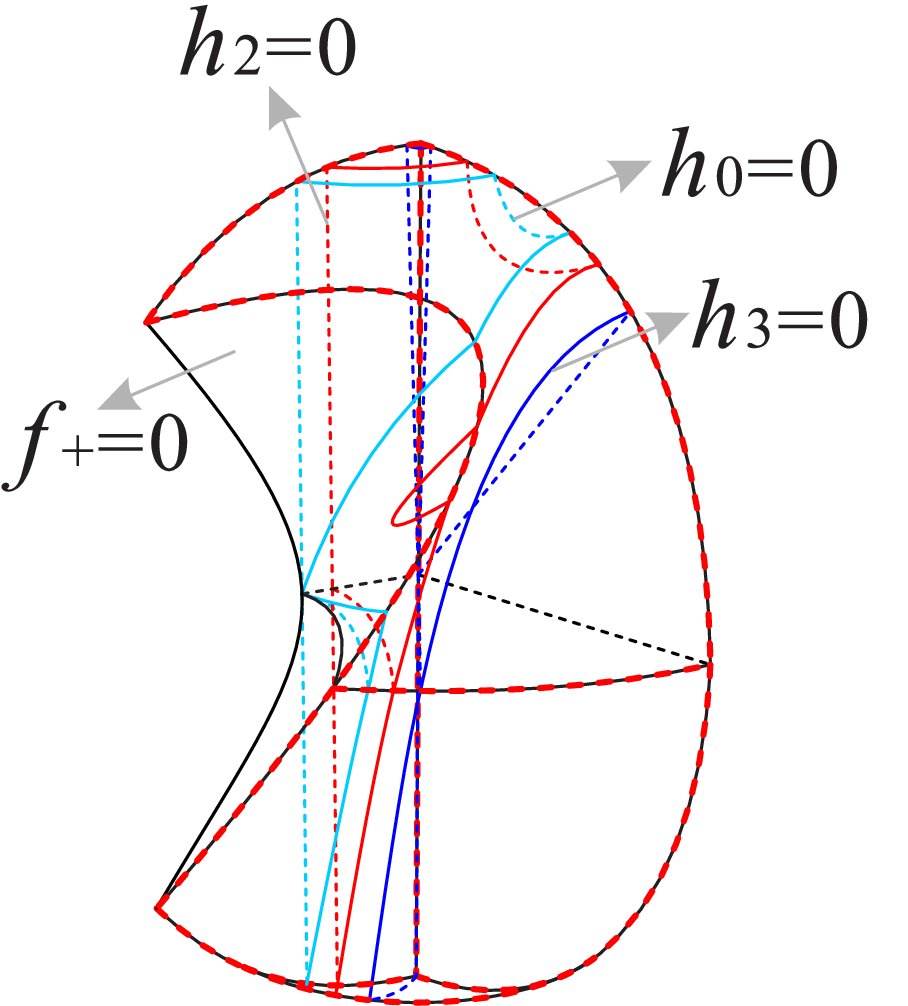}}
    \caption{The surfaces $h_0$, $h_2$ and $h_3$ restricted to the surface $f_+(x,z,u)= 0$ and the region $Q_1$ of the Poincar\'e ball.}
  \label{fig31}
  \end{minipage}
 \end{figure}

\begin{figure}[]
  \begin{minipage}{130mm}
\centering
\subfigure{\includegraphics[width=10cm]{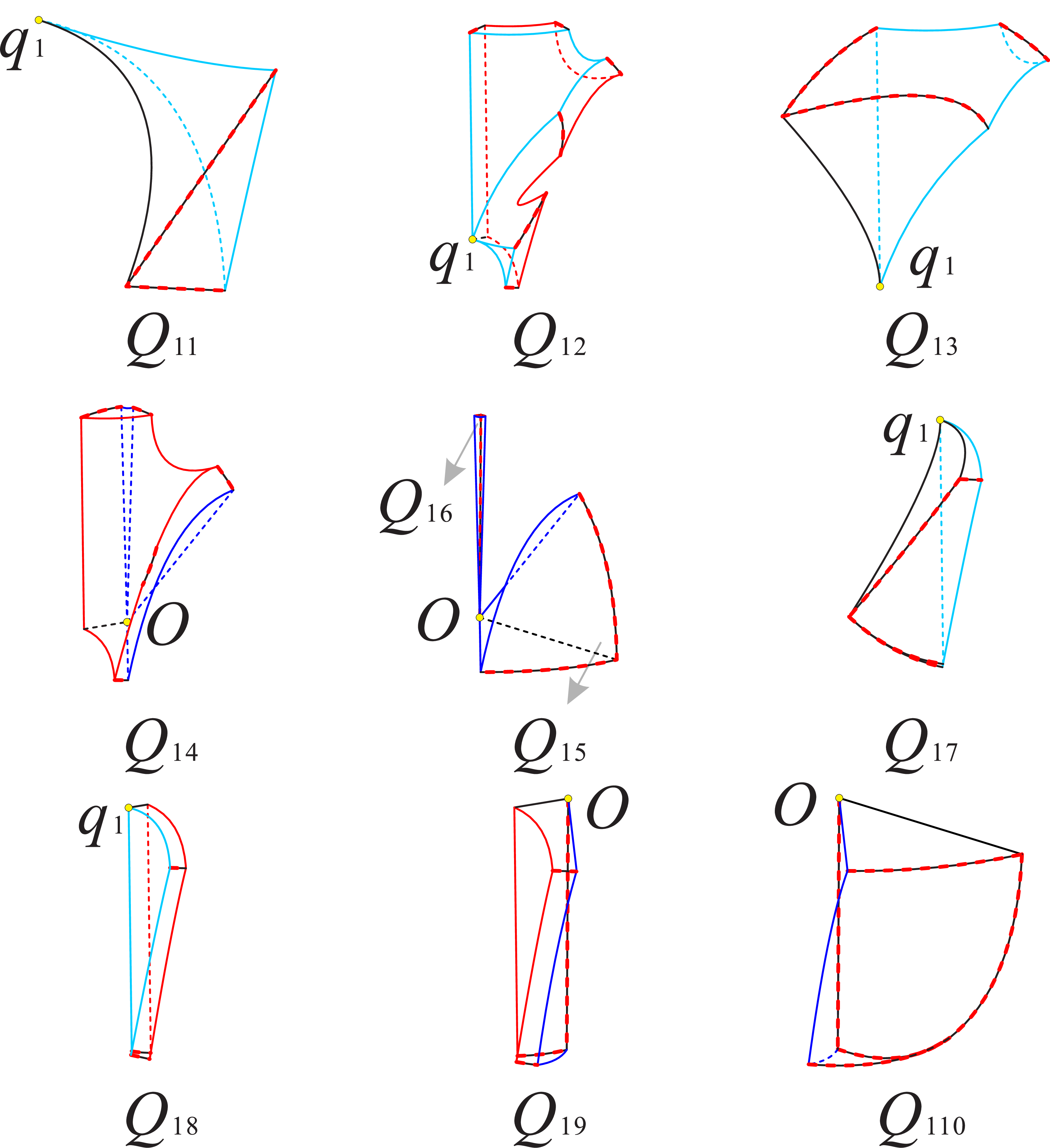}}\\
\subfigure{\includegraphics[width=2cm]{Fig14f15f16f17f21b23b24c26b30b32b33b34b_frame2_x_z_u.eps}}
    \caption{The ten subregions inside the region $Q_1$ of the Poincar\'e ball.}
  \label{fig32}
  \end{minipage}
 \end{figure}

The dynamics of system (\ref{eq31}) inside the region $Q_1$ is governed by the behavior of the orbits in the planes and surfaces as follows 
\begin{equation*}
\begin{array}{rl}
&x=0,\ z=0,\ u=0,\ f_+(x,z,u)=0,\\
&h_0(x,z,u)=0,\ h_2(x,z,u)=0,\ h_3(x,z,u)=0,
\end{array}
\end{equation*}
where $h_0(x,z,u)=h_2(x,z,u)-1=h_3(x,z,u)-3=3 x^2 + 2 (u - z) z - 3$. Then the interior of region $Q_1$ is separated into ten subregions $Q_{1n},\ n=(1,\dots,10)$ by the above planes and surfaces, from these there are six subregions above the invariant plane $u=0$ and four subregions below. See Figures \ref{fig31} and \ref{fig32} for more details. The signs of the functions $h_0$, $h_2$ and $h_3$ in the subregion of $Q_1$ are given in Table \ref{table10}.

%Figure 31

%Figure 32

 %Table 10
\begin{table}[!htb]
\newcommand{\tabincell}[2]{\begin{tabular}{@{}#1@{}}#2\end{tabular}}
\centering
\caption{\label{opt}Signs of functions $h_0$, $h_2$ and $h_3$ in the subregion of $Q_1$.}
\footnotesize
\rm
\centering
\begin{tabular}{@{}*{12}{l}}
\specialrule{0em}{2pt}{2pt}
 \toprule
\hspace{2mm}\textbf{Functions}&\textbf{Positive}&\textbf{Negative}\\
\specialrule{0em}{2pt}{2pt}
\toprule
\tabincell{l}{\hspace{8mm}$h_0$}&\tabincell{l}{$Q_{11}, Q_{13}, Q_{17}$}&\tabincell{l}{$Q_{12}, Q_{14}, Q_{15}, Q_{16}, Q_{18}, Q_{19}, Q_{110}$}\\
\specialrule{0em}{2pt}{2pt}
\hline
\specialrule{0em}{2pt}{2pt}
\tabincell{l}{\hspace{8mm}$h_2$}&\tabincell{l}{$Q_{11}, Q_{12}, Q_{13}, Q_{17}, Q_{18}$}&\tabincell{l}{$Q_{14}, Q_{15}, Q_{16}, Q_{19}, Q_{110}$}\\
\specialrule{0em}{2pt}{2pt}
\hline
\specialrule{0em}{2pt}{2pt}
\tabincell{l}{\hspace{8mm}$h_3$}&\tabincell{l}{$Q_{11}, Q_{12}, Q_{13}, Q_{14}, Q_{17}, Q_{18}, Q_{19}$}&\tabincell{l}{$Q_{15}, Q_{16}, Q_{110}$}\\
\specialrule{0em}{2pt}{2pt}
 \toprule
\label{table10}
\end{tabular}
\end{table}

According to Figure \ref{fig32} it can be seen that the bottom plane of the subregion $Q_{11}$ is contained in the invariant plane $u=0$, the left surface is contained in the invariant surface $f_+(x,z,u)= 0$, the right surface is contained in the surface $h_1=0$, and the front surface is contained in the Poincar\'e sphere of the subregion. Table \ref{table11} shows that the orbits of system (\ref{eq31}) increase monotonically along the positive directions of the three coordinate axes in this subregion, so the orbits in this subregion start at the finite equilibrium point $q_1$ and then finally tend to the equilibrium points at infinity of Poincar\'e sphere.

 %Table 11
\begin{table}[!htb]
\newcommand{\tabincell}[2]{\begin{tabular}{@{}#1@{}}#2\end{tabular}}
\centering
\caption{\label{opt}Dynamical behavior in the ten subregions.}
\footnotesize
\rm
\centering
\begin{tabular}{@{}*{12}{l}}
\specialrule{0em}{2pt}{2pt}
 \toprule
\hspace{2mm}\textbf{Subregions}&\textbf{Corresponding Region}&\textbf{Increase or decrease}\\
\specialrule{0em}{2pt}{2pt}
\toprule
\tabincell{l}{\hspace{8mm}$Q_{11}$}&\tabincell{l}{$h_0>0,\ h_2>0,\ h_3>0,\ x>0, \ z>0,\ u>0$}&\tabincell{l}{$\dot{x}>0,\ \dot{z}>0,\ \dot{u}>0$}\\
\specialrule{0em}{2pt}{2pt}
\hline
\specialrule{0em}{2pt}{2pt}
\tabincell{l}{\hspace{8mm}$Q_{12}$}&\tabincell{l}{$h_0<0,\ h_2>0,\ h_3>0,\ x>0, \ z>0,\ u>0$}&\tabincell{l}{$\dot{x}<0,\ \dot{z}>0,\ \dot{u}>0$}\\
\specialrule{0em}{2pt}{2pt}
\hline
\specialrule{0em}{2pt}{2pt}
\tabincell{l}{\hspace{8mm}$Q_{13}$}&\tabincell{l}{$h_0>0,\ h_2>0,\ h_3>0,\ x>0, \ z>0,\ u>0$}&\tabincell{l}{$\dot{x}>0,\ \dot{z}>0,\ \dot{u}>0$}\\
\specialrule{0em}{2pt}{2pt}
\hline
\specialrule{0em}{2pt}{2pt}
\tabincell{l}{\hspace{8mm}$Q_{14}$}&\tabincell{l}{$h_0<0,\ h_2<0,\ h_3>0, \ x>0, \ z>0,\ u>0$}&\tabincell{l}{$\dot{x}<0,\ \dot{z}<0,\ \dot{u}>0$}\\
\specialrule{0em}{2pt}{2pt}
\hline
\specialrule{0em}{2pt}{2pt}
\tabincell{l}{\hspace{8mm}$Q_{15}$}&\tabincell{l}{$h_0<0,\ h_2<0,\ h_3<0,\ x>0, \ z>0,\ u>0$}&\tabincell{l}{$\dot{x}<0,\ \dot{z}<0,\ \dot{u}<0$}\\
\specialrule{0em}{2pt}{2pt}
\hline
\specialrule{0em}{2pt}{2pt}
\tabincell{l}{\hspace{8mm}$Q_{16}$}&\tabincell{l}{$h_0<0,\ h_2<0,\ h_3<0,\ x>0, \ z>0,\ u>0$}&\tabincell{l}{$\dot{x}<0,\ \dot{z}<0,\ \dot{u}<0$}\\
\specialrule{0em}{2pt}{2pt}
\hline
\specialrule{0em}{2pt}{2pt}
\tabincell{l}{\hspace{8mm}$Q_{17}$}&\tabincell{l}{$h_0>0,\ h_2>0,\ h_3>0,\ x>0, \ z>0,\ u<0$}&\tabincell{l}{$\dot{x} >0,\ \dot{z}>0,\ \dot{u}<0$}\\
\specialrule{0em}{2pt}{2pt}
\hline
\specialrule{0em}{2pt}{2pt}
\tabincell{l}{\hspace{8mm}$Q_{18}$}&\tabincell{l}{$h_0<0,\ h_2>0,\ h_3>0,\ x>0, \ z>0,\ u<0$}&\tabincell{l}{$\dot{x} <0,\ \dot{z}>0,\ \dot{u}<0$}\\
\specialrule{0em}{2pt}{2pt}
\hline
\specialrule{0em}{2pt}{2pt}
\tabincell{l}{\hspace{8mm}$Q_{19}$}&\tabincell{l}{$h_0<0,\ h_2<0,\ h_3>0,\ x>0, \ z>0,\ u<0$}&\tabincell{l}{$\dot{x}<0,\ \dot{z}<0,\ \dot{u}<0$}\\
\specialrule{0em}{2pt}{2pt}
\hline
\specialrule{0em}{2pt}{2pt}
\tabincell{l}{\hspace{8mm}$Q_{110}$}&\tabincell{l}{$h_0<0,\ h_2<0,\ h_3<0,\ x>0, \ z>0,\ u<0$}&\tabincell{l}{$\dot{x}<0,\ \dot{z}<0,\ \dot{u}>0$}\\
\specialrule{0em}{2pt}{2pt}
 \toprule
\label{table11}
\end{tabular}
\end{table}

Similarly the bottom plane, the back-left plane, and the right-back plane of the subregion $Q_{12}$ are contained in the invariant planes $u=0$, $z=0$, and $x=0$, respectively. The back surface is contained in the surface $h_2=0$, the top-back surface and the front quadrilateral surface are contained in the Poincar\'e sphere, the upper and lower surfaces on the left side are contained in the surface $h_1=0$ and in the invariant plane $f_+ =0$, respectively. Then the orbits monotonically decrease along the positive direction of the $x$-axis in this subregion, and monotonically increase along the positive direction of the other two coordinate axes, that is, the orbits start at the finite equilibrium point $q_1$ and then go to the subregions $Q_{13}$ and $Q_{14}$.

The front surface of the subregion $Q_{13}$ is contained in the invariant plane $f_+(x,z,u)= 0$, the top surface is contained in the Poincar\'e sphere, the left and right planes of the back are contained in the invariant planes $z=0$ and $x=0$, respectively, and the middle surface of the back is contained in the surface $h_1=0$. Table \ref{table11} shows that the orbits in this subregion increase monotonically along the positive direction of the three coordinate axes, this means that the orbits in this subregion start from the finite equilibrium point $q_1$, and may also come from the adjacent subregion $Q_{12}$. They tend to the Poincar\'e sphere at infinity.

The left and right planes of the back of the subregion $Q_{14}$ are contained in the invariant planes $z=0$ and $x=0$ respectively, the back surface is contained in the surface $h_3=0$, the top and front surfaces are contained in the Poincar\'e sphere, and the left surface is contained in the surface $h_2=0$ and in the invariant surface $f_+(x,z,u)= 0$, the right side surface is contained in the surface $h_3=0$, and the bottom plane is contained in the invariant plane $u=0$. Note Table \ref{table11} states that the orbits in this subregion decrease monotonically along the positive direction of the $x$-axis and $z$-axis, and increase monotonically along the positive direction of the $u$-axis, so the orbits in this subregion start from the infinite equilibrium points in the Poincar\'e sphere or in the subregion $Q_{12}$, and then go to the subregion $Q_{16}$ covering the $u$-axis, because the entire $u$-axis is filled with the equilibrium points of system \eqref{eq31}.

The left side surface of the subregion $Q_{15}$ is contained in the surface $h_3=0$, the back plane and the bottom plane are contained in the invariant planes $x=0$ and $u=0$, respectively, and the front surface is contained in the Poincar\'e sphere. The front surface of the subregion $Q_{16}$ is contained in the surface $h_3=0$, and the left and right planes of the back are contained in the invariant planes $z=0$ and $x=0$, respectively, and the section line of the invariant planes $z=0$ and $x=0$ in this subregion is the $u$-axis. The subregions $Q_{15}$ and $Q_{16}$ are connected together through the origin of system \eqref{eq31}. According to Table \ref{table11} the orbits in the two subregions decrease monotonically along the positive directions of the three coordinate axes, indicating that the orbits in the subregion $Q_{15}$ start in the Poincar\'e sphere at infinity and after enter in the subregion $Q_{14}$ or tend to the origin $O$. The orbits in the subregion $Q_{16}$ come from the subregion $Q_{14}$, and finally go to the finite and infinite equilibrium points of the $u$-axis.

The right and left surfaces of the subregion $Q_{17}$ are contained in the surface $h_1=0$ and in the invariant surface $f_+(x,z,u)= 0$, the top plane is contained in the invariant plane $u=0$, the front and bottom surfaces are contained in the Poincar\'e sphere, and the back plane is contained in the invariant plane $z=0$. Table \ref{table11} shows that the orbits in this subregion increase monotonically along the positive direction of the $x$-axis and $z$-axis, and decrease monotonically along the positive direction of the $u$-axis, which indicate that the orbits actually originate from the finite equilibrium point $q_1$ and then run towards the equilibrium points at infinity on the Poincar\'e sphere.

The subregion $Q_{18}$ has the same composition as in $Q_{17}$ except that the left and right surfaces are contained in the surfaces $h_1=0$ and $h_2=0$, respectively. We find from Table \ref{table11} that the orbits monotonically increase along the positive direction of the $z$-axis and decrease monotonically along the positive directions of the other two coordinate axes. The orbits start at the finite equilibrium point $q_1$ in this subregion and finally enter in the subregion $Q_{19}$.

The subregion $Q_{19}$ also has the same composition as $Q_{17}$ except that the left and right surfaces are contained in the surfaces $h_2=0$ and $h_3=0$, respectively. Table \ref{table11} shows that the orbits monotonically decrease along the positive directions of the three coordinate axes. The orbits begin at the infinite equilibrium points on the Poincar\'e sphere or come from the subregion $Q_{18}$, and then tend to the equilibrium points on the $u$-axis.

In the subregion $Q_{110}$ the left and right surfaces are contained in the surface $h_3=0$ and in the invariant plane $x=0$, the top plane is contained in the invariant plane $u=0$, and the front and bottom surfaces are contained in the Poincar\'e sphere. It is noted from Table \ref{table11} that the orbits in this subregion decrease monotonically along the positive direction of the $x$-axis and $z$-axis, and increase monotonically along the positive direction of the $u$-axis, indicating that the orbits actually start at the equilibrium points on the Poincar\'e sphere at infinity, and eventually tend to the equilibrium points on the $u$-axis.

In summary the dynamics of the orbits in the ten subregions inside the region $Q_1$ studied above can be sketched in the following flow chart
$$\begin{tikzpicture}[->, thick]
 \node (Q12) at (0, 0) [] {$Q_{12}$};
  \node (Q18) at (-1.3, -1.1) [] {$Q_{18}$};
  \node (q1) at (-1.3, 0) [] {$q_1$};
   \node (Q17) at (-2.6, -1.1) [] {$Q_{17}$};
    \node (Q13) at (-1.3, 1.1) [] {$Q_{13}$};
      \node (PSQ13) at (-3.3, 1.1) [] {PS in $Q_{13}$};    
\node (Q11) at (-2.6, 0) [] {$Q_{11}$};
  \node (PSQ17) at (-4.6, -1.1) [] {PS in $Q_{17}$};   
  \node (PSQ11) at (-4.6, 0) [] {PS in $Q_{11}$};   
\node (Q14) at (1.42, 0) [] {$Q_{14}$};
 \node (Q19) at (0.12, -1.1) [] {$Q_{19}$};
   \node (PSQ19) at (0.12, -2.2) [] {PS in $Q_{19}$};   
  \node (PSQ14) at (1.42, 1.1) [] {PS in $Q_{14}$};   
\node (Q16) at (3.42, -1.1) [] {$Q_{16}$};
 \node (Q15) at (2.84, 0) [] {$Q_{15}$};
  \node (u) at (1.78, -1.1) [] {$u$-axis};
    \node (PSQ15) at (4.87, 0) [] {PS in $Q_{15}$.};  
      \node (O) at (2.84, 1.1) [] {$O$};
\draw (Q12) -- (Q14) ;
\draw (Q12) -- (Q13) ;
\draw (Q14) -- (Q16) ;
\draw (PSQ14) -- (Q14) ;
\draw (Q16) -- (u) ;
\draw (Q15) -- (Q14) ;
\draw (Q15) -- (O) ;
\draw (PSQ15) -- (Q15) ;
\draw (q1) -- (Q12) ;
\draw (q1) -- (Q13) ;
\draw (q1) -- (Q11) ;
\draw (q1) -- (Q17) ;
\draw (Q17) -- (PSQ17) ;
\draw (q1) -- (Q18) ;
\draw (Q18) -- (Q19) ;
\draw (Q19) -- (u) ;
\draw (PSQ19) -- (Q19) ;
\draw (Q11) -- (PSQ11) ;
\draw (Q13) -- (PSQ13) ;
\draw (Q12) -- (Q13) ;
\end{tikzpicture}$$

The above flow chart indicates that the orbits of system (\ref{eq31}) contained in the region $Q_1$ admit an $\alpha$-limit at the finite equilibrium point $q_1$. Moreover the orbits also have an $\alpha$-limit at the equilibrium points in the subregions $Q_{14}$, $Q_{15}$ and $Q_{19}$ when they are restricted to the Poincar\'e sphere at infinity. In addition the orbits have an $\omega$-limit at the equilibrium points located on the $u$-axis and at the infinite equilibrium points restricted to the subregions $Q_{11}$, $Q_{13}$ and $Q_{17}$ on the Poincar\'e sphere.

%section IV
\section{Case IV: $s=0, k=-1$}

In this case system (\ref{eq31}) has the same phase portraits on the three invariant planes $x = 0$, $z = 0$ and $u = 0$ as in sections 5.1.1-5.1.3. In addition the local phase portraits of the finite and infinite equilibrium points on the Poincar\'e sphere are consistent in sections 5.1.4 and 5.1.5 respectively. However the physical region of interest $f_-(x,z,u) = 0$ is no more an invariant surface.

Here we again divide the Poincar\'e ball into four regions restricted to the region $x^2-(u+z)^2\leq1$ as follows
\begin{equation*}
\begin{array}{rl}
T_1:\ x\geq0,\ z\geq0.\ \ \
T_2:\ x\leq0,\ z\geq0.\\
T_3:\ x\leq0,\ z\leq0.\ \ \
T_4:\ x\geq0,\ z\leq0.
\end{array}
\end{equation*}
We only need to examine the phase portrait of system (\ref{eq31}) in the region $T_1$ taking into account the symmetries $(x,z,u)\mapsto(-x,z,u)$, $(x,z,u)\mapsto(x,-z,-u)$ and $(x,z,u)\mapsto(-x,-z,-u)$. Then system (\ref{eq31}) has the same phase portrait on the boundaries of this region as in section 5.2 except on the non-invariant boundary surface $f_-(x,z,u) = 0$.

\subsection{Dynamics in the interior of the region $T_1$}

\begin{figure}[]
  \begin{minipage}{130mm}
\centering
\subfigure{\includegraphics[width=5.5cm]{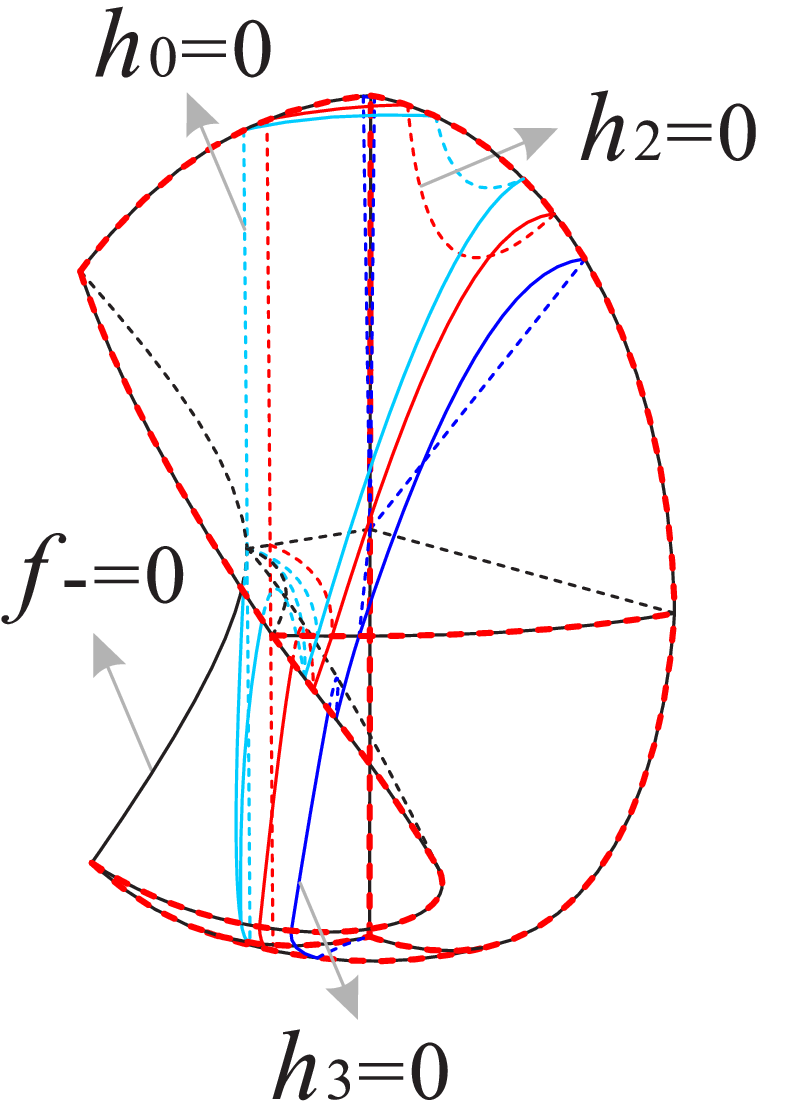}}
\subfigure{\includegraphics[width=2cm]{Fig14f15f16f17f21b23b24c26b30b32b33b34b_frame2_x_z_u.eps}}
    \caption{The surfaces $h_0$, $h_2$ and $h_3$ restricted to the surface $f_-(x,z,u)=0$ and the region $T_1$ of the Poincar\'e ball.}
  \label{fig33}
  \end{minipage}
 \end{figure}

\begin{figure}[]
  \begin{minipage}{130mm}
\centering
\subfigure{\includegraphics[width=10cm]{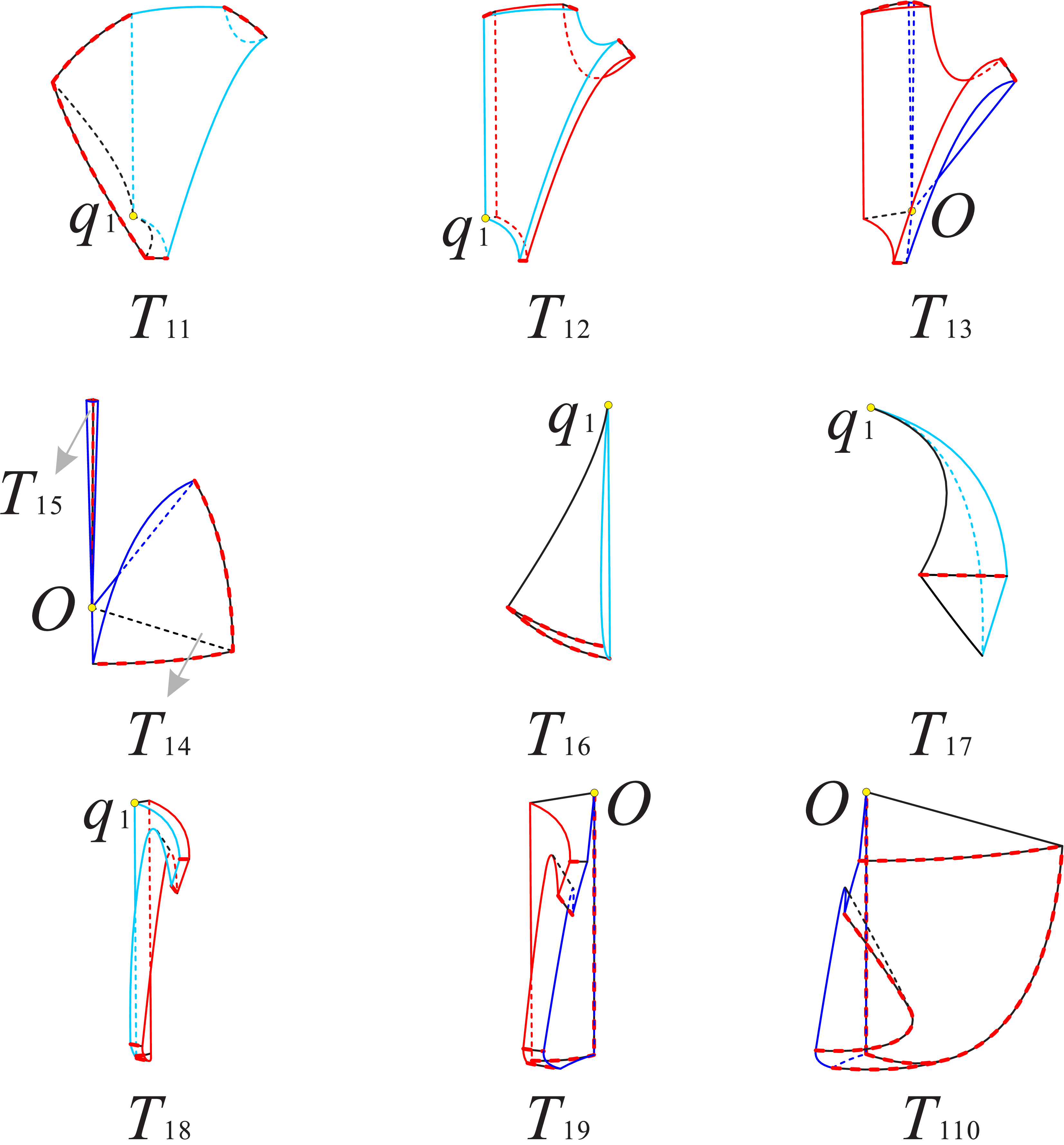}}\\
\subfigure{\includegraphics[width=2cm]{Fig14f15f16f17f21b23b24c26b30b32b33b34b_frame2_x_z_u.eps}}
    \caption{The ten subregions inside the region $T_1$ of the Poincar\'e ball.}
  \label{fig34}
  \end{minipage}
 \end{figure}

In order to examine the orbital behavior inside the region $T_1$, we note that the invariant planes $x=0$, $z=0$, $u=0$, and the surfaces $h_0$, $h_2$, $h_3$ as well as $f_-(x,z,u)=0$ divide the region into ten subregions $T_{11},\ T_{12},\ \cdots, T_{110}$, see Figures \ref{fig33}-\ref{fig34} for more details. The signs of the functions $h_0$, $h_2$ and $h_3$ in these subregion of $T_1$ are shown in Table \ref{table12}.

%Figure 33

%Figure 34

 %Table 12
\begin{table}[!htb]
\newcommand{\tabincell}[2]{\begin{tabular}{@{}#1@{}}#2\end{tabular}}
\centering
\caption{\label{opt}Signs of functions $h_0$, $h_2$ and $h_3$ in the subregion of $T_1$.}
\footnotesize
\rm
\centering
\begin{tabular}{@{}*{12}{l}}
\specialrule{0em}{2pt}{2pt}
 \toprule
\hspace{2mm}\textbf{Functions}&\textbf{Positive}&\textbf{Negative}\\
\specialrule{0em}{2pt}{2pt}
\toprule
\tabincell{l}{\hspace{8mm}$h_0$}&\tabincell{l}{$T_{11}, T_{12}, T_{16}, T_{17}$}&\tabincell{l}{$T_{13}, T_{14}, T_{15}, T_{18}, T_{19}, T_{110}$}\\
\specialrule{0em}{2pt}{2pt}
\hline
\specialrule{0em}{2pt}{2pt}
\tabincell{l}{\hspace{8mm}$h_2$}&\tabincell{l}{$T_{11}, T_{12}, T_{16}, T_{17}, T_{18}$}&\tabincell{l}{$T_{13}, T_{14}, T_{15}, T_{19}, T_{110}$}\\
\specialrule{0em}{2pt}{2pt}
\hline
\specialrule{0em}{2pt}{2pt}
\tabincell{l}{\hspace{8mm}$h_3$}&\tabincell{l}{$T_{11}, T_{12}, T_{13}, T_{16}, T_{17}, T_{18}, T_{19}$}&\tabincell{l}{$T_{14}, T_{15}, T_{110}$}\\
\specialrule{0em}{2pt}{2pt}
 \toprule
\label{table12}
\end{tabular}
\end{table}

The left and right surfaces of the subregion $T_{11}$ in Figure \ref{fig34} are contained in the surfaces $f_-(x,z,u)= 0$ and $h_0=0$, the left and right planes of the back are contained in the invariant planes $z=0$ and $x=0$, respectively, and the bottom plane is contained in the invariant plane $u=0$, the front surface is contained in the Poincar\'e sphere. Table \ref{table13} shows that the orbits in this subregion increase monotonically along the positive directions of the three coordinate axes, indicating that the orbits start at the finite equilibrium point $q_1$ or from the adjacent subregion $T_{12}$, and then go to the infinite equilibrium points on the Poincar\'e sphere.

The left and right surfaces of the subregion $T_{12}$ are contained in the surfaces $h_0=0$ and $h_2=0$, respectively, and the left back plane, right back plane, and bottom plane are contained in the invariant plane $z=0$, $x=0$, and $u= 0$ respectively, the top surface and the front surface are contained in the Poincar\'e sphere. The orbits increase monotonically along the positive direction of the three coordinate axes, indicating that the orbits start at the finite equilibrium point $q_1$ and finally enter the subregion $T_{11}$ and tend to the infinity equilibrium points on the Poincar\'e sphere.

The composition structure of the subregion $T_{13}$ is the same as $Q_{14}$ in Figure \ref{fig32} except that the left surface in this subregion is completely contained in the surface $h_2=0$. In addition the subregions $T_{14}$ and $T_{15}$ have the same structure as the subregions $Q_{15}$ and $Q_{16}$ in Figure \ref{fig32}, respectively. Moreover the dynamic behavior of the orbits in the subregions $T_{13}$, $T_{14}$, and $T_{15}$ is the same as that in the subregions $Q_{14}$, $Q_{15}$, and $Q_{16}$, respectively. That is, the orbits in the subregions $T_{13}$ and $T_{14}$ originate from their respective infinite equilibrium points on the Poincar\'e sphere, and then the orbits in $T_{13}$ enter the subregion $T_{15}$ and eventually run to the equilibrium points located on the $u$-axis, and the orbits in $T_{14}$ go to the origin $O$ or enter the subregion $T_{13}$.

In the subregion $T_{16}$ the front surface is contained in the surface $f_-(x,z,u)= 0$, the right surface is contained in the surface $h_0=0$, the back surface is contained in the invariant plane $z=0$, and the bottom surface is contained in the Poincar\'e sphere. Table \ref{table13} implies that the orbits increase monotonically along the positive direction of the $x$-axis and $z$-axis, and decrease monotonically along the positive direction of the $u$-axis, so that the orbits in this subregion start from the finite equilibrium point $q_1$ and finally run to the equilibrium points on the Poincar\'e shpere on the sphere at infinity, or enter in the outer space through the surface $f_-(x,z,u)= 0$.

For the subregion $T_{17}$ the left side surface is contained in the surface $f_-(x,z,u)= 0$, the right side surface is contained in the surface $h_0=0$, the top plane is contained in the invariant plane $u=0$, and the front surface is contained in the Poincar\'e sphere. However it can be followed from Table \ref{table13} that the dynamic behavior of the orbits in this subregion is consistent with that in the subregion $T_{16}$.

In the subregion $T_{18}$ the left side surface and the right side surface are contained in the surfaces $h_0=0$ and $h_2=0$, respectively, the top plane is contained in the invariant plane $u=0$, and the surfaces on the front part and the bottom part are contained in the Poincar\'e sphere, the middle part of the front surface is contained in the surface $f_-(x,z,u)= 0$. In this subregion the orbits monotonically decrease along the positive direction of the $x$-axis and $z$-axis, and increase monotonically along the positive direction of the third axis, which means that the orbits start from the finite equilibrium point $q_1$ and then enter the subregion $T_{19}$.

The left side surface and the right side surface in the subregion $T_{19}$ are contained in the surfaces $h_2=0$ and $h_3=0$, respectively, and the composition structure of the remaining part is the same as the corresponding part of the subregion $T_{18}$. The orbits in this subregion monotonically decrease along the positive direction of the three coordinate axes, implying that the orbits start at the infinity equilibrium points on the Poincar\'e sphere or come from the subregion $T_{18}$, and finally tend to the equilibrium points on the $u$-axis.

The front and back parts of the left surface of the subregion $T_{110}$ are contained in the surfaces $h_3=0$ and $f_-(x,z,u)= 0$, respectively, the right and top planes are contained in the invariant planes $x=0$ and $u=0$, respectively, and the front and bottom surfaces are contained in the Poincar\'e sphere. Note that Table \ref{table13} states that the orbits in this subregion decrease monotonically along the positive direction of the $x$-axis and $z$-axis, and increase monotonically along the positive direction of the other coordinate axis. Then the orbits in this subregion start at the infinite equilibrium points on the Poincar\'e sphere and eventually tend to the finite and infinite equilibrium points on the $u$-axis.

%table 13
\begin{table}[!htb]
\newcommand{\tabincell}[2]{\begin{tabular}{@{}#1@{}}#2\end{tabular}}
\centering
\caption{\label{opt}Dynamical behavior in the ten subregions.}
\footnotesize
\rm
\centering
\begin{tabular}{@{}*{12}{l}}
\specialrule{0em}{2pt}{2pt}
 \toprule
\hspace{2mm}\textbf{Subregions}&\textbf{Corresponding Region}&\textbf{Increase or decrease}\\
\specialrule{0em}{2pt}{2pt}
\toprule
\tabincell{l}{\hspace{8mm}$T_{11}$}&\tabincell{l}{$h_0>0,\ h_2>0,\ h_3>0,\ x>0,\ z>0,\ u>0$}&\tabincell{l}{$\dot{x}>0,\ \dot{z}>0,\ \dot{u}>0$}\\
\specialrule{0em}{2pt}{2pt}
\hline
\specialrule{0em}{2pt}{2pt}
\tabincell{l}{\hspace{8mm}$T_{12}$}&\tabincell{l}{$h_0>0,\ h_2>0,\ h_3>0,\ x>0,\ z>0,\ u>0$}&\tabincell{l}{$\dot{x}>0,\ \dot{z}>0,\ \dot{u}>0$}\\
\specialrule{0em}{2pt}{2pt}
\hline
\specialrule{0em}{2pt}{2pt}
\tabincell{l}{\hspace{8mm}$T_{13}$}&\tabincell{l}{$h_0<0,\ h_2<0,\ h_3>0,\ x>0,\ z>0,\ u>0$}&\tabincell{l}{$\dot{x}<0,\ \dot{z}<0,\ \dot{u}>0$}\\
\specialrule{0em}{2pt}{2pt}
\hline
\specialrule{0em}{2pt}{2pt}
\tabincell{l}{\hspace{8mm}$T_{14}$}&\tabincell{l}{$h_0<0,\ h_2<0,\ h_3<0,\ x>0,\ z>0,\ u>0$}&\tabincell{l}{$\dot{x}<0,\ \dot{z}<0,\ \dot{u}<0$}\\
\specialrule{0em}{2pt}{2pt}
\hline
\specialrule{0em}{2pt}{2pt}
\tabincell{l}{\hspace{8mm}$T_{15}$}&\tabincell{l}{$h_0<0,\ h_2<0,\ h_3<0,\ x>0,\ z>0,\ u>0$}&\tabincell{l}{$\dot{x}<0,\ \dot{z}<0,\ \dot{u}<0$}\\
\specialrule{0em}{2pt}{2pt}
\hline
\specialrule{0em}{2pt}{2pt}
\tabincell{l}{\hspace{8mm}$T_{16}$}&\tabincell{l}{$h_0>0,\ h_2>0,\ h_3>0,\ x>0,\ z>0,\ u<0$}&\tabincell{l}{$\dot{x}>0,\ \dot{z}>0,\ \dot{u}<0$}\\
\specialrule{0em}{2pt}{2pt}
\hline
\specialrule{0em}{2pt}{2pt}
\tabincell{l}{\hspace{8mm}$T_{17}$}&\tabincell{l}{$h_0>0,\ h_2>0,\ h_3>0,\ x>0,\ z>0,\ u<0$}&\tabincell{l}{$\dot{x} >0,\ \dot{z}>0,\ \dot{u}<0$}\\
\specialrule{0em}{2pt}{2pt}
\hline
\specialrule{0em}{2pt}{2pt}
\tabincell{l}{\hspace{8mm}$T_{18}$}&\tabincell{l}{$h_0<0,\ h_2>0,\ h_3>0,\ x>0,\ z>0,\ u<0$}&\tabincell{l}{$\dot{x} <0,\ \dot{z}>0,\ \dot{u}<0$}\\
\specialrule{0em}{2pt}{2pt}
\hline
\specialrule{0em}{2pt}{2pt}
\tabincell{l}{\hspace{8mm}$T_{19}$}&\tabincell{l}{$h_0<0,\ h_2<0,\ h_3>0,\ x>0,\ z>0,\ u<0$}&\tabincell{l}{$\dot{x}<0,\ \dot{z}<0,\ \dot{u}<0$}\\
\specialrule{0em}{2pt}{2pt}
\hline
\specialrule{0em}{2pt}{2pt}
\tabincell{l}{\hspace{8mm}$T_{110}$}&\tabincell{l}{$h_0<0,\ h_2<0,\ h_3<0,\ x>0,\ z>0,\ u<0$}&\tabincell{l}{$\dot{x}<0,\ \dot{z}<0,\ \dot{u}>0$}\\
\specialrule{0em}{2pt}{2pt}
 \toprule
\label{table13}
\end{tabular}
\end{table}

In short the dynamic behavior of the orbits of system (\ref{eq31}) in the region $T_1$ can be summarized as follows
$$\begin{tikzpicture}[->, thick]
  \node (u) at (0, 0) [] {$u$-axis}; 
    \node (T110) at (0, 1.1) [] {$T_{110}$};
    \node (PST110) at (2.07,1.1) [] {PS in $T_{110}$};  
  \node (T15) at (1.6, 0) [] {$T_{15}$}; 
    \node (T13) at (2.95, 0) [] {$T_{13}$};  
      \node (PST13) at (2.95, -1.1) [] {PS in $T_{13}$};  
      \node (T14) at (4.3, 0) [] {$T_{14}$};
      \node (O) at (4.3, -1.1) [] {$O$.};
        \node (PST14) at (4.3, 1.1) [] {PS in $T_{14}$};   
  \node (T19) at (-1.6, 0) [] {$T_{19}$};  
                 \node (PST19) at (-1.6, 1.1) [] {PS in $T_{19}$};   
    \node (T18) at (-2.95, 0) [] {$T_{18}$}; 
        \node (T17) at (-2.84, -1.1) [] {$T_{17}$};  
                 \node (PST17) at (-0.94, -1.1) [] {PS in $T_{17}$};  
      \node (q1) at (-4.2, 0) [] {$q_1$};
            \node (OS) at (-4.2, -1.1) [] {$OS$};
            \node (T12) at (-4.2, 1.1) [] {$T_{12}$}; 
               \node (PST12) at (-6.1, 1.1) [] {PS in $T_{12}$}; 
        \node (T11) at (-5.45, 0) [] {$T_{11}$};  
                \node (T16) at (-5.57, -1.1) [] {$T_{16}$};
                          \node (PST16) at (-7.47, -1.1) [] {PS in $T_{16}$}; 
          \node (PST11) at (-7.35, 0) [] {PS in $T_{11}$};
\draw (PST14) -- (T14);
\draw (T14) -- (T13);
\draw (T14) -- (O);
\draw (T13) -- (T15);
\draw (T15) -- (u);
\draw (T19) -- (u);
\draw (T18) -- (T19);
\draw (q1) -- (T18);
\draw (q1) -- (T11);
\draw (T11) -- (PST11);
\draw (T13) -- (PST13);
\draw (T110) -- (u);
\draw (PST110) -- (T110);
\draw (PST19) -- (T19);
\draw (q1) -- (T12);
\draw (q1) -- (T16);
\draw (q1) -- (T17);
\draw (T16) -- (PST16);
\draw (T17) -- (PST17);
\draw (T16) -- (OS);
\draw (T17) -- (OS);
\draw (T12) -- (T11);
\draw (T12) -- (PST12);
\end{tikzpicture}$$

The above flow chart in the region $T_1$ indicates that the orbits of system (\ref{eq31}) have an $\alpha$-limit at the finite equilibrium point $q_1$, and at the infinite equilibrium points located on the Poincar\'e sphere restricted to the subregions $T_{13}$, $T_{14}$, $T_{19}$ and $T_{110}$. Furthermore the orbits have an $\omega$-limit at the equilibrium points on the $u$-axis, as well as the equilibrium points at infinity of the Poincar\'e sphere in the subregions $T_{11}$, $T_{12}$, $T_{16}$ and $T_{17}$.

\section{Conclusions}

By taking advantage of the fact that the cosmological equations (\ref{1}) remain unchanged under the symmetry $(x, z, u)\mapsto(x, -z, -u)$when $s\neq0$, and it remains unchanged under the additional two symmetries $(x, z, u)\mapsto(-x, -z, -u)$, $(x, z, u)\mapsto(-x, z, u)$ when $s=0$. Then for a wide range of $s$ in the present paper we completely describe the global phase portrait of Ho\v{r}ava-Lifshitz cosmology in the non-flat universe in the case of non-zero cosmological constant, all of these are located in the physical region of interest $G$ with the invariant boundary surface $f_+(x, z, u) = 0$ and non-invariant boundary surface $f_-(x, z, u) = 0$, respectively.

For the case $s=0$ the phase portrait of system (\ref{eq31}) restricted to the region $G$ shows that the orbits ultimately move to the finite and infinite equilibrium points located on the $u$-axis or tend to the infinite equilibrium points on the Poincar\'e sphere in the subregions $Q_{11}$, $Q_{13}$ and $Q_{17}$ when the universe is closed. Moreover the orbits of system (\ref{eq31}) finally go to the equilibrium points that lie on the $u$-axis or shift to the infinite equilibrium points restricted to the subregions $T_{11}$, $T_{12}$, $T_{16}$ and $T_{17}$ on the Poincar\'e sphere when the universe is open. Furthermore we apply the aforementioned symmetries of system (\ref{eq31}) to perform simple calculations and find that the unstable equilibrium points $q_1$ and $q_2$ correspond to the universe ruled by dark matter.

For the case $s\neq0$ in addition to the initial conditions on the invariant planes $z = 0$ and $u=0$, the phase portrait shows that the orbital evolution of system (\ref{1}) in region $G$ eventually tends to the equilibrium points at infinity, which are restricted to the subregions $R_{11}$, $R_{19}$, $R_{110}$ and $R_{111}$, $R_{21}$ and $R_{29}$ on the Poincar\'e sphere when the universe is closed. For an open universe the orbits of system (\ref{1}) will go to the infinite equilibrium points restricted to the subregions $S_{11}$, $S_{19}$, $S_{21}$, $S_{22}$, $S_{27}$, $S_{29}$ and $S_{210}$ on the Poincar\'e sphere.

For the studied non-flat universe due to some simple calculation combined with the analysis of the phase portrait of system (\ref{1}) in the previous sections, we conclude from the perspective of cosmology that unstable or non-hyperbolic finite equilibrium points $p_1$, $p_2$, $p_3$, $p_4$ and $p_5$ correspond to the universe dominated by dark matter. We note that there are two finite equilibrium points $P_{17}$ and $P_{18}$, that were considered as non-physical points in \cite{Leon2009}-\cite{Leon2012}, because the value of their corresponding dark matter equation-of-state parameter $\omega_M$ is $2$, but we found that this value is $1/3$ and with the notation of our paper corresponds to the equilibrium points $p_3$ and $p_4$, and of course $1/3$ satisfies the physical range $(0,1)$. Furthermore the previous flow charts show that the orbits of the cosmological model in the region $G$ tend to the equilibrium points at infinity, which can be the late-time state of the universe. Moreover the finite equilibrium point $p_5$ can also be the late-time state of the universe when the initial conditions are on the invariant boundary plane $u=0$. Besides the orbits will spend a finite lapse of time near the finite equilibrium point $p_4$ on the invariant plane $u = 0$ before reaching the late-time state $p_5$ or the infinite equilibrium points on the Poincar\'e sphere.

Based on the Ho\v{r}ava-Lifshitz gravity in the non-flat FLRW spacetime with $\Lambda\neq0$, equations (\ref{6}) shows that the Hubble parameter $H$ tends to $0$ in the forward time in this cosmological model. For the late-time state $p_5$ of the universe on the invariant plane $u=0$, the equations (\ref {6}), (\ref {7}) and $ H = \dot a(t) / a(t) $ indicate that the Hubble parameter $H$ is an exponential function, and that the scale factor $a(t)$ of the expanding universe is a double exponential function with respect to the time $t$, which expands much more quickly than an usual exponential function.

 \section*{Appendix A}

We recall the results on normally hyperbolic submanifold according to the monograph of Hirsch et al. \cite{Hirsch}.\\
\textbf{Definition 1}. It is assumed that $\phi_t$ is a smooth flow on a manifold $M$ and $C$ is a submanifold of $M$, where $C$ is completely composed of the equilibrium points of the flow. If the tangent bundle of $M$ on $C$ is divided into three invariant subbundles $TC$, $E^s$, $E^u$ under $\phi_t$ satisfying the conditions \\
\indent (A1) $d \phi_t$ contracts $E^s$ exponentially,\\
\indent (A2) $d \phi_t$ expands $E^u$ exponentially,\\
\indent (A3) $TC = $ tangent bundle of $C$,\\
then $C$ is called \textit{normally hyperbolic submanifold}.

For normally hyperbolic submanifolds, one usually have smooth stable and unstable manifolds as well as the permanence of these invariant manifolds under small perturbations. To be more precise, we present the following theorem.

\textbf{Theorem 1}. If $C$ is a normally hyperbolic submanifold consisting of equilibrium points for a smooth flow $\phi_t$, then there are smooth stable and unstable manifolds, which tangent along $C$ to $E^s\oplus TC$ and $E^u\oplus TC$ respectively. In addition, the submanifold $C$ as well as the stable and unstable manifolds are persist under small perturbations of the flow.

\section*{Acknowledgments}

The first author gratefully acknowledges the support of the National Natural Science Foundation of China (NSFC) through grant Nos. 12172322, 11672259, and the China Scholarship Council (CSC) through grant No. 201908320086.

The second author gratefully acknowledges the support of the Ministerio de Econom$\acute{\i}$a, Industria y Competitividad, Agencia Estatal de Investigaci\'on grants MTM2016-77278-P (FEDER), the Ag$\grave{\text{e}}$ncia de Gesti\'o d'Ajuts Universitaris i de Recerca grant 2017SGR1617, and the H2020 European Research Council grant MSCA-RISE-2017-777911.

\end{document}